\documentclass[12pt]{article}
\usepackage{amsmath,amssymb,amsfonts,color,graphicx,cite,color,psfrag}
\input paperdef

\graphicspath{{figs/}}

\evensidemargin \oddsidemargin
\allowdisplaybreaks

\begin{document}
\thispagestyle{empty}

\def\thefootnote{\fnsymbol{footnote}}

\begin{flushright}
DCPT/08/40\\ 
IPPP/08/20 \\
arXiv:0805.2359 [hep-ph]
\end{flushright}

\vspace{1cm}

\begin{center}
{\large\sc {\bf \boldmath{$B$}-Physics Observables and 
Electroweak Precision Data\\[.5em]
in the CMSSM, mGMSB and mAMSB}}

\vspace{1cm}

{\sc
S.~Heinemeyer$^1$%
\footnote{email: Sven.Heinemeyer@cern.ch}%
, X.~Miao$^{2}$%
\footnote{email: miao@physics.arizona.edu}%
, S.~Su$^{2}$%
\footnote{email: shufang@physics.arizona.edu}%
~and G.~Weiglein$^{3}$}%
\footnote{email: Georg.Weiglein@durham.ac.uk}

\vspace*{1cm}

{\sl
$^1$Instituto de Fisica de Cantabria (CSIC-UC), 
Santander,  Spain 

\vspace{0.2cm}

$^2$Dept.\ of Physics, University of Arizona, Tucson, AZ 85721, USA

\vspace{0.2cm}

$^3$IPPP, University of Durham, Durham DH1~3LE, UK
}

\end{center}

\vspace{1cm}

\begin{abstract}
\noindent
We explore electroweak precision observables (EWPO) and $B$-physics
observables (BPO) in the CMSSM, the mGMSB and the mAMSB.
We perform a $\chi^2$ analysis based on the combination of current EWPO
and BPO data. 
For the first time this allows the comparison of the mGMSB and mAMSB in
terms of EWPO and BPO with the CMSSM.
We find that relatively low mass scales in all three scenarios are
favored. However, the current data from EWPO and BPO can hardly exclude
any parameters at 
the level of $\De\chi^2 = 9$. Remarkably the mAMSB scenario, despite
having one free GUT scale parameter less than the other two scenarios,
has a somewhat lower total minimum $\chi^2$.  
We present predictions for the lightest Higgs boson mass, based on the
$\chi^2$ analysis of current data, where
relatively good compatibility with the bounds from Higgs searches at LEP
is found. We also present the predictions for other
Higgs sector parameters and SUSY mass scales, allowing to compare the
reach of the LHC and the ILC in the three scenarios.
We furthermore explore the future sensitivities of the EWPO and BPO for the
current best-fit results and for a hypothetical point with somewhat
higher mass scales that results in a
similar Higgs and SUSY spectrum in the three scenarios. We find that 
the future improvement of the accuracy of the EWPO and BPO will lead to
a significant gain in the indirect parameter determination.
The improvement is similar in the CMSSM, mGMSB and mAMSB and 
will yield constraints to the parameter space even for
heavy Higgs and SUSY mass scales.
\end{abstract}

\def\thefootnote{\arabic{footnote}}
\setcounter{page}{0}
\setcounter{footnote}{0}

\newpage


\section {Introduction}

The dimensionality of the parameter space of the minimal supersymmetric
extension of the Standard Model (MSSM)~\cite{susy,susy2} is so high that
phenomenological analyses often make simplifying assumptions 
that reduce drastically the number of parameters. 
One assumption that is frequently employed is
that (at least some of) the soft SUSY-breaking parameters are universal
at some high input scale, before renormalization. 
One model based on this simplification is the 
constrained MSSM (CMSSM), in which all the soft SUSY-breaking scalar
masses $m_0$ are assumed to be universal at the GUT scale, as are the
soft SUSY-breaking gaugino masses $m_{1/2}$ and trilinear couplings
$A_0$. The assumption that squarks and sleptons with the same gauge
quantum numbers have the same masses is motivated by the absence of
identified supersymmetric contributions to flavor-changing neutral
interactions and rare decays (see \citere{hfag} and references therein).
Universality between squarks and 
sleptons with different gauge interactions may be motivated by some GUT
scenarios~\cite{GUTs}. 
Other ``simplified'' versions of the MSSM that are based on (some)
unification at a higher scale are (minimal) Gauge mediated
SUSY-breaking (mGMSB)~\cite{oldGMSB,newGMSB,GR-GMSB} and
(minimal) Anomaly mediated SUSY-breaking
(mAMSB)~\cite{lr,giudice,wells}. 

One approach to analyze the reduced
parameter spaces of the CMSSM, mGMSB, mAMSB or other GUT-based models
is a combined $\chi^2$~analysis of electroweak precision
observables (EWPO) and of $B$-physics observables (BPO).
Those analyses have yet been restricted to the CMSSM or the
non-universal Higgs mass (NUHM)
model~\cite{ehow3,ehow4,ehoww,ehhow,ehow5,other,LSPlargeTB,masterfit} 
(see also \citeres{fut,nazilla2,nmssmBPO,MSSM11fit}).
In these analyses also 
the cold dark matter density constraint imposed by WMAP and other
cosmological data~\cite{WMAP} has been taken into account. In this case
the lightest SUSY particle (LSP), assumed to be the lightest neutralino,
is required to give rise to the correct amount of cold dark matter (CDM). 

The aim of this paper is to perform a $\chi^2$~analysis to compare the
predictions of the CMSSM, mGMSB and mAMSB. The mechanisms to
fulfill the CDM constraints are less clear in mGMSB and mAMSB as
compared to the CMSSM. In order to treat the three soft SUSY-breaking
scenarios on the same footing,  we do not impose the CDM constraint in
our analysis and scan over the full parameter space of the three models. 
Concerning the impact of CDM constraints, it should be kept in mind that small
modifications of the physics scenario that concern neither the theory
basis nor the collider phenomenology could have a strong impact on the
CDM derived bounds. If the amount of CDM appears to be too
small, other DM candidates can provide the necessary amount to reach the
measured density (see also \citere{nazilla} for a recent analysis). 
If, on the other hand, the CDM density appears to be too
large, a small amount of $R$-parity violation~\cite{herbi}, not affecting the
collider phenomenology, could remove the CDM bound completely. Other
possibilities not invoking $R$-parity violation are 
``thermal inflation''~\cite{thermalinf} or ``late-time entropy
injection''~\cite{latetimeentropy}. They could offer a mechanism
for bringing a high CDM density into agreement with the WMAP
measurements. Applying the WMAP constraints always assumes ``standard
cosmology''. Therefore the choice of not imposing the CDM constraints,
as we do, can be motivated in the wider class of models under
investigation here. 
For the CMSSM we have checked that including the CDM constraint previous
results could be reproduced.

The set of EWPO included in our
analysis is the $W$~boson mass $\MW$,  the effective leptonic weak
mixing angle $\sweff$, the anomalous magnetic moment of the  muon
$(g-2)_\mu$, and the mass of the lightest $\cp$-even MSSM Higgs boson mass
$\Mh$. In addition, we also include two BPO:  the branching ratios
$\br(b \to s \ga)$ and $\br(B_s \to \mu^+ \mu^-)$. Other BPO such as
$\br(B_u \to \tau \nu_\tau)$ and the $B_s$ mass mixing parameter
$\De M_{B_s}$ have shown to possess only a low sensitivity with the
current precision in this kind of $\chi^2$~analysis~\cite{ehoww}. 
For the evaluation of the BPO we assume minimal flavor violation (MFV)
at the electroweak scale. Non-minimal flavor violation (NMFV) effects
can be induced by RGE running from the high scale, see
e.g.~\citere{nmfv}, that may amount to $\sim 10\%$ of the SUSY
corrections. These additional contributions are neglected throughout the
paper. 
For each observable, we construct the $\chi^2$~function
including both theoretical and experimental systematic uncertainties, as
well as statistical errors. 
Our analysis should be seen as an exploratory study, with the main goal
to compare the three soft SUSY-breaking scenarios. A more elaborate
investigation using more precision data and a refined $\chi^2$ analysis,
see e.g.\ \citere{masterfit}, can be performed in a later stage and is
beyond the scope of this paper. 

The rest of the paper is organized as follows.
We first briefly review the three soft SUSY-breaking scenarios and the
investigated parameter space. 
In \refse{sec:expdata} we shortly describe the current status of the
EWPO and BPO that we use, our treatment of the available theoretical
calculations and their uncertainties, as well as their present
experimental values. 
The analysis within the three soft SUSY-breaking scenarios using current
experimental data can be found in \refse{sec:asbsana}. 
In a final step we assume an improvement of the various EWPO and BPO
accuracies from future experimental data and theory calculations and
analyze in \refse{sec:future} the improvement in the parameter
determination. 
The conclusions can be found in \refse{sec:conclusions}.


\section{The soft SUSY-breaking scenarios}
\label{sec:asbs}

The fact that no SUSY partners of the SM particles have so far been
observed means that low-energy SUSY cannot be realized as an unbroken
symmetry in nature, and SUSY models thus have to incorporate
additional Supersymmetry breaking interactions. 
This is achieved by adding to the Lagrangian (defined by the 
${\rm SU(3)}_C\times {\rm SU(2)}_L \times  {\rm U(1)}_Y$ gauge
symmetry and the superpotential $W$)
some further interaction terms that respect the gauge symmetry but break 
Supersymmetry (softly, i.e.\ no quadratic divergences appear), so
called ``soft SUSY-breaking'' (SSB) terms.
Assuming that the $R$-parity symmetry~\cite{herbi}
is conserved, which we do in this paper
for all SUSY breaking scenarios, reduces the amount of 
new soft terms allowed in the Lagrangian.
Choosing a particular soft SUSY-breaking pattern allows further
reduction of the number of free parameters and the construction
of predictive models. The three most prominent scenarios for such
models are

\begin{itemize}

\item 
{\bf CMSSM} (constrained Minimal Supersymmetric Standard 
Model)~\cite{Hall,mSUGRArev}:\\
Apart from the SM parameters (for the experimental values of the SM 
input parameters we use~\citere{pdg}), 
4~parameters and a sign are required to define the CMSSM scenario:
\BE
\{\; m_0\;,\;m_{1/2}\;,\;A_0\;,\;\tb \;,\; {\rm sign}(\mu)\; \} \;.
\label{mSUGRAparams}
\EE
While $m_0$, $m_{1/2}$ and $A_0$ define the scalar and fermionic
masses and the trilinear couplings at the GUT scale 
($\sim 10^{16} \gev$), $\tb$ (the ratio of the two vacuum expectation
values) and the sign($\mu$) ($\mu$ is the supersymmetric Higgs mass
parameter) are defined at the low-energy scale. 
For our numerical analyses, see \refses{sec:asbsana}~and~\ref{sec:future}, 
we have scanned 
over the following parameter space
\BEA
50~{\rm GeV} \le &m_0& \le 2~{\rm TeV} \;, \nonumber \\
50~{\rm GeV} \le  &m_{1/2}& \le 2~{\rm TeV} \;, \nonumber \\
-3~{\rm TeV} \le &A_0& \le 3~{\rm TeV} \;, \non \\
1.5 \le &\tan\beta & \le 60 \;, \non \\
 &{\rm sign}\, \mu& = \pm 1 .
\label{msugraparam}
\EEA

\medskip
\item 
{\bf mGMSB} (minimal Gauge Mediated SUSY-Breaking)~\cite{GR-GMSB}:\\
A very promising alternative to the CMSSM is based on the hypothesis
that the soft SUSY-breaking occurs at relatively low energy scales 
and is mediated mainly by gauge interactions through the so-called
``messenger sector''~\cite{oldGMSB,newGMSB,GR-GMSB,GMSBrecent,GMSBrecent2}. 
Also in this scenario, the low-energy parameters depend on
4~parameters and a sign,
\BE
\{\; M_{\rm mess}, \; N_{\rm mess}, \; \Lambda, \; 
     \tb, \; {\rm sign}(\mu) \; \} \;,
\label{eq:pars}
\EE
where $M_{\rm mess}$ is the overall messenger mass scale; 
$N_{\rm mess}$ is a number called the 
messenger index, parameterizing the structure of the messenger
sector; $\Lambda$ is the universal soft SUSY-breaking mass scale felt by the
low-energy sector.
The phenomenology of mGMSB is characterized by the presence of a very
light gravitino  $\tilde{G}$ with mass given by  
$m_{3/2} = m_{\tilde{G}} = \frac{F}{\sqrt{3}M'_P} \simeq 
\left(\frac{\sqrt{F}}{100 \tev}\right)^2 2.37 \; {\rm eV}$~\cite{Fayet},  
where $\sqrt{F} (\sim M_{\rm mess})$ is the fundamental scale of SSB and 
$M'_P = 2.44 \times 10^{18} \gev$ is the reduced Planck mass.
Since $\sqrt{F}$ is typically of order 100 TeV, the $\tilde{G}$ is always the 
LSP in these theories. 
The numerical analysis in \refses{sec:asbsana}~and~\ref{sec:future} is
based on the following scatter ranges:
\BEA
10^4 \gev \le &\La& \le 2\,\times\,10^5 \gev \;, \non \\
1.01\,\La \le &M_{\rm mess}& \le 10^5\,\La \;, \nonumber \\
1 \le  &N_{\rm mess}& \le 8 \;, \nonumber \\
1.5 \le &\tan\beta & \le 60 \;, \non \\
 &{\rm sign}\, \mu& = \pm 1 .
\label{gmsbparam}
\EEA
Values of $N_{\rm mess}$ larger than~$\sim 8$ result in problems with
perturbativity of the gauge interactions at very high scales~\cite{GR-GMSB}. 

\item
{\bf mAMSB} (minimal Anomaly Mediated SUSY-Breaking)~\cite{lr,giudice,wells}:\\
In this model, SUSY breaking happens on a separate brane and is  
communicated to the visible world via the super-Weyl anomaly. 
The particle spectrum is determined by 3~parameters and a sign:  
\BE
\{m_{\rm aux},\ m_{0},\ \tb,\ {\rm sign}(\mu) \} .
\label{amsbparams}
\EE
The overall scale of SUSY particle masses is set by $m_{\rm aux}$, 
which is the vacuum expectation value 
of the auxiliary field in the supergravity multiplet.
$m_0$ is introduced as a phenomenological parameter 
to avoid negative slepton mass squares, for other
approaches to this problem see \citeres{lr,negative,clm,kss,jjw}.
The scatter parameter space for the numerical analysis in
\refses{sec:asbsana}~and~\ref{sec:future} is chosen to be
\BEA
20 \tev \le &m_{\rm aux}& \le 200 \tev , \non \\
0 \le &m_0& \le 2 \tev , \non \\
1.5 \le &\tb& \le 60 , \non \\
 &{\rm sign}\, \mu& = \pm 1 . 
\label{amsbparam}
\EEA

\end{itemize}
The upper bound on $m_0$ has been chosen in agreement with the CMSSM
scenario. Concerning $m_{\rm aux}$, being linked to the SUSY-breaking
scale, we have chosen the upper bound of $200 \tev$, which should be
sufficient to cover the essential 
features of the low-energy spectrum of mAMSB.

The low-energy spectra for all soft SUSY-breaking scenarios have been
evaluated with the program {\tt SoftSUSY}~\cite{softsusy} 
(version {\tt 2.0}), taking into
account the experimental constraints from SUSY particle
searches~\cite{pdg}. 
The parameter ranges have been sampled by a random scan over the
four- (three-)dimensional space of the free parameters in the CMSSM and
mGMSB (in mAMSB). The sign of $\mu$ has been treated as another free
parameter. For each soft SUSY-breaking scenario about $\sim 10^5$
random points have been generated. This large number ensures that all
regions of the four- (three-)dimensional hypercube of free parameters are
reached.


\section{The Precision Observables}
\label{sec:expdata}

The considered data set includes four EWPO~\cite{PomssmRep}:
the mass of the $W$~boson, $\MW$, the effective leptonic
weak mixing angle, $\sweff$, 
the anomalous magnetic moment of the muon, $(g-2)_\mu$, and the mass of
the lightest $\cp$-even MSSM Higgs boson, $\Mh$.
Another EWPO, the total $Z$~boson width, $\Ga_Z$, has shown to have
little sensitivity to SUSY corrections~\cite{ehoww,ZOpope}. 
In addition, we include two BPO:
the branching ratios $\br(b \to s \ga)$ and $\br(B_s \to \mu^+ \mu^-)$.
Other BPO such as $\br(B_u \to \tau \nu_\tau)$ and the $B_s$~mass-mixing
parameter 
$\De M_{B_s}$ with their current experimental and theoretical precision
have only a small sensitivity to SUSY corrections~\cite{ehoww}.

In this Section we start our analysis by recalling the current
precisions of the experimental results and the theoretical predictions
for all these observables. In the following, we refer to the theoretical
uncertainties from unknown 
higher-order corrections as `intrinsic' theoretical uncertainties and
to the uncertainties induced by the experimental errors of the SM input
parameters as `parametric' theoretical uncertainties.
We do not discuss here the theoretical uncertainties in the
renormalization-group running between the high-scale input parameters
and the weak scale. 
At present, these uncertainties are less important than the experimental
and theoretical uncertainties in the precision observables.

Assuming that the six observables listed above are
uncorrelated, a $\chi^2$ fit 
has been performed with

\BE
\chi^2 \equiv \sum_{n=1}^{4} \KL
              \frac{R_n^{\rm exp} - R_n^{\rm theo}}{\si_n} \KR^2
                                               + \chi^2_{\Mh}
                                               + \chi^2_{B_s}.
\label{eq:chi2}
\EE
Here $R_n^{\rm exp}$ denotes the experimental central value of the
$n$th observable ($\MW$, $\sweff$, \mbox{$(g-2)_\mu$} and
$\br(b \to s \ga)$),
$R_n^{\rm theo}$ is the corresponding MSSM prediction and $\si_n$
denotes the combined error, as specified below. 
$\chi^2_{\Mh}$ and $\chi^2_{B_s}$ denote the $\chi^2$ contribution
coming from the experimental limits 
on the lightest $\cp$-even MSSM Higgs boson mass and on 
$\br(B_s \to \mu^+\mu^-)$, 
respectively, which are also described below.
In \refse{sec:future} we assume a future measurement of $\Mh$ and use
$\chi^2_{\Mh} = ((\Mh^{\rm exp} - \Mh^{\rm theo})/\si_{\Mh})^2$. 

We also list below the parametric uncertainties in the predictions on the
observables induced by the experimental uncertainties of all relevant SM
input parameters. 
These parametric uncertainties are then added to the other errors
(intrinsic and experimental) of the observables as described in
the text below.
A particularly important input parameter in this respect is the
top-quark mass. 
We evaluate the SUSY spectrum and the observables for
each data point for the nominal value,
$\mt = 171.4 \gev$~\cite{mt1714} but include the error induced by the
experimental uncertainty of 
$\de\mt^{\rm exp} = 2.1 \gev$.%
\footnote{Using the most recent experimental value, 
$\mt = 172.6 \gev$, including the experimental error of 
$\de\mt^{\rm exp} = 1.4 \gev$~\cite{mt1726}, see below, 
would have a relatively small impact on our analysis, see also the
discussion at the end of \refse{sec:analow}.}%
%


\subsection{The $W$~Boson Mass}
\label{subsec:mw}

The $W$~boson mass can be evaluated from
\BE
\MW^2 \KL 1 - \frac{\MW^2}{\MZ^2}\right) = 
\frac{\pi \al}{\sqrt{2} \GF} \left(1 + \De r\KR ,
\label{eq:delr}
\EE
where $\al$ is the fine structure constant and $\GF$ the Fermi constant.
The radiative corrections are summarized 
in the quantity $\De r$~\cite{sirlin}.
The prediction for $\MW$ within the SM
or the MSSM is obtained by evaluating $\De r$ in these models and
solving \refeq{eq:delr} for $\MW$. 

We include the complete \onel\ result in the
MSSM~\cite{deltarMSSM1lA,deltarMSSM1lB} as well as higher-order QCD
corrections of SM type that are of 
\order{\al\als}~\cite{drSMgfals,deltarSMgfals}
and \order{\al\als^2}~\cite{drSMgfals2,drSMgfals2LF}. Furthermore, we
incorporate 
supersymmetric corrections of \order{\al\als}~\cite{dr2lA} and of
\order{\al_t^2}~\cite{drMSSMal2B,drMSSMal2} to the quantity 
$\De\rho$, which involves the leading universal corrections induced by
the mass splitting between fields in an isospin doublet~\cite{rho}.%
\footnote{
A recent re-evaluation of $\MW$~\cite{MWpope} shows good agreement with
the values used here. 
}%

The remaining intrinsic theoretical uncertainty in the prediction for
$\MW$ within the MSSM is still significantly larger than in the SM. For
typical parameters (based on \citere{drMSSMal2}) we estimate the
current and future intrinsic uncertainties to
\BE
\De\MW^{\rm intr,current} \lsim 10 \mev~, 
\quad
\De\MW^{\rm intr,future} = 2 \mev~, 
\EE
depending on the mass scale of the supersymmetric particles.
The parametric uncertainties are dominated by the experimental error of
the top-quark mass
and the hadronic contribution to the shift in the
fine structure constant. Their current errors induce the following
parametric uncertainties~\cite{ehoww,PomssmRep}
\BEA
\de\mt^{\rm current} = 2.1 \gev &\Rightarrow&
\De\MW^{{\rm para},\mt, {\rm current}} \approx 13 \mev,  \\[.3em]
\de(\De\al_{\rm had}^{\rm current}) = 35 \times 10^{-5} &\Rightarrow&
\De\MW^{{\rm para},\De\al_{\rm had}, {\rm current}} \approx 6.3 \mev~.
\EEA
At the ILC, the top-quark mass will be measured with an accuracy of
about 100~MeV~\cite{mtdet1,mtdet2}. The parametric uncertainties induced
by the 
future experimental errors of $\mt$ and $\De\al_{\rm had}$~\cite{fredl}
will then be~\cite{deltamt}
\BEA
\de\mt^{\rm future} = 0.1 \gev &\Rightarrow&
\De\MW^{{\rm para},\mt, {\rm future}} \approx 1 \mev,  \\[.3em]
\de(\De\al_{\rm had}^{\rm future}) = 5 \times 10^{-5} &\Rightarrow&
\De\MW^{{\rm para},\De\al_{\rm had}, {\rm future}} \approx 1 \mev .
\EEA
The present experimental value of $\MW$ 
is~\cite{lepewwg,LEPEWWG,TEVEWWG,MWcdf,MWworld}, see also \citere{ssdd}.
\BE
\MW^{\rm exp,current} = 80.398 \pm 0.025 \gev.
\label{mwexp}
\EE
With the GigaZ option of the ILC (i.e.\ high-luminosity running at the
$Z$~resonance and the $WW$ threshold) the $W$-boson mass will be
determined with an accuracy of about~\cite{mwgigaz,blueband}
\BE
\de\MW^{\rm exp,future} = 7 \mev .
\label{mwexpfuture}
\EE
We add the experimental and theoretical errors for $\MW$ (for the current
situation as well as for the future estimates) 
in quadrature in our analysis.

\begin{figure}[htb!]
\begin{center}
\includegraphics[width=.7\textwidth,height=10cm]{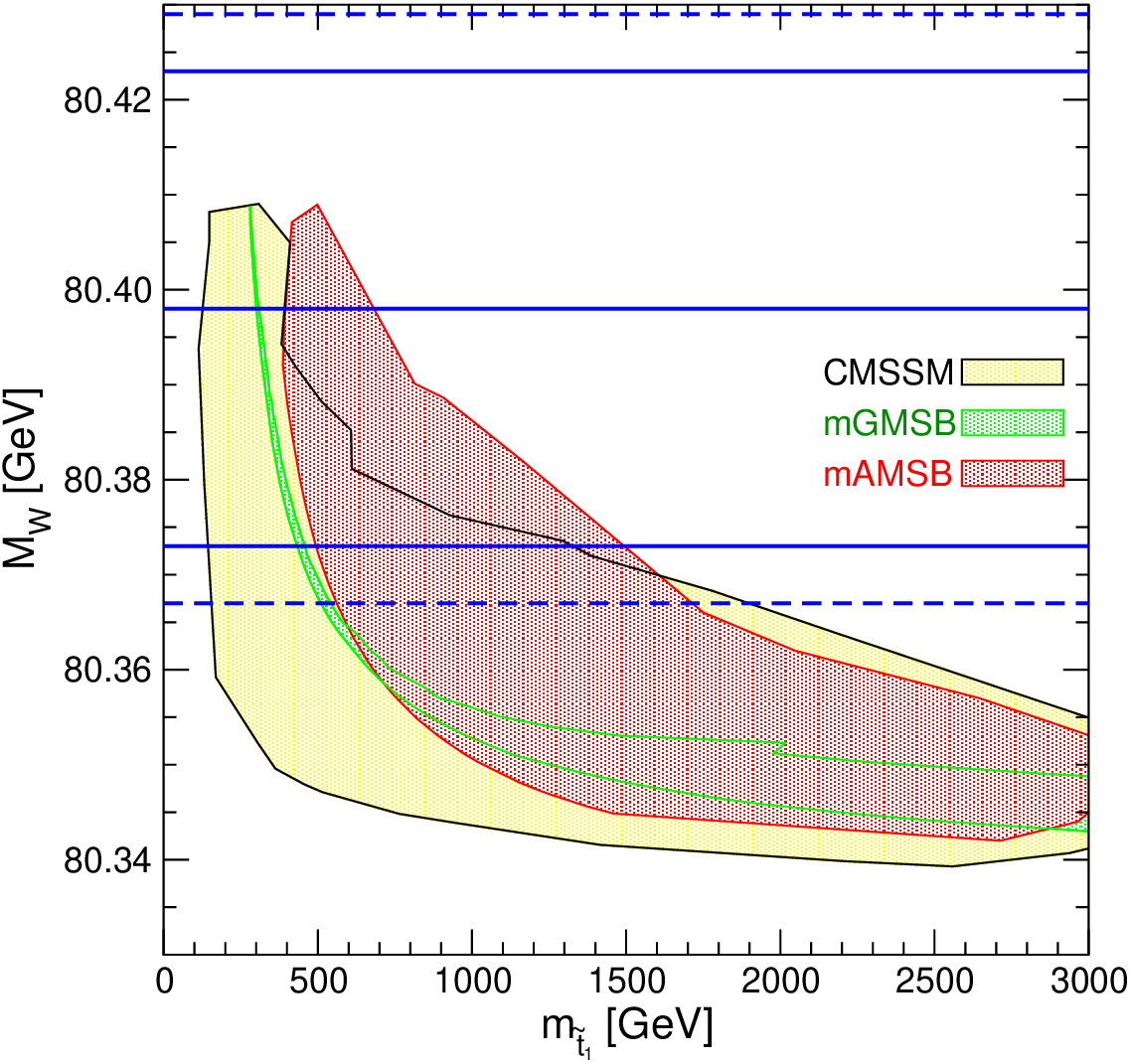}
\vspace{-1.0em}
\caption{%
The predictions for $\MW$ as obtained from the parameter scan are shown
as a function of $\mste$ for the three 
soft SUSY-breaking scenarios for $\mu > 0$. 
The top quark mass has been fixed to 
$\mt = 171.4 \gev$. The solid (dashed) lines indicate the currently
allowed $1\,\si$ interval from the experimental uncertainty (including
also theoretical uncertainties). 
}
\label{fig:MW}
\end{center}
\vspace{-1.5em}
\end{figure}

The predictions for $\MW$ in the three scenarios are compared with each
other in \reffi{fig:MW} (for $\mu > 0$, see \refse{sec:g-2}), where the
$W$~boson mass is shown as a function of the 
lighter scalar top quark mass, $\mste$. 
The shown areas are obtained as the borders of the scan over the
parameters as specified in \refeqs{msugraparam}, (\ref{gmsbparam}) and
(\ref{amsbparam}).
The upper limit of $\mste$ reached in the three scenarios is 
similar in the CMSSM and in mAMSB (related to the upper bounds on $m_{1/2}$
and $m_{\rm aux}$), whereas the allowed area for $\mste$ is somewhat
larger in mGMSB. Since these upper bounds depend on the chosen ranges
for the high-energy scale parameters, they should be considered to be
artificial and it does not make sense to compare the three soft
SUSY-breaking scenarios in these terms. Consequently, we have 
truncated the plot at $\mste = 3 \tev$. 
The range of the
$\MW$ prediction is very similar in the three scenarios.
The solid (dashed) lines represent the 
currently allowed $1\,\si$ interval from the experimental uncertainty
(including also theoretical uncertainties). 
This indicates that at the current level of accuracy all three models
agree similarly well with the experimental measurement. A preference for
relatively low values of $\mste$ is visible, which is most prominent in
mGMSB.

\subsection{The Effective Leptonic Weak Mixing Angle}
\label{subsec:sweff}

The effective leptonic weak mixing angle at the $Z$~boson peak
can be written as
\BE
 \sweff = \frac{1}{4} \, \left( 1 - \re \frac{v_{\rm eff}}{a_{\rm eff}}  
\right) \ ,
\EE
where $v_{\rm eff}$ and $a_{\rm eff}$ 
denote the effective vector and axial couplings
of the $Z$~boson to charged leptons.
Our theoretical prediction for $\sweff$ contains the same class of
higher-order contributions as described in \refse{subsec:mw}, supplemented
with a small correction based on \citere{ZOpope}, see the evaluation in 
\citere{ehoww}. 

For the intrinsic theoretical uncertainty in the prediction for $\sweff$
we use an estimate (based on \citeres{drMSSMal2,ehoww,sw2eff2l}) of
\BE
\De\sweff^{\rm intr,current} \lsim 12 \times 10^{-5}~, 
\quad
\De\sweff^{\rm intr,future} \lsim 2 \times 10^{-5}~.
\EE
The experimental errors of $\mt$ and $\De\al_{\rm had}$
induce the following parametric uncertainties~\cite{ZOpope}\\
\BEA
\de\mt^{\rm current} = 2.1 \gev &\Rightarrow&
\De\sweff^{{\rm para},\mt, {\rm current}} \approx 6.3 \times 10^{-5}, \\[.3em]
\de(\De\al_{\rm had}^{\rm current}) = 35 \times 10^{-5} &\Rightarrow&
\De\sweff^{{\rm para},\De\al_{\rm had}, {\rm current}} \approx 
12 \times 10^{-5} .
\EEA

\bigskip
\noindent
For the future accuracies we assume
\BEA
\de\mt^{\rm future} = 0.1 \gev &\Rightarrow&
\De\sweff^{{\rm para},\mt, {\rm future}} \approx 0.4 \times 10^{-5},  \\[.3em]
\de(\De\al_{\rm had}^{\rm future}) = 5 \times 10^{-5} &\Rightarrow&
\De\sweff^{{\rm para},\De\al_{\rm had}, {\rm future}} \approx 
1.8 \times 10^{-5} .
\EEA
The experimental value is~\cite{lepewwg,LEPEWWG}%
\footnote{It should be noted that this value is determined mostly by
  two measurements that are only marginally compatible: the forward-backward
  asymmetry for $b$~quarks $A_{\rm FB}^b$, and the left-right asymmetry
  for electrons $A_{\rm LR}^e$~\cite{lepewwg}.
}%
\BE
\sweff^{\rm exp,current} = 0.23153 \pm 0.00016~.
\label{swfit}
\EE
The experimental accuracy will improve to
about
\BE
\de\sweff^{\rm \,exp,future} = 1.3 \times 10^{-5} .
\label{swexpfuture}
\EE
at GigaZ~\cite{ewpo:gigaz2} (see also \citere{gigaz} for a corresponding
discussion). 
We add the experimental and theoretical errors for $\sweff$ 
in quadrature in our analysis.

\begin{figure}[htb!]
\begin{center}
\includegraphics[width=.7\textwidth,height=10cm]{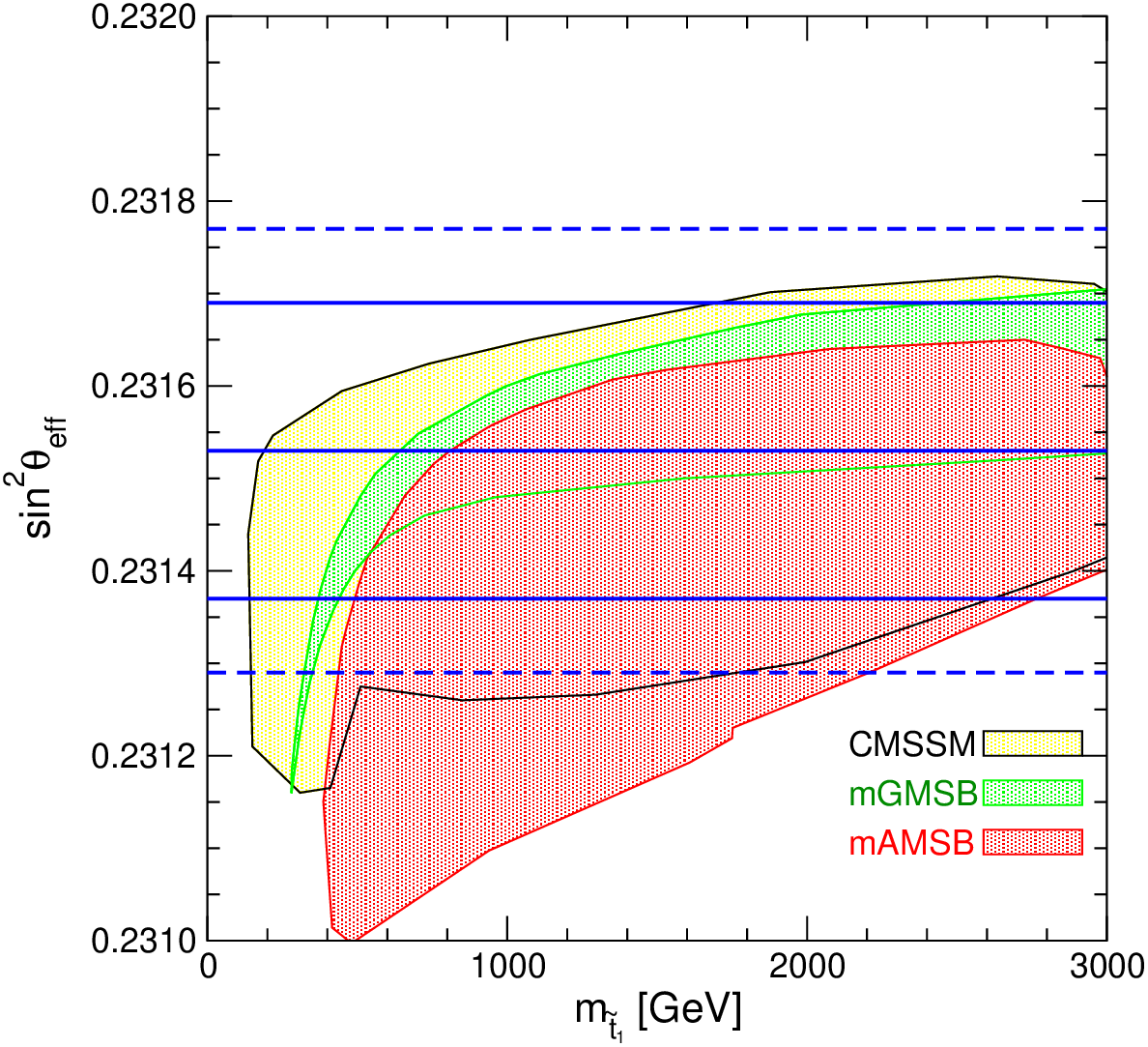}
\caption{%
The predictions for $\sweff$ as obtained from the parameter scan are
shown as a function of $\mste$ for the three 
soft SUSY-breaking scenarios for $\mu > 0$. The top quark mass has been
fixed to  
$\mt = 171.4 \gev$. The solid (dashed) lines indicate the currently
allowed $1\,\si$ interval from the experimental uncertainty (including
also theoretical uncertainties). 
}
\label{fig:SW}
\end{center}
\vspace{-1em}
\end{figure}

The predictions for $\sweff$ in the three scenarios are compared with
each other in 
\reffi{fig:SW} (for $\mu > 0$, see \refse{sec:g-2}), where the effective
weak mixing angle is shown as a 
function of the lighter scalar top quark mass, $\mste$ (truncated at 
$\mste = 3 \tev$).
As for $\MW$, the range of the $\sweff$ prediction is very 
similar in the three scenarios. Smallest values are reached in mAMSB.
The solid (dashed) lines indicate the
currently allowed $1\,\si$ interval from the experimental uncertainty
(including also theoretical uncertainties).
This indicates, as for $\MW$, that at the current level of accuracy all
three models agree equally well with the experimental data, where no
preference for $\mste$ can be deduced.

\subsection{The Anomalous Magnetic Moment of the Muon}
\label{sec:g-2}

The SM prediction for the anomalous magnetic moment of 
the muon, $\amu = \frac{1}{2} (g-2)_\mu$, 
(see~\citeres{g-2review,g-2review2,g-2reviewDS,g-2reviewMRR,g-2reviewFJ,g-2reviewPMS}
for reviews) depends in particular on the evaluation of QED contributions (see
\citeres{Kinoshita,g-2QEDmassdep,g-2QEDrecent} for recent updates), the
hadronic vacuum polarization and light-by-light (LBL) contributions. The
former  have been evaluated
in~\citeres{DEHZ,g-2HMNT,g-2HMNT2,Jegerlehner,g-2reviewFJ,Yndurain,DDDD} 
and the latter in~\citeres{LBLrev,LBL,LBLnew,LBLnew2}. 
The evaluations of the 
hadronic vacuum polarization contributions using $e^+ e^-$ and $\tau$ 
decay data give somewhat different results. In view of the fact that
recent $e^+ e^-$ measurements tend to confirm earlier results, whereas
the correspondence between previous $\tau$ data and preliminary data
from BELLE~\cite{belle} is not so clear, and also in view of the additional
uncertainties associated with  
the isospin transformation from $\tau$ decay (see \citere{g-2tauiso}), 
we use here the latest
estimate based on $e^+e^-$ data~\cite{DDDD}:
\BE
\amutheo = 
(11\, 659\, 180.5 \pm 4.4_{\rm had} \pm 3.5_{\rm LBL} \pm 0.2_{\rm QED+EW})
 \times 10^{-10},
\label{eq:amutheo}
\EE
where the source of each error is labeled. We note that the more recent 
$e^+e^-$ data sets of~\citeres{KLOE,CMD2,SND,KLOEeps} have 
been partially included in the updated estimate of $(g - 2)_\mu$. 

The SM prediction is to be compared with
the final result of the Brookhaven $(g-2)_\mu$ experiment 
E821~\cite{g-2exp,g-2exp2}, namely:
\BE
\amuexp = (11\, 659\, 208.0 \pm 6.3) \times 10^{-10},
\label{eq:amuexp}
\EE
leading to an estimated discrepancy~\cite{DDDD,g-2SEtalk}
\BE
\amuexp-\amutheo = (27.5 \pm 8.4) \times 10^{-10},
\label{delamu}
\EE
equivalent to a 3.3-$\sigma$ effect%
\footnote{Three other recent evaluations yield slightly different
  numbers~\cite{g-2reviewFJ,g-2HMNT2,g-2reviewMRR}, 
  but similar discrepancies with the SM prediction.}%
.~While it would be premature to regard this deviation as a firm
evidence for new physics, within the context of SUSY, it does indicate a
preference for a non-zero contribution from superpartners.

Concerning the MSSM contribution, the complete one-loop
result was evaluated over a decade ago~\cite{g-2MSSMf1l}. 
In view of the correlation between the signs of $(g - 2)_\mu$ and of
$\mu$~\cite{correlation}, variants of the MSSM with
$\mu < 0$ (or more precisely a positive $\mu \cdot M_2$,
where we use the convention of positive $M_2$ for the three
scenarios) are already severely challenged by the
present data on $\amu$. However, as indicated in \refse{sec:asbs}, we
have analyzed both signs of $\mu$, and correspondingly find a strong
preference for $\mu > 0$, see \reffi{fig:AMU} below.
Therefore, in the other plots shown here we focus on the case $\mu > 0$.

In addition to the full one-loop contributions, the leading QED
two-loop corrections have also been
evaluated~\cite{g-2MSSMlog2l}. Further corrections at the two-loop
level have been obtained more recently~\cite{g-2FSf,g-2CNH}, 
leading to corrections to the one-loop result that are $\lsim 10\%$. These
corrections are taken into account in our analysis according to the
approximate formulas given in~\citeres{g-2FSf,g-2CNH}.

The current intrinsic uncertainties in the SUSY
contributions to $\amu$ can be estimated to be 
$\lsim 1 \times 10^{-10}$~\cite{g-2reviewDS}. 
We assume that in the future the uncertainty in \refeq{delamu} will be
reduced by a factor two. All errors are added in quadrature.

\begin{figure}[htb!]
\begin{center}
\includegraphics[width=.7\textwidth,height=10cm]{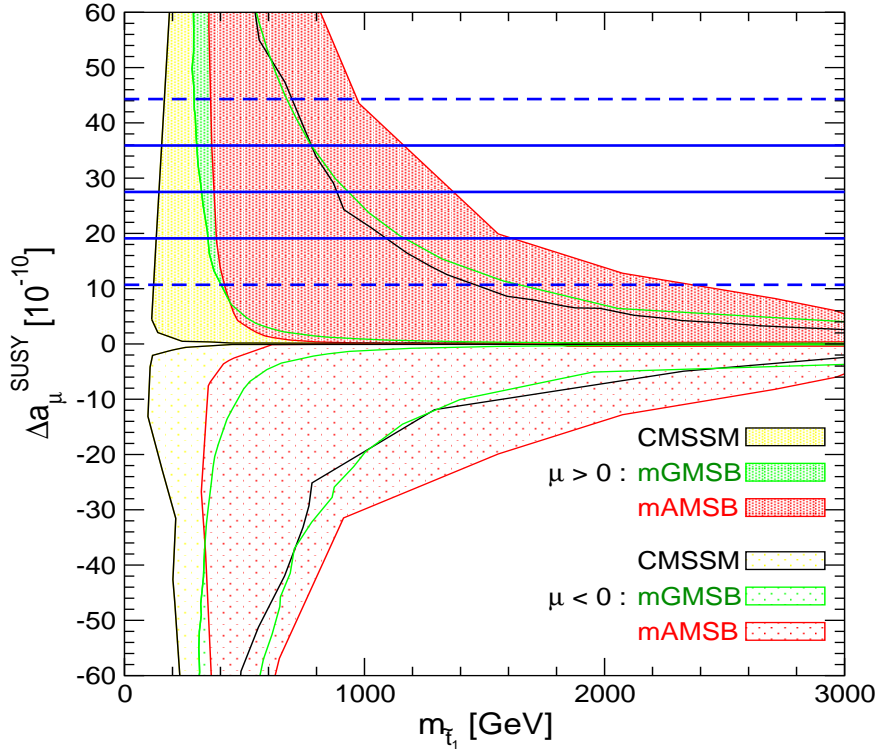}
\vspace{-1em}
\caption{%
The predictions for $\De\amu^{\rm SUSY}$ as obtained from the parameter
scan are shown as a function of $\mste$ for the three 
soft SUSY-breaking scenarios. The full (dot) shaded areas are obtained for
$\mu > (<) 0$, resulting in $\De\amu^{\rm SUSY} > (<) 0$. The top quark
mass has been 
fixed to $\mt = 171.4 \gev$. The solid (dashed) lines indicate the currently
allowed $1 (2)\,\si$ intervals of the experimental uncertainty.
}
\label{fig:AMU}
\end{center}
\vspace{-1em}
\end{figure}

The predictions for $\De\amu^{\rm SUSY}$ in the three scenarios are
compared with each other in \reffi{fig:AMU}, where the anomalous
magnetic moment of the muon is shown as a function of the lighter scalar
top quark mass, $\mste$ (truncated at $\mste = 3 \tev$). 
The full (dot) shaded areas are obtained for 
$\mu > (<) 0$, resulting in $\De\amu^{\rm SUSY} > (<) 0$. 
The range of the $\amu$ prediction is very 
similar in the three scenarios.
The solid (dashed) lines indicate the
currently allowed $1 (2)\,\si$ intervals of the experimental uncertainty.
It becomes apparent that points with $\mu < 0$ are strongly disfavored by
the analysis of $(g-2)_\mu$. Furthermore, 
at the $2\,\si$ level stop masses heavier than $\sim 2 \tev$ are clearly
disfavored.


\subsection{The Mass of the Lightest \boldmath{$\cp$}-even 
  MSSM Higgs Boson}
\label{sec:mh}

The mass of the lightest $\cp$-even MSSM Higgs boson can be predicted in 
terms of the other MSSM parameters. At the tree level, the two
$\cp$-even Higgs  boson masses are obtained as functions of $\MZ$, the
$\cp$-odd Higgs boson mass $\MA$, and $\tb$, whereas other parameters enter
into the loop corrections. 
We employ the Feynman-diagrammatic method~\cite{mhiggsf1lC,mhiggsletter} 
for the theoretical prediction of $\Mh$, using the code 
{\tt FeynHiggs}~\cite{feynhiggs,mhiggslong,mhiggsAEC,mhcMSSMlong},
which includes all numerically relevant known higher-order corrections.
The status of these results 
can be summarized as follows. For the
one-loop part, the complete result within the MSSM is 
known~\cite{ERZ,mhiggsf1lC,mhiggsf1lB}. 
Computation of the two-loop
effects is quite advanced: see~\citere{mhiggsAEC} and
references therein. These include the strong corrections
at \order{\al_t\als} and Yukawa corrections at \order{\al_t^2}
to the dominant one-loop \order{\al_t} term, and the strong
corrections from the bottom/sbottom sector at \order{\al_b\als}. 
In the case of the $b/\Sbot$~sector
corrections, an all-order resummation of the $\Tb$-enhanced terms,
\order{\al_b(\als\tb)^n}, is also known~\cite{deltamb,deltamb1}.
More recently, the \order{\al_t \al_b} and \order{\al_b^2} corrections
have been derived~\cite{mhiggsEP5}~%
\footnote{
A two-loop effective potential calculation has been presented 
in~\citere{fullEP2l}, including now even the leading three-loop
corrections~\cite{mhiggs3l}, but no public code based on this result
is currently available. Most recently another leading three-loop
calculation, valid for certain SUSY mass combinations, became
available~\cite{mhiggsFD3l}. 
}%
. The current and future intrinsic error of $\Mh$ due to
unknown higher-order corrections has been estimated to 
be~\cite{mhiggsAEC,mhiggsFDalbals,PomssmRep,mhiggsWN}
\BE
\label{Mh:intrcurrent}
\De\Mh^{\rm intr,current} = 3 \gev~,
\quad
\De\Mh^{\rm intr,future} = 0.5 \gev~.
\EE
The current uncertainty we interpret effectively as a $\sim 95~\%$
confidence level limit in the evaluation of the $\chi^2$ contribution,
see below. 

The by far largest parametric uncertainty is induced by the error in
$\mt$~\cite{mt1714} (also slightly depending on the SUSY parameters) see
\citeres{PomssmRep,tbexcl} for details, 
\begin{align}
{\rm CMSSM} &: \de\mt^{\rm current} = 2.1 \,(1.4) \gev ~\Rightarrow~ 
               \De\Mh^{{\rm para},\mt,{\rm current}} = 1.4 \,(0.9) \gev~, \non \\
{\rm mGMSB} &: \de\mt^{\rm current} = 2.1 \,(1.4) \gev ~\Rightarrow~ 
               \De\Mh^{{\rm para},\mt,{\rm current}} = 1.5 \,(1.0) \gev~, \\
{\rm mAMSB} &: \de\mt^{\rm current} = 2.1 \,(1.4) \gev ~\Rightarrow~ 
               \De\Mh^{{\rm para},\mt,{\rm current}} = 1.2 \,(0.8) \gev~. \non
\end{align}
This is already substantially below the current intrinsic uncertainty.
The numbers in brackets correspond to the latest $\mt$
measurement~\cite{mt1726} and are given for the sake of comparison.

It should be noted that, for the unconstrained MSSM with small values
of $\MA$ and values of $\tb$ which 
are not too small, a significant
suppression of the $hZZ$ coupling can occur compared to the SM value, in
which case the experimental lower bound on $\Mh$ may be more than $20 \gev$
below the SM value~\cite{LEPHiggsMSSM} (for the MSSM with real parameters). 
However, it had been checked that within the 
CMSSM, mGMSB and mAMSB the $hZZ$ coupling is
always very close to the SM value. Accordingly, the
bounds from the SM Higgs search at LEP~\cite{LEPHiggsSM}
can be taken over directly (see \citeres{asbs1,ehow1}).

Concerning the $\chi^2$ analysis, we 
use the complete likelihood information available from LEP.
We evaluate the $\Mh$ contribution to the overall $\chi^2$ function
exactly as outlined in Sect.~2.6 of \citere{ehoww}. This evaluation
takes into account the intrinsic uncertainty given in \refeq{Mh:intrcurrent}.
The $\chi^2$ contribution is then combined with the corresponding
quantities for the other observables we consider, see \refeq{eq:chi2}. 

For the analysis of future sensitivities, see \refse{sec:future}, we
assume a measurement of the lightest MSSM Higgs boson mass with a
precision of~\cite{teslatdr,orangebook,acfarep,Snowmass05Higgs}
\BE
\De\Mh^{\rm exp,future} = 50 \mev~.
\EE
The future parametric uncertainties are expected to be
\BEA
\de\mt^{\rm future} = 0.1 \gev &\Rightarrow&
\De\Mh^{{\rm para},\mt, {\rm future}} \approx 0.1 \gev ,  \\[.3em]
\de\als^{\rm future} = 0.001 &\Rightarrow&
\De\Mh^{{\rm para},\als, {\rm future}} \approx 0.1 \gev .
\EEA
Thus, the intrinsic error, \refeq{Mh:intrcurrent}, would be the dominant
source of uncertainty in 
the future. The errors are added in quadrature, yielding $\si_{\Mh}$, 
and we use for the analysis of the future sensitivities
$\chi^2_{\Mh} = ((\Mh^{\rm exp} - \Mh^{\rm theo})/\si_{\Mh})^2$.

\begin{figure}[htb!]
\begin{center}
\includegraphics[width=.7\textwidth,height=10cm]{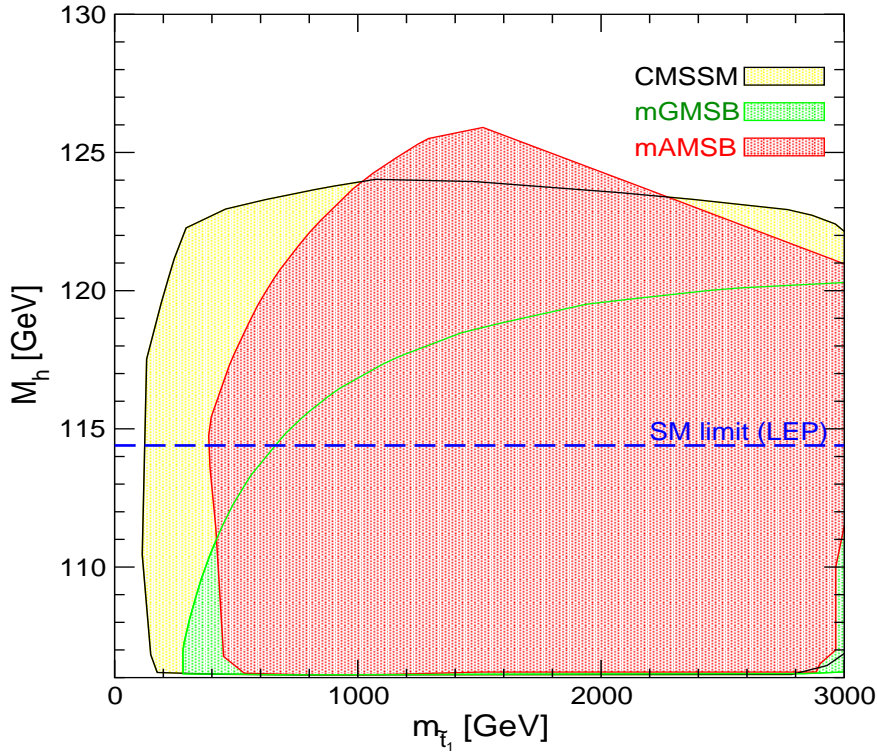}
\caption{%
The predictions for $\Mh$ as obtained from the parameter scan are shown
as a function of $\mste$ for the three 
soft SUSY-breaking scenarios for $\mu > 0$. The top quark mass has been
fixed to $\mt = 171.4 \gev$. The SM lower limit of 
$114.4 \gev$ obtained at LEP is indicated with a dashed (blue) line.
}
\label{fig:Mh}
\end{center}
\vspace{-1em}
\end{figure}

The predictions for $\Mh$ in the three scenarios are compared with each
other in \reffi{fig:Mh} (for $\mu > 0$, see \refse{sec:g-2}), where the
lightest $\cp$-even Higgs boson mass is shown as a 
function of the lighter scalar top quark mass, $\mste$ (truncated at
$\mste = 3 \tev$). The SM limit of 
$114.4 \gev$ obtained at LEP is indicated with a dashed (blue) line.
In each scenario the SM bound from Higgs searches at LEP of 
$\Mh > 114.4 \gev$ results in important constraints. On the other hand,
the bound is still fulfilled for large parts of the parameter space.
No preference for any $\mste$ can be found.


\subsection{The decay \boldmath{$b \to s \ga$}}
\label{sec:bsg}

Since this decay occurs at the loop level in the SM, the MSSM 
contribution might {\it a priori} be of similar magnitude. A
recent
theoretical estimate of the SM contribution to the branching ratio at
the NNLO QCD level is~\cite{bsgtheonew}
\BE
\br( b \to s \ga ) = (3.15 \pm 0.23) \times 10^{-4}~.
\label{bsga}
\EE
We record that the error estimate for $\br(b \to s \ga)$ is still under
debate~\cite{hulupo}, and that other SM contributions to 
$b \to s \ga$ have been calculated~\cite{bsgneubert}. These
corrections are small compared with the theoretical uncertainty quoted
in \refeq{bsga}. 

For comparison, the present experimental 
value estimated by the Heavy Flavour Averaging Group (HFAG)
is~\cite{bsgexp,hfag}
\BE
\br(b \to s \ga) = (3.55 \pm 0.24 {\;}^{+0.09}_{-0.10} \pm 0.03) 
                   \times 10^{-4},
\label{bsgaexp}
\EE
where the first error is the combined statistical and uncorrelated
systematic uncertainty, and the other
two errors are correlated systematic theoretical uncertainties
and corrections, respectively.

Our numerical results have been derived with the 
$\br(b \to s \ga)$ evaluation provided in \citeres{bsgGH,ali,ali2},
incorporating also the latest SM corrections provided in~\citere{bsgtheonew}. 
The calculation has been checked against other
codes~\cite{bsgMicro,bsgKO1,bsgKO2}.
For the evaluation of the $\br(b \to s \ga)$, we assume minimal flavor
violation (MFV) at the electroweak scale and neglect NMFV effects that 
can be induced by RGE running from the high scale, see
e.g.~\citere{nmfv}, that may amount to $\sim 10\%$ of the SUSY
corrections. 

Concerning the total error in a conservative approach we add linearly
the errors of \refeqs{bsga} and (\ref{bsgaexp}) as well an intrinsic
SUSY error of $0.15 \times 10^{-4}$~\cite{ehoww}, except the statistical
error that is then added in quadrature.
For the analysis of the future sensitivities in \refse{sec:future} we
assume that the total error will be reduced by a factor of~3.

\begin{figure}[htb!]
\begin{center}
\includegraphics[width=.7\textwidth,height=10cm]{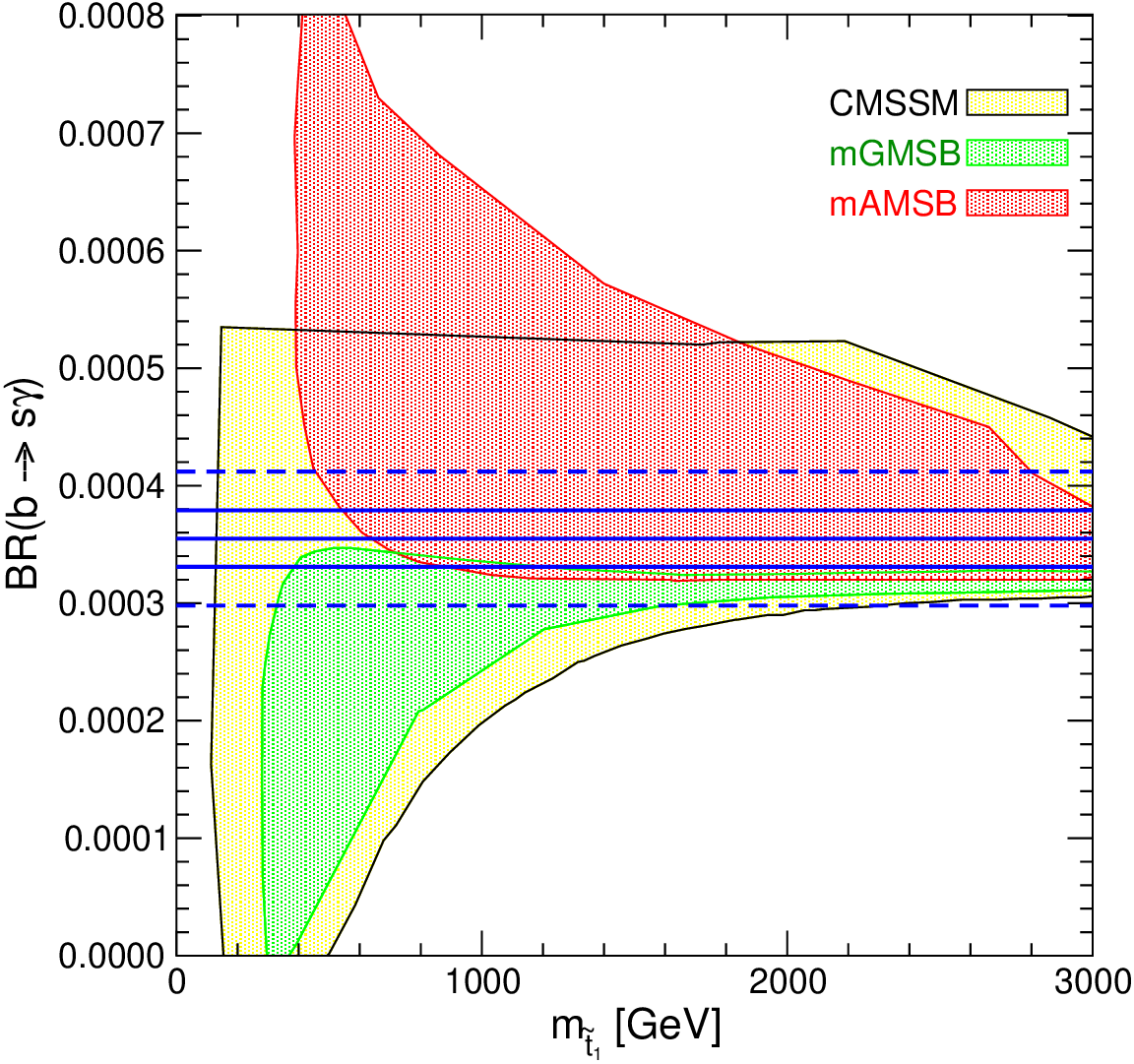}
\caption{%
The predictions for $\br(b \to s \ga)$ as obtained from the parameter
scan are shown as a function of $\mste$ 
for the three soft SUSY-breaking scenarios for $\mu > 0$. The top quark
mass has been fixed to  
$\mt = 171.4 \gev$. The solid (dashed) lines indicate the currently
allowed $1\,\si$ interval from the experimental uncertainty (including
also theoretical uncertainties, which are added linearly). 
}
\label{fig:BSG}
\end{center}
\vspace{-1em}
\end{figure}

The predictions for $\br(b \to s \ga)$ in the three scenarios are
compared with each other in 
\reffi{fig:BSG} (for $\mu > 0$, see \refse{sec:g-2}), where the
branching ratio is shown as a function of the lighter scalar top quark mass,
$\mste$ (truncated at $\mste = 3 \tev$). 
The solid (dashed) lines indicate the currently
allowed $1\,\si$ interval from the experimental uncertainty (including
also theoretical uncertainties, which are added linearly, see above). 
In all three scenarios large parts of the parameter space lie 
within the $1\,\si$ interval. However, for small mass scales 
$\br(b \to s \ga)$ provides important constraints on the three models.
While the CMSSM and mGMSB can have very small values of $\br(b \to s
\ga)$ for small $\mste$%
\footnote{
Where the $\br(b \to s \ga)$ becomes close to zero the calculation of
the SUSY corrections is not reliable anymore. However, these parts of
the parameter space anyhow result in an experimentally excluded value for
$\br(b \to s \ga)$.
}%
,~mAMSB has typically large values of the BR. The
reason can be traced back to the 
fact that the sign of the stop mixing angle $\tst$ comes out with
a positive sign in mAMSB, whereas it is negative in the CMSSM and mGMSB
(as output and in the conventions of {\tt SoftSUSY}). This different
sign, in combination 
with a positive $\mu$, results in a positive SUSY contribution to 
$\br(b \to s \ga)$ within mAMSB and (for most values of the other
parameters) a negative contribution in the CMSSM
and mGMSB, see also the discussion in the beginning of \refse{sec:asbsana}.


\subsection{The Branching Ratio for \boldmath{$B_s \to \mu^+\mu^-$}}
\label{subsec:bsmm}

The SM prediction for this branching ratio is 
$(3.4 \pm 0.5) \times 10^{-9}$~\cite{bsmmtheosm}, and 
the present experimental upper limit from the Fermilab Tevatron collider
is $5.8 \times 10^{-8}$ at the $95\%$ C.L.~\cite{bsmmexp}, 
still providing 
room for the MSSM to dominate the SM contribution. The current Tevatron
sensitivity is based on an integrated luminosity of about 2~\ifb\
collected at CDF. 
For the $\chi^2$ contribution, in order to incorporate the latest
Tevatron bound, we use a smoothed step function, penalizing data points with 
$\br(B_s \to \mu^+\mu^-) > 5.8 \times 10^{-8}$ and preferring lower BRs.

The Tevatron sensitivity is expected to improve significantly in the
future. The limit that could be reached at the end of Run~II is 
$\sim 2 \times 10^{-8}$ assuming 8~\ifb\ collected with each
detector~\cite{bsmmexpfuture}. 
A sensitivity even down to the SM value can be
expected at the LHC. Assuming the SM value, i.e.\
$\br(B_s \to \mu^+ \mu^-) \approx 3.4 \times 10^{-9}$, it has been
estimated~\cite{lhcb} that LHCb can observe 33~signal events
over 10~background events within 3~years of low-luminosity
running. Therefore this process offers good prospects for probing the MSSM.

For the theoretical prediction we use results from~\citere{bsmumu}, 
which are in good agreement with \citere{ourBmumu}. This calculation
includes the full one-loop evaluation and the leading two-loop 
QCD corrections.
As in \refse{sec:bsg}, we neglect any NMFV effects from RGE running.
We do not include $\br(B_s \to \mu^+\mu^-)$ in our analysis of the
future sensitivities (but still require agreement with the current
bound), because its impact will strongly depend on the
value realized in Nature.

\begin{figure}[htb!]
\begin{center}
\includegraphics[width=.7\textwidth,height=10cm]{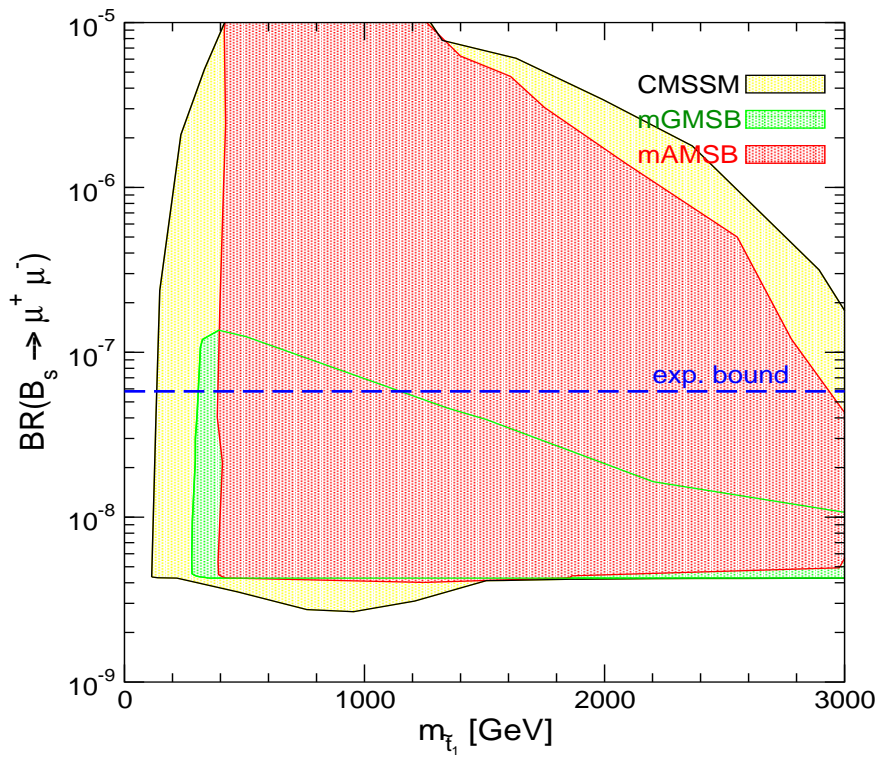}
\vspace{-1em}
\caption{%
The predictions for $\br(B_s \to \mu^+\mu^-)$ as obtained from the
parameter scan are shown as a function of 
$\mste$ for the three soft SUSY-breaking scenarios for $\mu > 0$. The
top quark mass has been fixed to $\mt = 171.4 \gev$. The current upper 
limit of $5.8 \times 10^{-8}$ is indicated by a dashed (blue) line.
}
\label{fig:BMM}
\end{center}
\vspace{-1.5em}
\end{figure}

The predictions for $\br(B_s \to \mu^+\mu^-)$ in the three scenarios are
compared with each other in \reffi{fig:BMM} (for $\mu > 0$, see
\refse{sec:g-2}), where the BR is shown as a 
function of the lighter scalar top quark mass, $\mste$ (truncated at
$\mste = 3 \tev$). The current experimental 
limit of $5.8 \times 10^{-8}$ is indicated by a dashed (blue) line.
Each scenario has large parts of the parameter space with 
$\br(B_s \to \mu^+\mu^-) < 5.8 \times 10^{-8}$, 
where no limit on $\mste$ is provided by the upper limit on the BR.
Within the mGMSB scenario, due to its generally larger $\MA$ values (see
below), hardly any points are ruled out by the current upper bound on
the BR, while for the other two scenarios $\br(B_s \to \mu^+\mu^-)$
is already a strong constraint on the parameter space.
We have checked that including the CDM constraint and restricting to
values of $\tb \le 50$ the results for $\br(B_s \to \mu^+\mu^-)$ in
\citeres{ehow3,ehow4,ehoww} are reproduced.


\newpage
\section{\boldmath{$\chi^2$} analysis for CMSSM, 
        mGMSB, mAMSB}
\label{sec:asbsana}

In this section we present our numerical analysis, based on the $\chi^2$
evaluation as given in \refeq{eq:chi2}. The best fit point is given by
the lowest $\chi^2$ value. The sensitivities are shown as 
$\De\chi^2 = 1, 4, 9$, referred to as $\De_1$, $\De_4$ and $\De_9$,
respectively. They give an indication of the precision that has been
reached so far for 
the observables under investigation. Sometimes we refer to the $\De_4$
areas as `preferred' regions.
The lowest $\chi^2$ values for the three scenarios are given in
\refta{tab:chi2min}. Also shown are the individual contributions from
the precision observables. $\br(B_s \to \mu^+\mu^-)$
always gives a zero contribution, and we list the BR itself.

It is interesting to note that despite mAMSB has one parameter less, the
minimum $\chi^2$ value is lower by $\sim 1.5$--$2$ compared to the CMSSM and
mGMSB. 
The reason for the low $\chi^2$ values is a combination of two effects.
First, there is a good agreement of mAMSB with $(g-2)_\mu$ and 
$\br(b \to s \ga)$. The anomalous magnetic moment of the muon 
requires a positive $\mu$ (or more precisely a positive $\mu \cdot M_2$,
where we use the convention of positive $M_2$ for the three
scenarios, see the discussion above). $\br(b \to s \ga)$ 
on the other hand depends on the combinations of the stop masses, mixing
angle and $\mu$. The sign of the stop mixing angle $\tst$ comes out with
a positive sign in mAMSB, whereas it is negative in the CMSSM and mGMSB
(as output and in the convention of {\tt SoftSUSY}). This different
sign, in combination 
with a positive $\mu$, results in a positive SUSY contribution to 
$\br(b \to s \ga)$ within mAMSB and a (usually) negative contribution in
the CMSSM 
and mGMSB. In this way mAMSB can fulfill the $\br(b \to s \ga)$
constraint as well
as the other two scenarios (but with a best-fit value {\em above} the
experimental value). 
Second, due to the structure of the soft SUSY-breaking parameters in the
chargino/neutralino sector relatively light charginos are present in
mAMSB (where the lightest one is nearly mass degenerate with the
lightest neutralino). Thus a large contribution to $(g-2)_\mu$ and also
to $\MW$~\cite{MWpope} can be obtained for a relatively heavier spectrum
otherwise, resulting in an $\Mh$ value above $\sim 116 \gev$. 
The overall effect of this interplay is a total minimum $\chi^2$ value
of~2.9.

\begin{table}[htb!]
\renewcommand{\arraystretch}{1.3}
\BC
\begin{tabular}{|c||c|c|c|} 
\cline{2-4} \multicolumn{1}{c||}{}
                       & CMSSM & mGMSB & mAMSB \\
\hline\hline
$\chi^2_{\rm min}$       & 4.6 &  5.1   &  2.9 \\ \hline\hline
$\MW$                   & 1.7 &  2.1   &  0.6 \\ \hline
$\sweff$                & 0.1 &  0.0   &  0.8 \\ \hline
$(g-2)_\mu$              & 0.6 &  0.9   &  0.0 \\ \hline
$\br(b \to s \ga)$       & 1.1 &  2.0   & 1.5 \\ \hline
$\Mh$                    & 1.1 &  0.1   & 0.0 \\ \hline
$\br(B_s \to \mu^+\mu^-)$ & $4.5 \times 10^{-8}$ & $3.2 \times 10^{-8}$ & 
                            $0.4 \times 10^{-8}$ \\ \hline\hline
$\MA$ [GeV] (best-fit)   & 394 & 547 & 616 \\ \hline
$\tb$ (best-fit)         &  54 &  55 &   9 \\
\hline\hline
\end{tabular}
\EC
\vspace{-1em}
\caption{%
Minimum $\chi^2$ values for the three soft SUSY-breaking scenarios using
today's accuracies for the experimental and theoretical precisions.
We also show the individual contributions for $\MW$, $\sweff$,
$(g-2)_\mu$, $\br(b \to s \ga)$ and $\Mh$, as well as the value of
$\br(B_s \to \mu^+\mu^-)$. 
Shown in the last two rows are the best-fit values for the low-energy
parameters, $\MA$ and $\tb$, as analyzed in \refse{sec:analow}.
}
\label{tab:chi2min}
\renewcommand{\arraystretch}{1.0}
\end{table}

In the analysis presented below, in the first step we show the
three soft SUSY-breaking scenarios separately in terms of their
high-scale parameters. In a second step we compare their respective
predictions in terms of the low-scale parameters $\MA$ and $\tb$ and
other SUSY mass scales. In the
final step in \refse{sec:future} we assume future precisions for the
measurements and theory 
evaluations and compare the sensitivities the precision observables will
offer in the three scenarios.


\subsection{Analysis of high-scale parameters}
\label{sec:anahigh}

In the following subsections we analyze the CMSSM, mGMSB and mAMSB in
terms of their respective high-energy parameters, see \refse{sec:asbs}.

\subsubsection{CMSSM}

In \reffi{fig:cmssmHigh} we show the results for the $\De_{1,4,9}$ areas
in terms of the high-energy parameters, using the current experimental
and theoretical precisions as described in \refse{sec:expdata}. 
The $\De_1$ area is medium
shaded (green), the $\De_4$ are is dark shaded (red), and the $\De_9$ area
is light shaded (yellow). The rest of the scanned parameter space is
given in black shading. The best-fit point is marked with a circle.
Because of the contribution to $(g-2)_\mu$
only very few points with $\mu < 0$ have $\De\chi^2 < 9$, and we
concentrate here on the data with $\mu > 0$. For this sign of $\mu$ the
$\De_9$ area nearly covers the whole parameter space (in agreement with
the results presented in \citere{ehoww}). In terms of $m_{1/2}$
relatively low values are favored around $m_{1/2} = 500 \gev$, with the
$\De_4$ region extending up to $m_{1/2} = 1000 \gev$. For $m_0$, on the
other hand, hardly any bound is obtained, and values up to $2000 \gev$
are possible. Only at the $\De_1$ level a preference of the
allowed values for a light $m_0$ can be found. 
For $A_0$ a slight preference for positive values can be observed
(note the different sign convention here in comparison with
\citeres{ehow3,ehow4,ehoww,ehhow,ehow5}), 
and the $\De_4$ region extends from $-1000 \gev$ to about $+2500 \gev$.
The apparent differences to existing
analyses~\cite{ehow3,ehow4,masterfit} are due to the fact that the CDM
constraint has not been applied here, see the discussion below.

\begin{figure}[htb!]
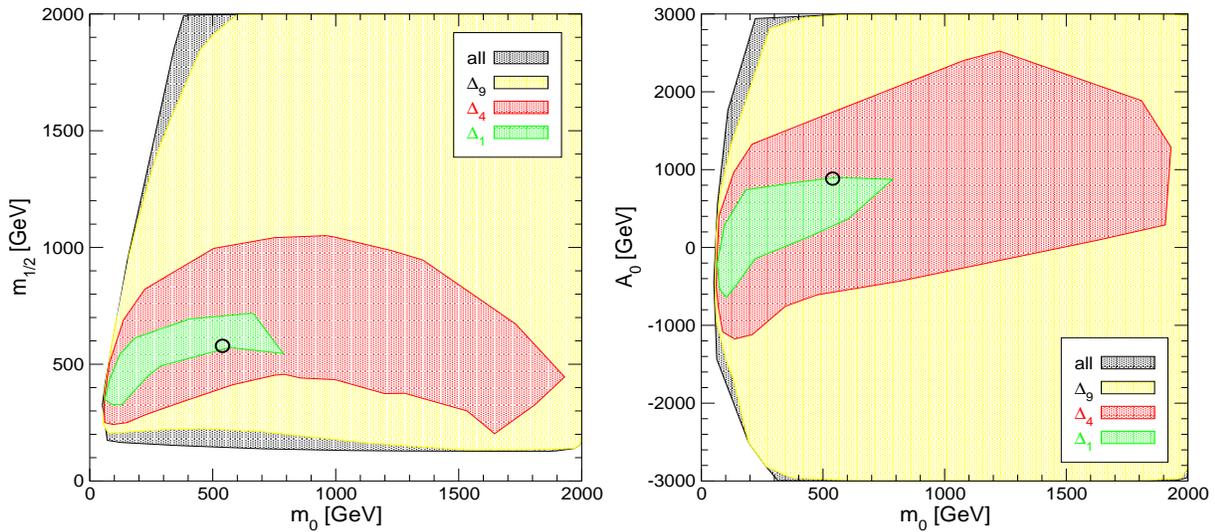

\begin{center}
\includegraphics[width=.48\textwidth,height=7cm]{asbs3_s_para01_cl}
\includegraphics[width=.48\textwidth,height=7cm]{asbs3_s_para02_cl}
\caption{%
The $\De_{1,4,9}$ regions in the $m_0$--$m_{1/2}$ plane (left) and in
the $m_0$--$A_0$ plane (right) in the CMSSM for $\mu > 0$. 
The $\De_1$ area is medium
shaded (green), the $\De_4$ area is dark shaded (red), and the $\De_9$ area
is light shaded (yellow). The rest of the scanned parameter space is
given in black shading. The best-fit point is marked with a circle.
}
\label{fig:cmssmHigh}
\end{center}
\vspace{-1em}
\end{figure}


\subsubsection{mGMSB}

In \reffis{fig:mgmsbHigh}, \ref{fig:mgmsbHigh2} we show the results for
the $\De_{1,4,9}$ areas in terms of the high-energy parameters, using
the current experimental 
and theoretical precisions as described in \refse{sec:expdata}. 
The color coding is as in \reffi{fig:cmssmHigh}.
As in the CMSSM, because of the contribution to $(g-2)_\mu$
only very few points with $\mu < 0$ have $\De\chi^2 < 9$, and we
concentrate here on the data with $\mu > 0$. 

The plots in \reffi{fig:mgmsbHigh} show the $\La$--$M_{\rm mess}$
plane for $N_{\rm mess} = 1 \ldots 8$ separately. The $\De\chi^2$ values are
obtained with respect to the overall best fit point, which is reached
for $N_{\rm mess} = 8$ (marked with a circle). 
The `preferred' $\La$ values depend on the choice of $N_{\rm mess}$,
going from $\sim 10^5 \gev$ at low $N_{\rm mess}$ down to 
$\sim 2 \times 10^4 \gev$ for large $N_{\rm mess}$.
However, the $\De_9$ region extend over large parts of the whole
parameter space. Furthermore no bound on $M_{\rm mess}$ can be set. 
Similar results are found in 
\reffi{fig:mgmsbHigh2}, where we show the $N_{\rm mess}$--$\La$ plane. The
lower $N_{\rm mess}$, the higher are the possible values for $\La$. 

In order to analyze the compatibility of the various $N_{\rm mess}$
values with the precision data, we show in \refta{tab:Nmesschi2} the
lowest $\chi^2_{{\rm min},N_{\rm mess}}$ values reached for each 
$N_{\rm mess}$. It can be seen that $\chi^2_{{\rm min},N_{\rm mess}}$
increases monotonically with decreasing $N_{\rm mess}$. 
In agreement with \reffis{fig:mgmsbHigh} and~\ref{fig:mgmsbHigh2} the
difference in the minimum $\chi^2$ between $N_{\rm mess} = 8$ and 
$N_{\rm mess} > 1$ is smaller than one, and only for $N_{\rm mess} = 1$
the difference exceeds one by $\sim 0.04$. 
Consequently no $\De_1$ region appears
in the $N_{\rm mess} = 1$ plots.

\begin{table}[htb!]
\renewcommand{\arraystretch}{1.3}
\BC
\begin{tabular}{|c|cccccccc|} 
\hline\hline
$N_{\rm mess}$ & 1 & 2    & 3    & 4    & 5    & 6    & 7    & 8 \\ \hline
$\chi^2_{{\rm min},N_{\rm mess}}$ & 
             6.17 & 5.53 & 5.45 & 5.25 & 5.25 & 5.20 & 5.16 & 5.13 \\
\hline\hline
\end{tabular}
\EC
\vspace{-1em}
\caption{%
Minimum $\chi^2$ values reached for each $N_{\rm mess}$.
}
\label{tab:Nmesschi2}
\renewcommand{\arraystretch}{1.0}
\end{table}

\begin{figure}[htb!]
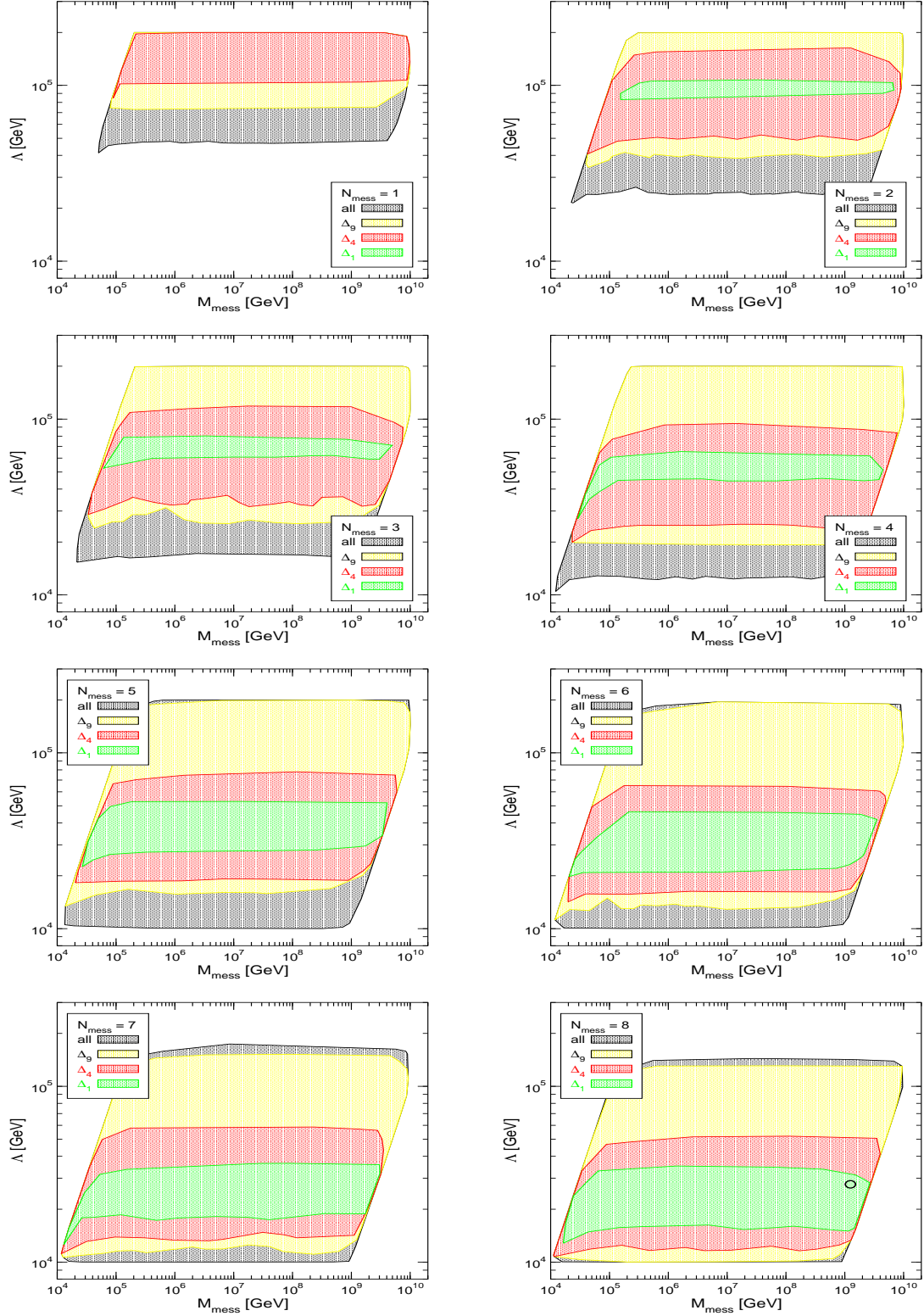

\begin{center}
\includegraphics[width=.4\textwidth,height=4.9cm]{asbs3_g_para01_cl}\hspace{1cm}
\includegraphics[width=.4\textwidth,height=4.9cm]{asbs3_g_para02_cl}\\[0.3cm]
\includegraphics[width=.4\textwidth,height=4.9cm]{asbs3_g_para03_cl}\hspace{1cm}
\includegraphics[width=.4\textwidth,height=4.9cm]{asbs3_g_para04_cl}\\[0.3cm]
\includegraphics[width=.4\textwidth,height=4.9cm]{asbs3_g_para05_cl}\hspace{1cm}
\includegraphics[width=.4\textwidth,height=4.9cm]{asbs3_g_para06_cl}\\[0.3cm]
\includegraphics[width=.4\textwidth,height=4.9cm]{asbs3_g_para07_cl}\hspace{1cm}
\includegraphics[width=.4\textwidth,height=4.9cm]{asbs3_g_para08_cl}
\caption{%
The $\De_{1,4,9}$ regions in the $M_{\rm mess}$--$\La$ plane for 
$N = 1 \ldots 8$ in the mGMSB for $\mu > 0$. 
The color coding is as in \reffi{fig:cmssmHigh}. 
The best fit point is realized for $N_{\rm mess} = 8$ and marked with a circle.
}
\label{fig:mgmsbHigh}
\end{center}
\vspace{-2em}
\end{figure}

\begin{figure}[htb!]
\begin{center}
\includegraphics[width=.60\textwidth]{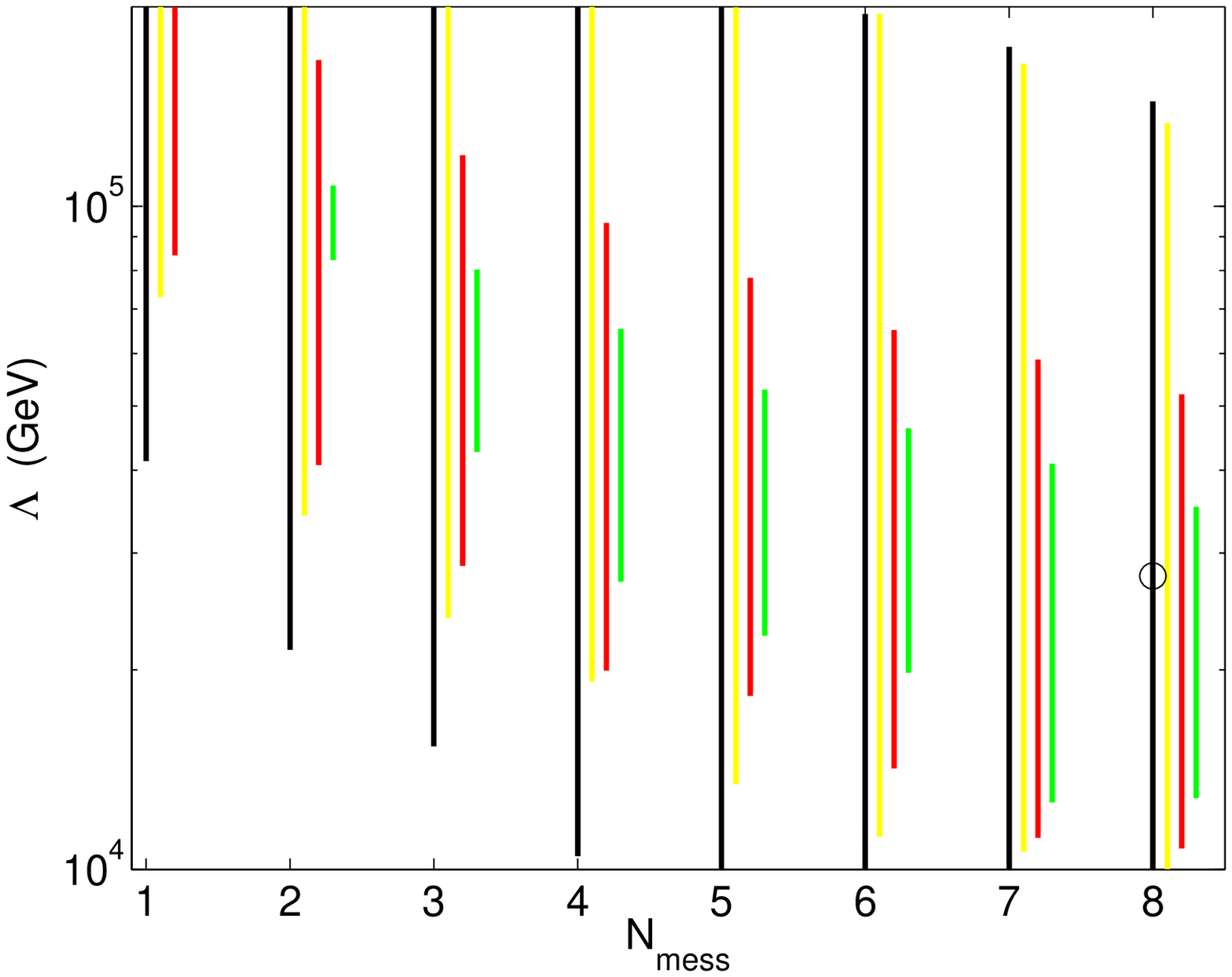}
\caption{%
The $\De_{1,4,9}$ regions in the $N_{\rm mess}$--$\La$ plane in the
mGMSB for $\mu > 0$. 
The color coding is as in \reffi{fig:cmssmHigh}.
Marked with a circle is the current best-fit point.
}
\label{fig:mgmsbHigh2}
\end{center}
\vspace{-1em}
\end{figure}


\subsubsection{mAMSB}
\label{sec:mamsbHigh}

In \reffi{fig:mamsbHigh} we show the only high-energy parameter plane in
the mAMSB, $m_{\rm aux}$ vs.\ $m_0$ for $\mu > 0$.
While nearly the whole parameter space is covered by the $\De_9$ area, the 
$\De_4$ and $\De_1$ regions are located at a relatively thin strip at
the lowest possible $m_0$ values with a width $\lsim 300 \gev$.
The precision observables clearly show a preference for a relatively
small scalar soft SUSY-breaking parameter $m_0$. This can be
traced back to the $\chi^2$ contribution to $(g-2)_\mu$ that requires
relatively light sleptons of the second generation. Since $m_0$ is
needed to prevent the tachyon problem within mAMSB, it controls to a
large extent the slepton masses. The strong bound from $(g-2)_\mu$ then
translates into a relatively strong bound on $m_0$.
On the other hand, $m_{\rm aux}$ is only mildly restricted. 
The lower absolute bound on $m_{\rm aux}$ is mainly due to the lower
experimental bound on the lightest chargino of $\sim 70 \gev$~\cite{pdg}.

\begin{figure}[htb!]
\begin{center}
\includegraphics[width=.6\textwidth,height=7cm]{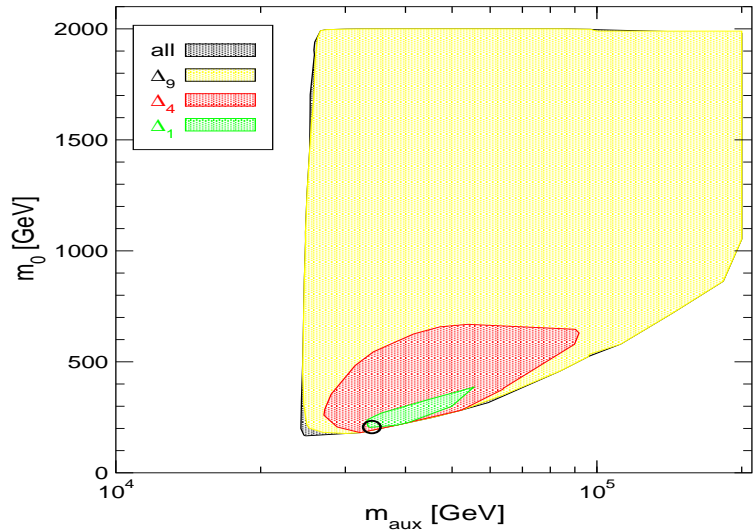}
\vspace{-1em}
\caption{%
The $\De_{1,4,9}$ regions in the $m_{\rm aux}$--$m_0$ plane
in the mAMSB for $\mu > 0$. 
The color coding is as in \reffi{fig:cmssmHigh}.
The best-fit point is marked with a circle.
}
\label{fig:mamsbHigh}
\end{center}
\vspace{-1em}
\end{figure}


\subsection{Low-energy analysis}
\label{sec:analow}

We now turn to the comparison of the three soft SUSY-breaking scenarios.
In \reffi{fig:asbsLow} we show the $\MA$--$\tb$ plane for the CMSSM
(top), mGMSB (middle) and mAMSB (bottom) with the same color coding as
in \reffi{fig:cmssmHigh}. As in \refse{sec:anahigh} we restrict
ourselves to $\mu > 0$. 
The allowed $\MA$--$\tb$ parameter space is somewhat different in the
three scenarios. While in mAMSB the parameters are restricted to 
$\MA \lsim 4 \tev$ and $\tb \lsim 50$, this extends to 
$\MA \lsim 4 \tev$ and $\tb \lsim 60$ (where we stopped our $\tb$ scan)
in the CMSSM, and within mGMSB $\MA$ values up to $6 \tev$ are
possible (not shown in the plot). 
The qualitative features of the $\De_{9,4,1}$ areas
are very similar for the three scenarios. 
The $\De_9$ area extends over large parts of the whole parameter space.
On the other hand, within all three scenarios, 
the $\De_4$ and even more the $\De_1$ areas are located at
relatively low $\MA$, extending up to $\MA \lsim 1000 \gev$ at the
$\De_4$ level in all three scenarios. 
The `preferred' $\tb$ regions, on the other hand, nearly span
the full possible range in the CMSSM and mGMSB, whereas in 
the mAMSB scenario the $\chi^2$ `preferred' areas are located at lower $\tb$
values, reaching up to $\tb \lsim 35$. The low value of
$\br(B_s \to \mu^+\mu^-)$ at the best-fit point in mAMSB is due to the
relatively low $\tb$ value.
However, in view of these ranges, the actual values of
the best-fit points for $\tb$ are not very significant, in accordance
with earlier analyses~\cite{ehow3,ehow4,ehoww,ehhow,other}.
In conclusion a preference for not too large $\MA$ values is clearly
visible as a common feature in all three scenarios.
Depending on the actual combination of $\MA$ and $\tb$, the LHC can
cover a large part of the `preferred' parameter space by searches for
the heavy Higgs
bosons~\cite{atlastdr,cmstdr,jakobs,schumi,cmshiggsOrg,cmsHiggs}.

\begin{figure}[htb!]
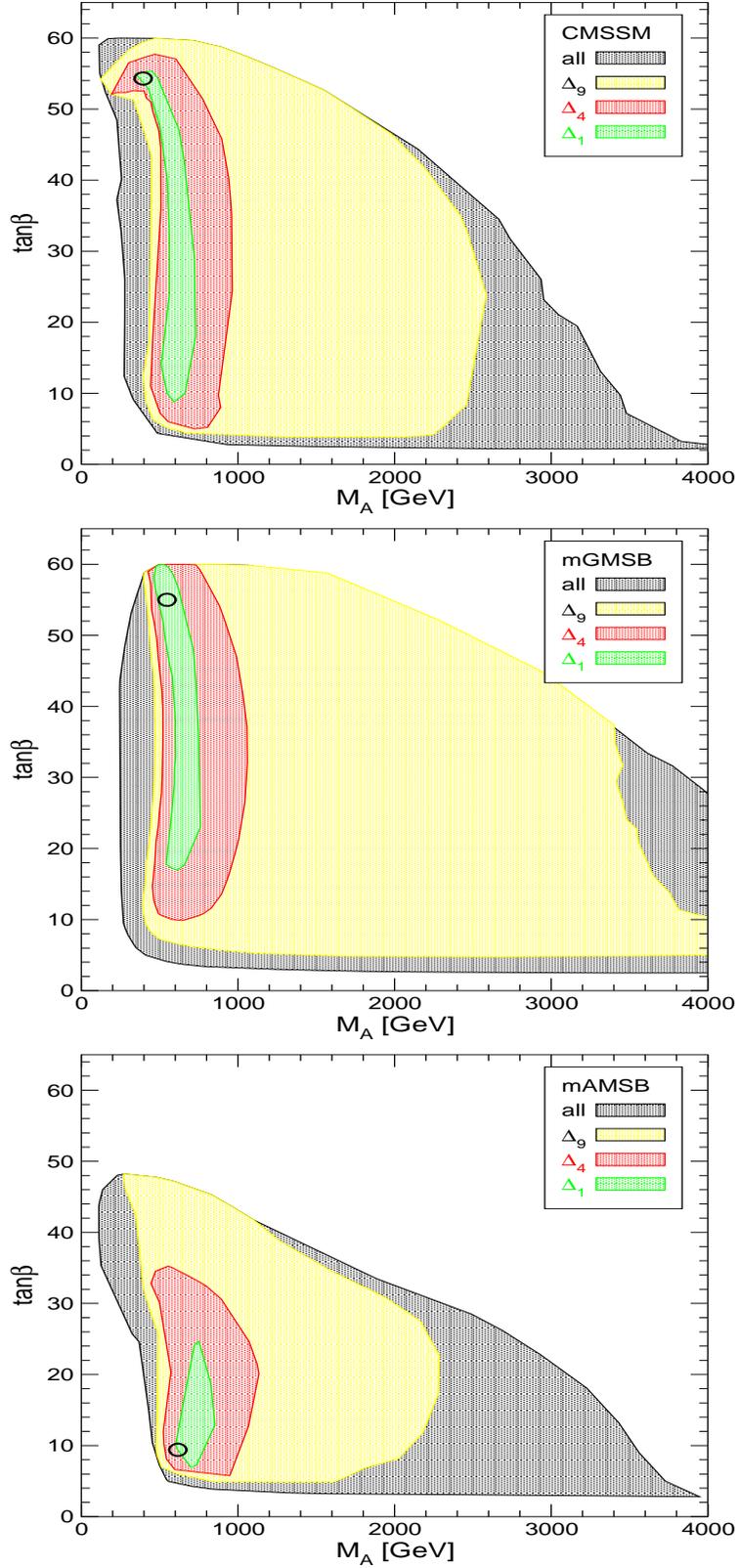

\begin{center}
\includegraphics[width=.6\textwidth,height=7.0cm]{asbs3_s_mass01_cl}\\[.3em]
\includegraphics[width=.6\textwidth,height=7.0cm]{asbs3_g_mass01_cl}\\[.3em]
\includegraphics[width=.6\textwidth,height=7.0cm]{asbs3_a_mass01_cl}
\caption{%
The $\De_{1,4,9}$ regions in the $\MA$--$\tb$ planes
in the CMSSM (top), mGMSB (middle) and mAMSB (bottom) for $\mu > 0$. 
The color coding is as in \reffi{fig:cmssmHigh}.
In each plot the best-fit point is marked with a circle.
}
\label{fig:asbsLow}
\end{center}
\vspace{-4em}
\end{figure}

We now turn to the analysis of various mass values in the three soft
SUSY-breaking scenarios. We start with the mass of the lightest $\cp$-even
Higgs boson, see \refse{sec:mh}, presented in \reffi{fig:asbs_mh}. 
$\Mh$ is shown in the CMSSM (top), mGMSB (middle) and mAMSB (bottom)
scenarios for $\mu > 0$ with the corresponding $\chi^2$, where the $\chi^2$
contribution of $\Mh$ itself has been left out. In this way the plot
shows the indirect predictions for $\Mh$ without imposing the bounds
from the Higgs boson searches at LEP.
In the CMSSM and in mGMSB the impact of dropping the $\chi^2$
contribution from $\Mh$ leads to a 
drastically lower total $\chi^2$ as compared to the case when the $\Mh$
bound is included, see \refta{tab:chi2min}. 
In these two scenarios the best-fit point changes to new points with
substantially lower $\Mh$ values (as discussed below). These new
best-fit points can also accomodate the other precision observables
better, thus leading to a reduction of $\chi^2_{\rm min}$ by more than
$\sim 3$ in the CMSSM and mGMSB. In the
mAMSB scenario, on the other hand, the effect is small, and the best-fit
point changes only slightly.
The color coding is as in \reffi{fig:cmssmHigh}. 

In all three scenarios
a shallow minimum can be observed. The $\De_1$ regions are in the
intervals of $\Mh = 98 \ldots 111 \gev$ (CMSSM), 
$97 \ldots 112 \gev$ (mGMSB) and $104 \ldots 122 \gev$ (mAMSB). In all
three scenarios the $\De_4$ regions extend beyond the LEP limit of 
$\Mh > 114.4 \gev$ at the 95\% C.L. shown as dashed (blue) line in
\reffi{fig:asbs_mh} (which is valid for the three soft
SUSY-breaking scenarios, see \citeres{asbs1,ehow1}). 
The analysis for the CMSSM can be compared with
\citeres{ehoww,masterfit}, where (among other contributions) also the
cold dark matter constraint had been included in the analysis. In
\citeres{ehoww,masterfit} best fit values of $\Mh = 110 \ldots 115 \gev$ 
(depending on $\tb$) had been observed, which is at the border of the
$\De_1$ region here. 
These results are well compatible with each other. The inclusion of the CDM
constraint yields the effect of cutting out a (thin) band in 
the $\Mh$--$\chi^2_{\rm tot}$ plane.
In conclusion all three scenarios have a
significant part of the parameter space with a relatively low total $\chi^2$
that is in agreement with the bounds from Higgs-boson searches at LEP.
Especially within the mAMSB scenario the $\De_1$ region extends beyond
the LEP bound of $114.4 \gev$.

\begin{figure}[htb!]
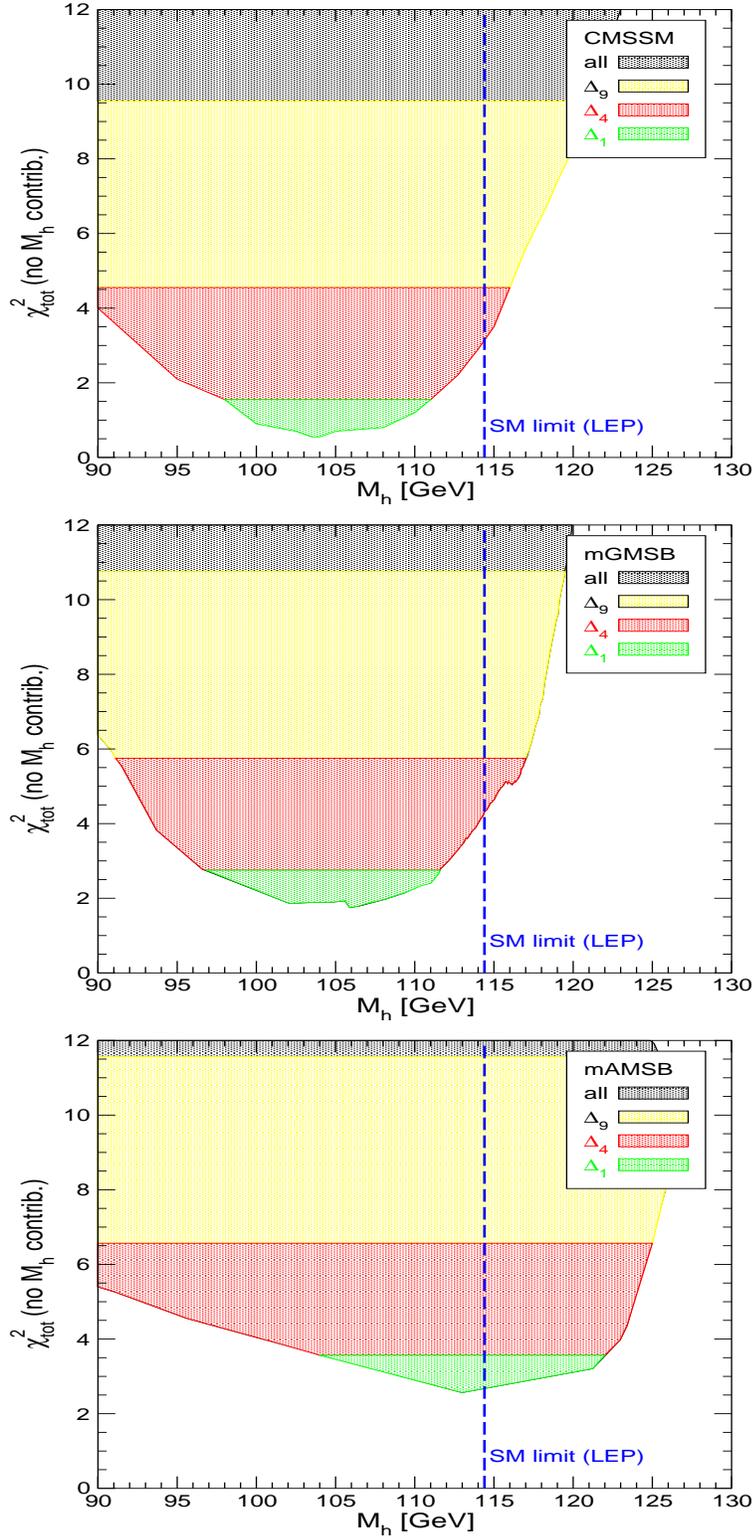

\begin{center}
\includegraphics[width=.6\textwidth,height=6.7cm]{asbs3_s_mass02_cl}\\[.3em]
\includegraphics[width=.6\textwidth,height=6.7cm]{asbs3_g_mass02_cl}\\[.3em]
\includegraphics[width=.6\textwidth,height=6.7cm]{asbs3_a_mass02_cl}
\caption{%
The $\Mh$ values in the CMSSM (top), mGMSB (middle) and mAMSB (bottom)
scenarios for $\mu > 0$ with their respective $\chi^2$, where the $\chi^2$
contribution of the $\Mh$ itself has been left out. 
The color coding is as in \reffi{fig:cmssmHigh}. The SM limit of 
$114.4 \gev$ obtained at LEP is indicated with a dashed (blue) line.
}
\label{fig:asbs_mh}
\end{center} 
\vspace{-4em}
\end{figure}

Next we turn to the prediction of the masses of various SUSY particles,
starting with $\mneu{1}$ (left) and $\mneu{2}$ (right) in
\reffi{fig:asbs_neu}. 
The masses are shown in the CMSSM (top), mGMSB (middle) and mAMSB (bottom)
scenarios for $\mu > 0$ with their respective total $\chi^2$, i.e.\
including the $\chi^2$ contribution of $\Mh$. 
The color coding is as in \reffi{fig:cmssmHigh}. 
The mGMSB shows for all masses (see below)
a local minimum at a lower value and an absolute minimum at a
somewhat higher mass value. The effect of having a minimum in
the $\chi^2$ plot can in general be understood by investigating the
$\chi^2$ contribution of $\Mh$ and of $(g-2)_\mu$. While the former
penalizes strongly a light spectrum (especially for the stops), the
latter penalizes a heavy spectrum (especially sleptons and
charginos/neutralinos). The appearance of the second local minimum at
lower mass values is a result from the interplay of several observables,
especially $\MW$ and $\Mh$. Going to a lighter spectrum improves
$\chi^2(\MW)$ more than it worsens $\chi^2(\Mh)$, while a very light
spectrum results in a very large $\chi^2$ contribution from $\Mh$,
yielding the local minimum in between.

In the three scenarios limited ranges can be observed for the $\De_1$ and
$\De_4$ regions, whereas the $\De_9$ regions extend to the highest
possible mass values. For the CMSSM and mAMSB the truncation of the parameter
space at high $m_{1/2}$, $m_{\rm aux}$ and $m_0$ is
clearly visible for some particle masses, e.g.\ in the left column of
\reffi{fig:asbs_neu}. The mass of the lightest neutralino
(the LSP) has  `preferred' values, $\De\chi^2 < 4$, ranging from about  
$100 \gev$ to values up to $500 \gev$, depending on the scenario.
Within the CMSSM and mAMSB the lightest neutralino, being stable, cannot
be observed via a decay to other particles, 
so that its detection has to rely on a `missing energy' signature.
In mGMSB the LSP is the gravitino, $\tilde G$, leading to
distinctive decay patterns of the $\neu{1}$ if it decays within the
detector. The decay BRs depend largely on the
mass pattern of the $\neu{1}$, $\staue$ and $\tilde G$. The `preferred'
mass values thus offer good prospects for the detection at the LHC and
excellent prospects for the ILC(1000) (i.e.\ with $\sqrt{s}$ up to 
$1 \tev$) in the case where the decay happens in the detector. At the
ILC also the process 
$e^+e^- \to \neu{1}\neu{1}\ga$ can in principle be observed, permitting
in this case the observation of the $\neu{1}$ in all the three scenarios
in the `preferred' mass ranges.

The second lightest neutralino, see the right column of \reffi{fig:asbs_neu},
can in principle be observed via its decay to a SM particle and the LSP
(or another SUSY particle if it is lighter than the $\neu{2}$, as e.g.\
the $\cha{1}$ in the case of the mAMSB). The best fit values vary around
$300 \gev$ to values above $550 \gev$, depending on the scenario. 
With these mass ranges the observation at the LHC will be very
challenging for the direct production, but might be better (depending on
SUSY mass patterns) for the production in cascades. 
At the ILC(1000)
one could search for the associated production of $e^+e^- \to \neu{1}\neu{2}$. 
The three soft SUSY-breaking scenarios show similar prospects for the
discovery, although mGMSB results in overall somewhat higher mass scales.

The predictions of the lightest chargino mass, $\mcha{1}$ (left), and the
gluino mass, $\mgl$ (right), are shown in 
\reffi{fig:asbs_cha}. 
As before, the masses are shown in the CMSSM (top), mGMSB (middle) and
mAMSB (bottom) scenarios for $\mu > 0$ with their respective total $\chi^2$.
The color coding is as in \reffi{fig:cmssmHigh}. 
In the three scenarios limited ranges can be observed for the $\De_1$ and
$\De_4$ regions, whereas the $\De_9$ regions extend to the highest
possible mass values. 
Within the CMSSM and mGMSB the light chargino mass ranges from about 
$100 \gev$ up to $\sim 900 \gev$ in the $\De_4$ area, whereas somewhat
higher masses are 
reached in mGMSB. Consequently only a part of the `preferred' parameter
space can be accessed at the LHC or the ILC(1000). 
Within the CMSSM and mGMSB the $\cha{1}$ and the $\neu{2}$ are nearly
mass degenerate, resulting in very similar results for the two particles
as can be seen in \reffis{fig:asbs_neu} and \ref{fig:asbs_cha}.
The situation concerning the observation of the $\cha{1}$ is much
more favorable in mAMSB, where much lighter masses, only up to about 
$300 \gev$ are preferred. This offers very good perspectives for its
production at the LHC and the ILC. However, it should be kept in mind
that in the mAMSB scenario the lightest chargino is only a few hundred
MeV heavier than the LSP, which poses certain problems for its
detection~\cite{AMSBpheno}.

The `preferred' gluino masses, as shown in the right column of
\reffi{fig:asbs_cha}, range from a few hundred GeV up to about 
$3 \tev$ in mGMSB, exhausting the accessible range at the LHC. In the
other two scenarios the $\De_4$ regions end at $\sim 2 \tev$ (mAMSB)
and $\sim 2.5 \tev$ (CMSSM), making them more easily accessible at the LHC
than in the mGMSB scenario.

We now turn to the scalar fermion sector.
The predictions for the two scalar tau masses, $\mstaue$ (left)
and $\mstauz$ (right), are shown in \reffi{fig:asbs_stau}. 
As before, the masses are shown in the CMSSM (top), mGMSB (middle) and
mAMSB (bottom) scenarios for $\mu > 0$ with their respective total $\chi^2$.
The color coding is as in \reffi{fig:cmssmHigh}. 
The light $\stau$ has its best-fit values at very low masses, and even
the $\De_4$ regions hardly exceed $\sim 500 \gev$ in mGMSB and mAMSB.
Therefore in these scenarios there are good prospects for the ILC(1000).
Also the LHC can be expected to  
cover large parts of the $\De_4$ mass intervals.
In the CMSSM scenario, on the other hand, the $\De_4$ region exceeds
$\sim 1 \tev$ such that only parts can be probed at the ILC(1000) and
the LHC. The `preferred' $\mstauz$
values, by construction larger than $\mstaue$, stay mostly below 
500, 1000, $1500 \gev$ for mAMSB, mGMSB and the CMSSM, respectively.

\begin{figure}[htb!]
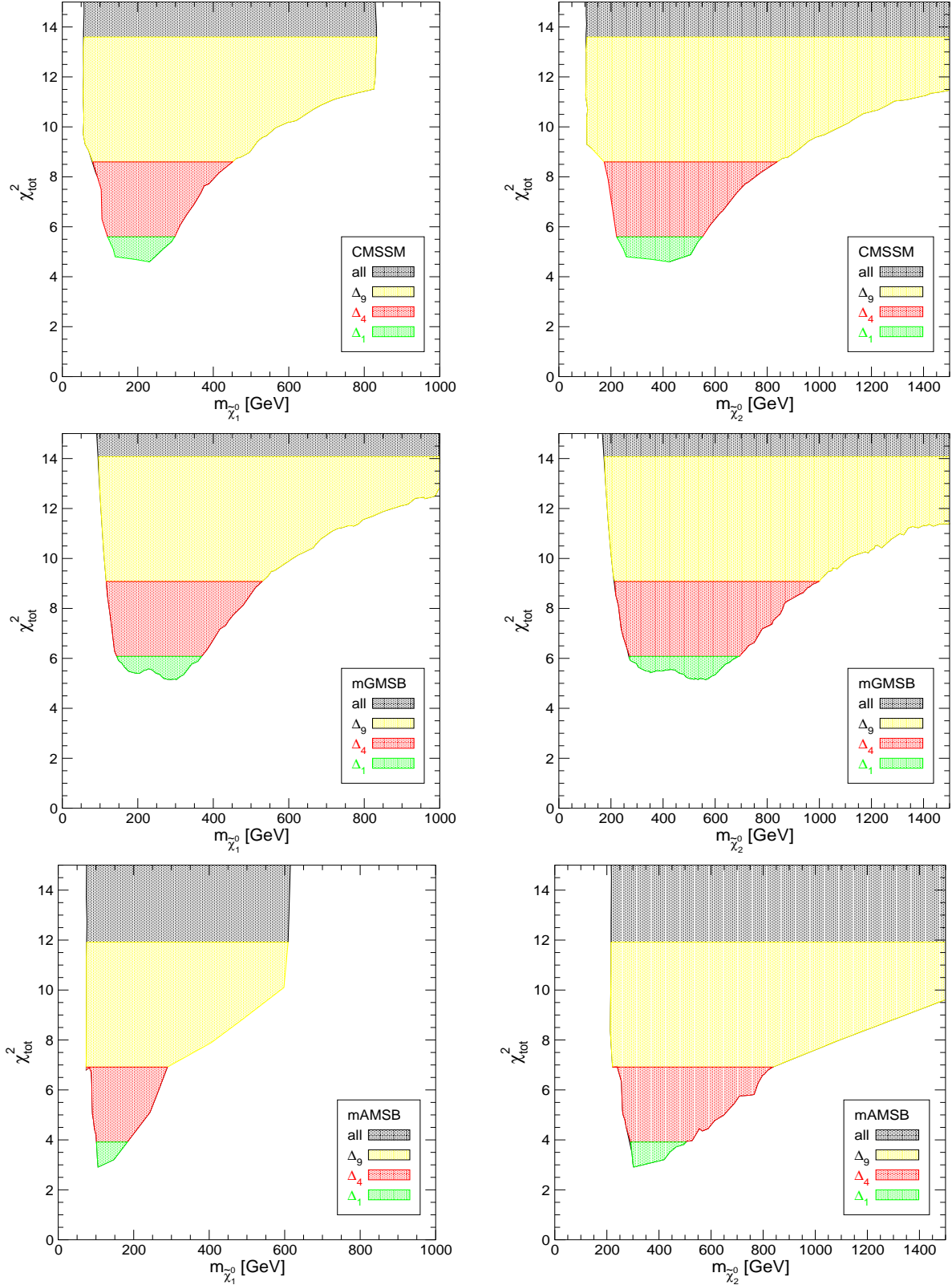

\begin{center}
\includegraphics[width=.45\textwidth,height=7.0cm]{asbs3_s_mass11_cl}
\hspace{0.5cm}
\includegraphics[width=.45\textwidth,height=7.0cm]{asbs3_s_mass12_cl}\\[.3em]
\includegraphics[width=.45\textwidth,height=7.0cm]{asbs3_g_mass11_cl}
\hspace{0.5cm}
\includegraphics[width=.45\textwidth,height=7.0cm]{asbs3_g_mass12_cl}\\[.3em]
\includegraphics[width=.45\textwidth,height=7.0cm]{asbs3_a_mass11_cl}
\hspace{0.5cm}
\includegraphics[width=.45\textwidth,height=7.0cm]{asbs3_a_mass12_cl}
\vspace{-1em}
\caption{%
$\mneu{1}$ (left) and $\mneu{2}$ (right) are 
shown in the CMSSM (top), mGMSB (middle) and mAMSB (bottom)
scenarios for $\mu > 0$ with their respective total $\chi^2$, i.e.\
including the $\chi^2$ contribution of $\Mh$. 
The color coding is as in \reffi{fig:cmssmHigh}. 
}
\label{fig:asbs_neu}
\end{center} 
\vspace{-3em}
\end{figure}

\begin{figure}[htb!]
\begin{center}
\includegraphics[width=.45\textwidth,height=7.0cm]{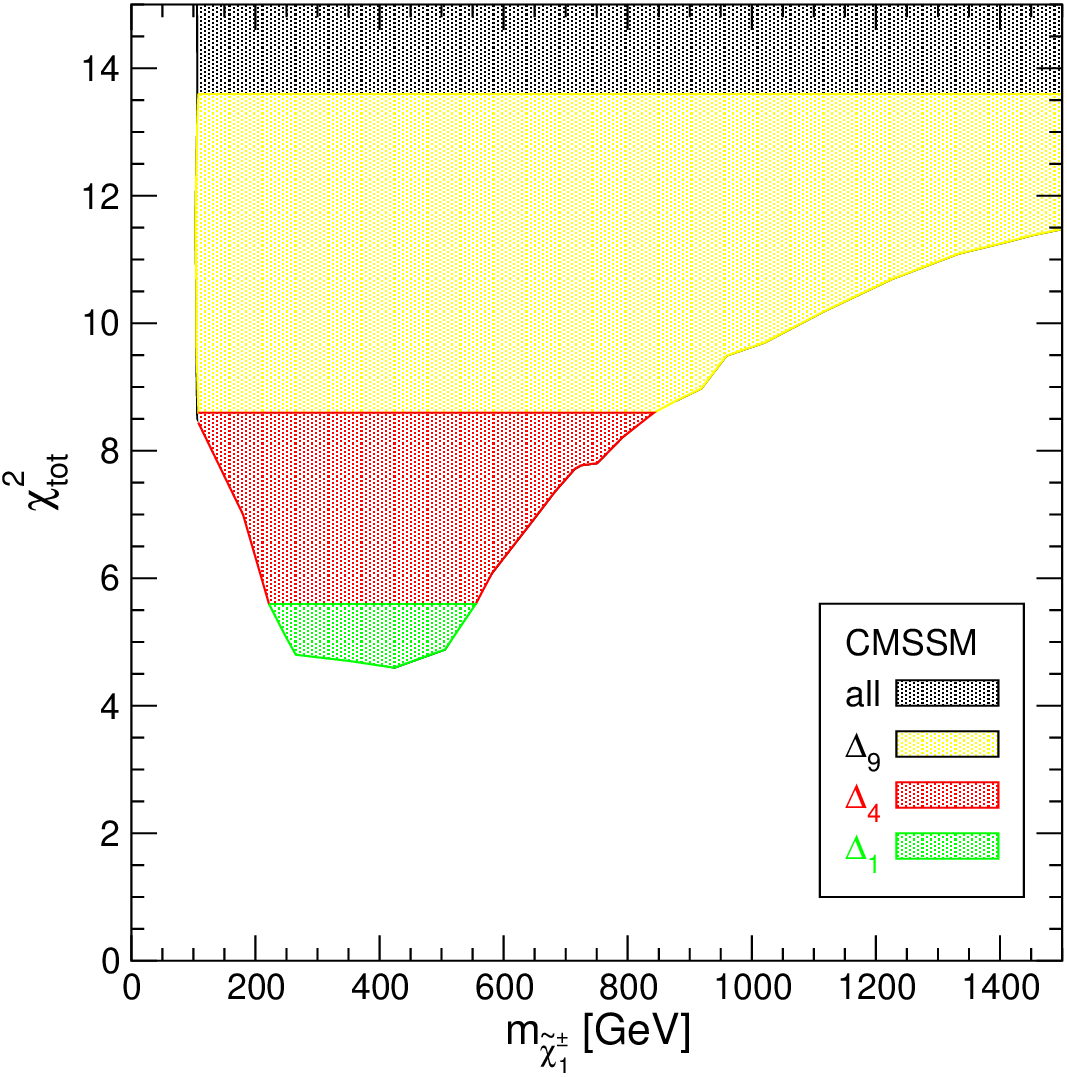}
\hspace{0.5cm}
\includegraphics[width=.45\textwidth,height=7.0cm]{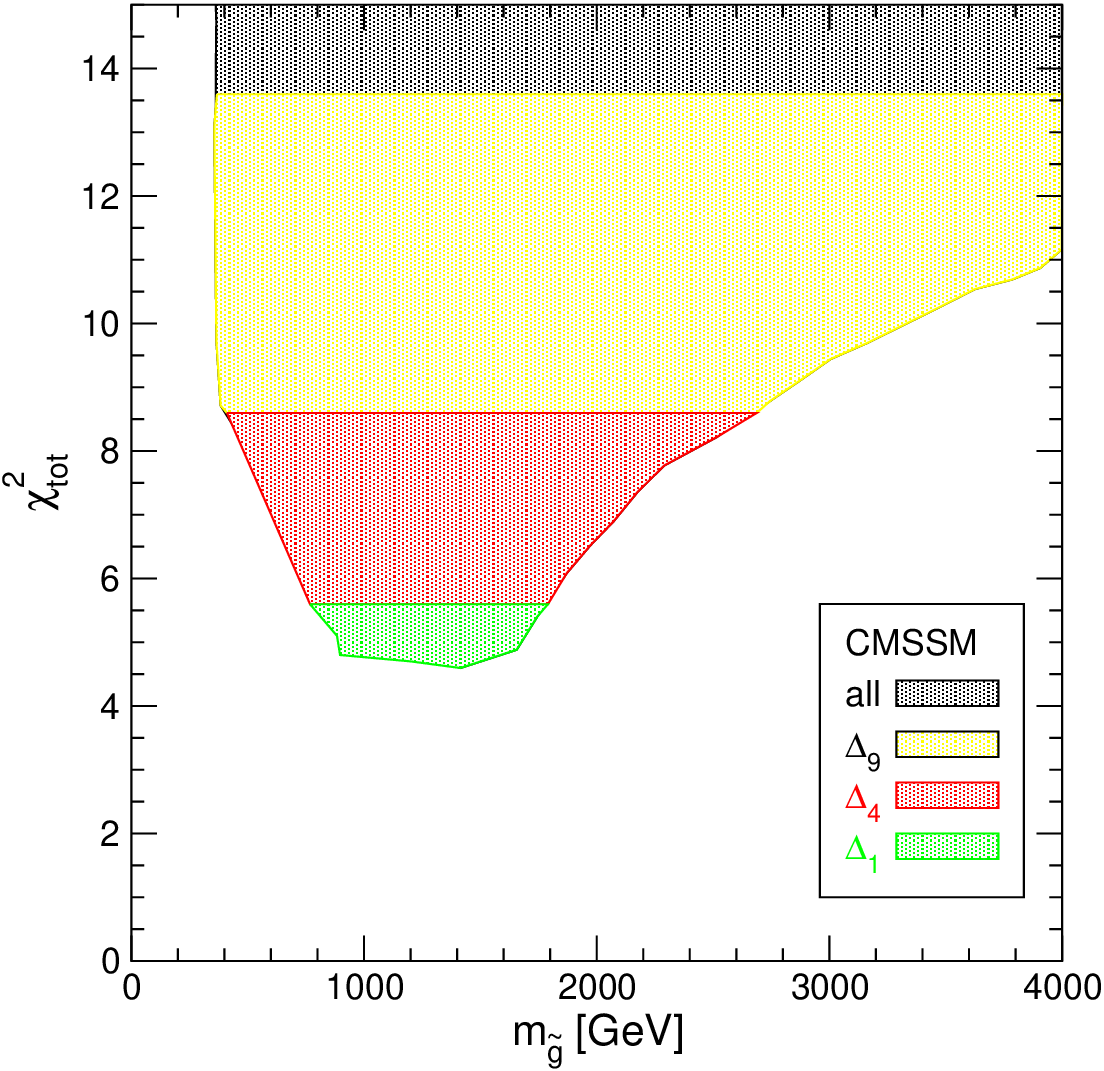}\\[.3em]
\includegraphics[width=.45\textwidth,height=7.0cm]{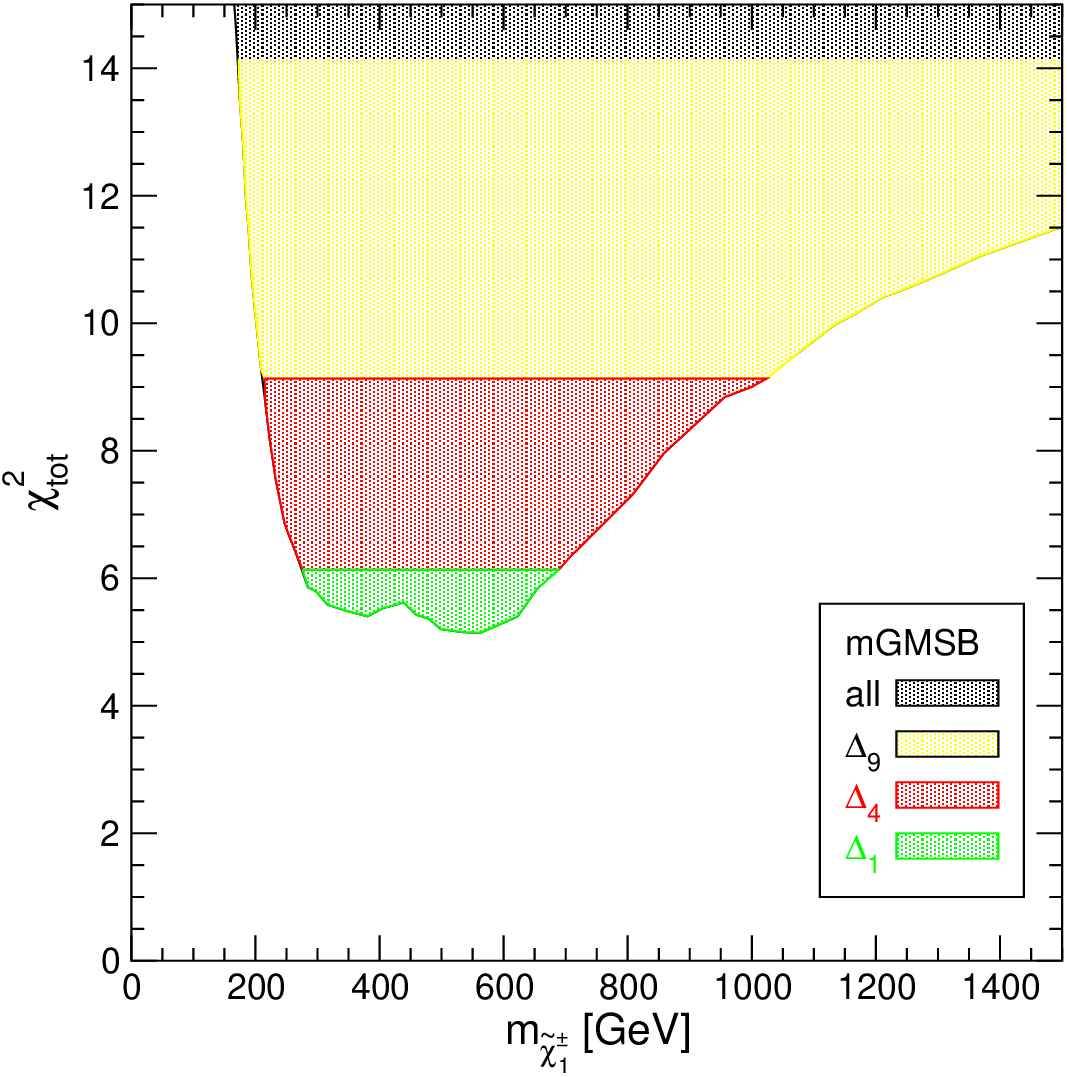}
\hspace{0.5cm}
\includegraphics[width=.45\textwidth,height=7.0cm]{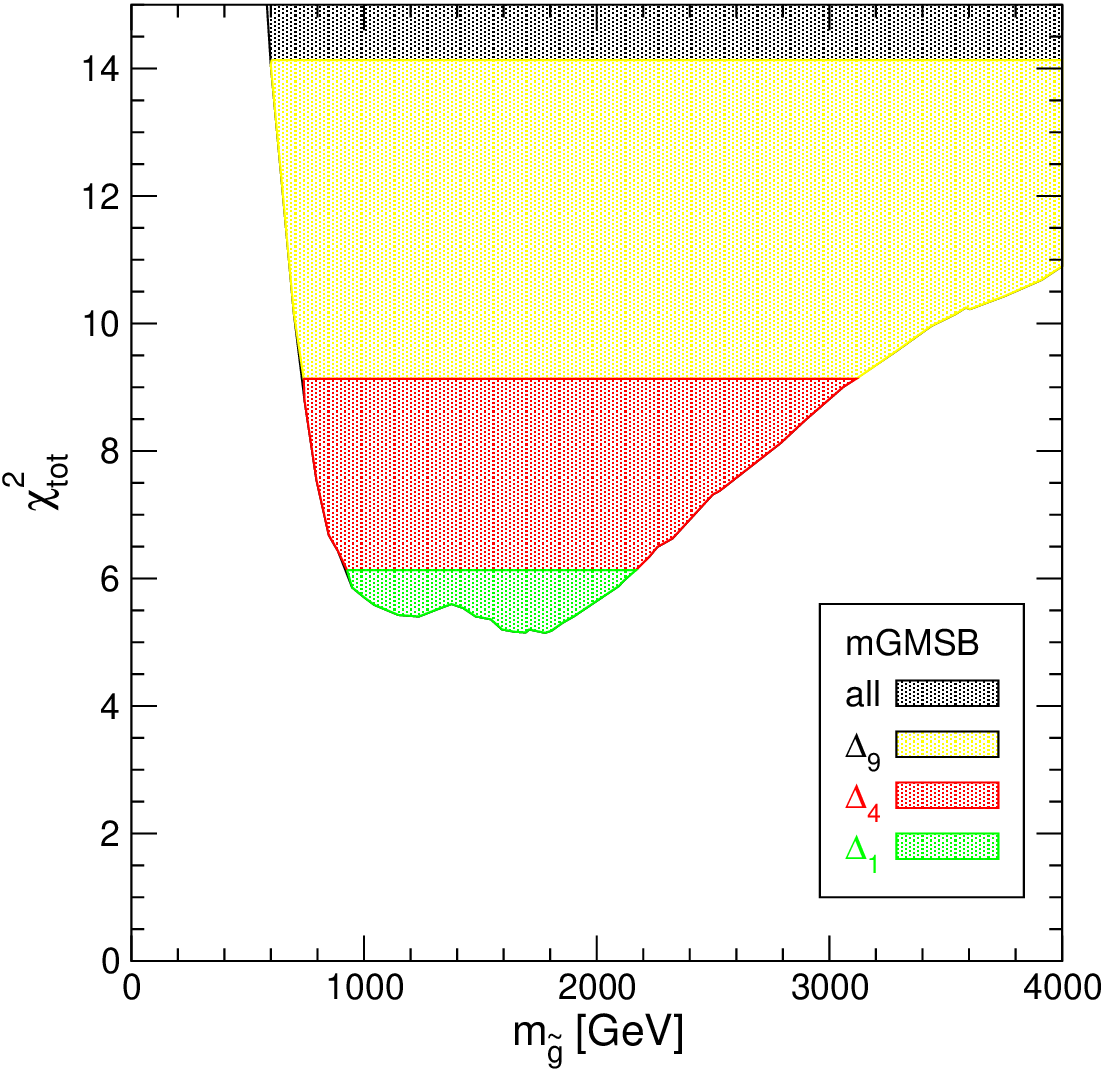}\\[.3em]
\includegraphics[width=.45\textwidth,height=7.0cm]{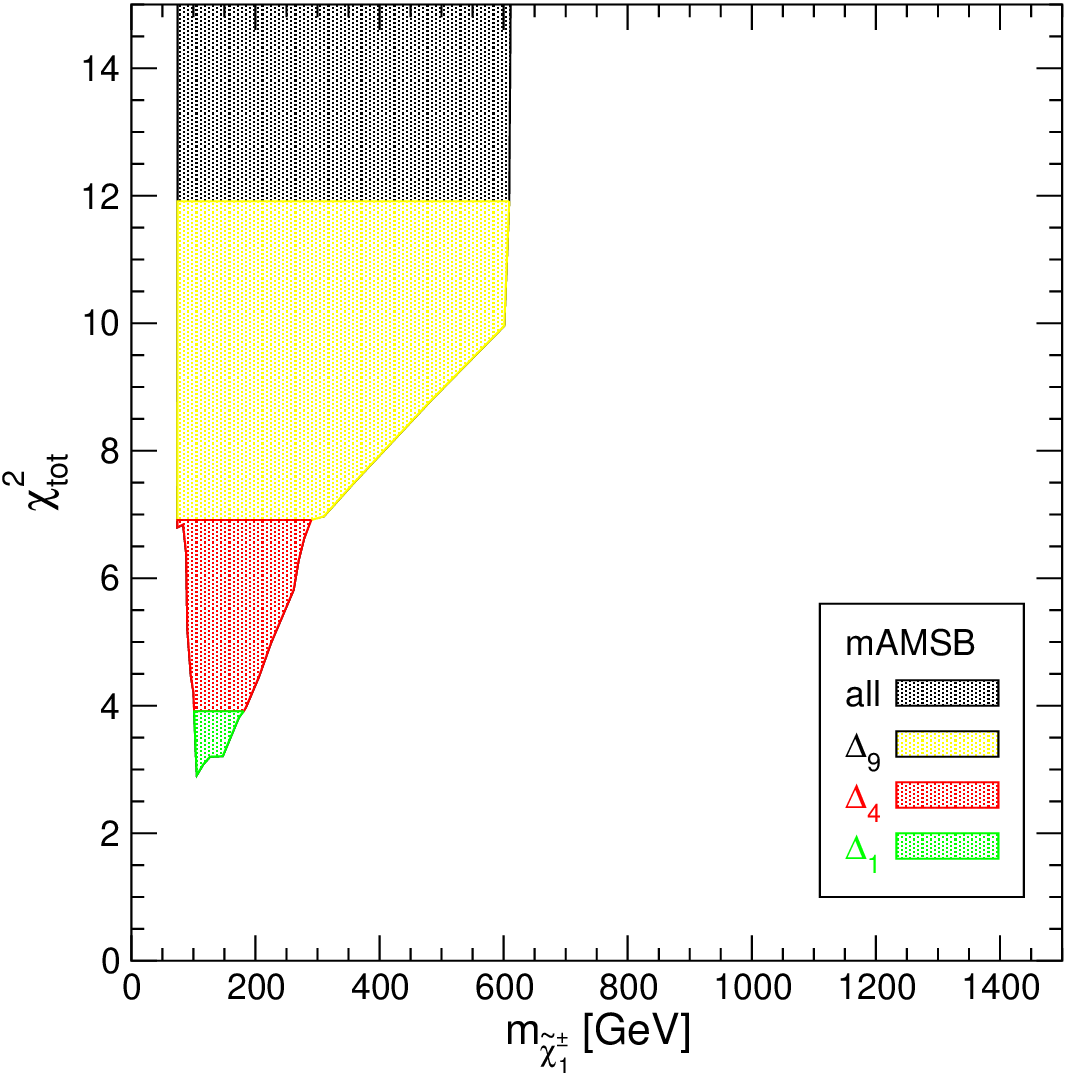}
\hspace{0.5cm}
\includegraphics[width=.45\textwidth,height=7.0cm]{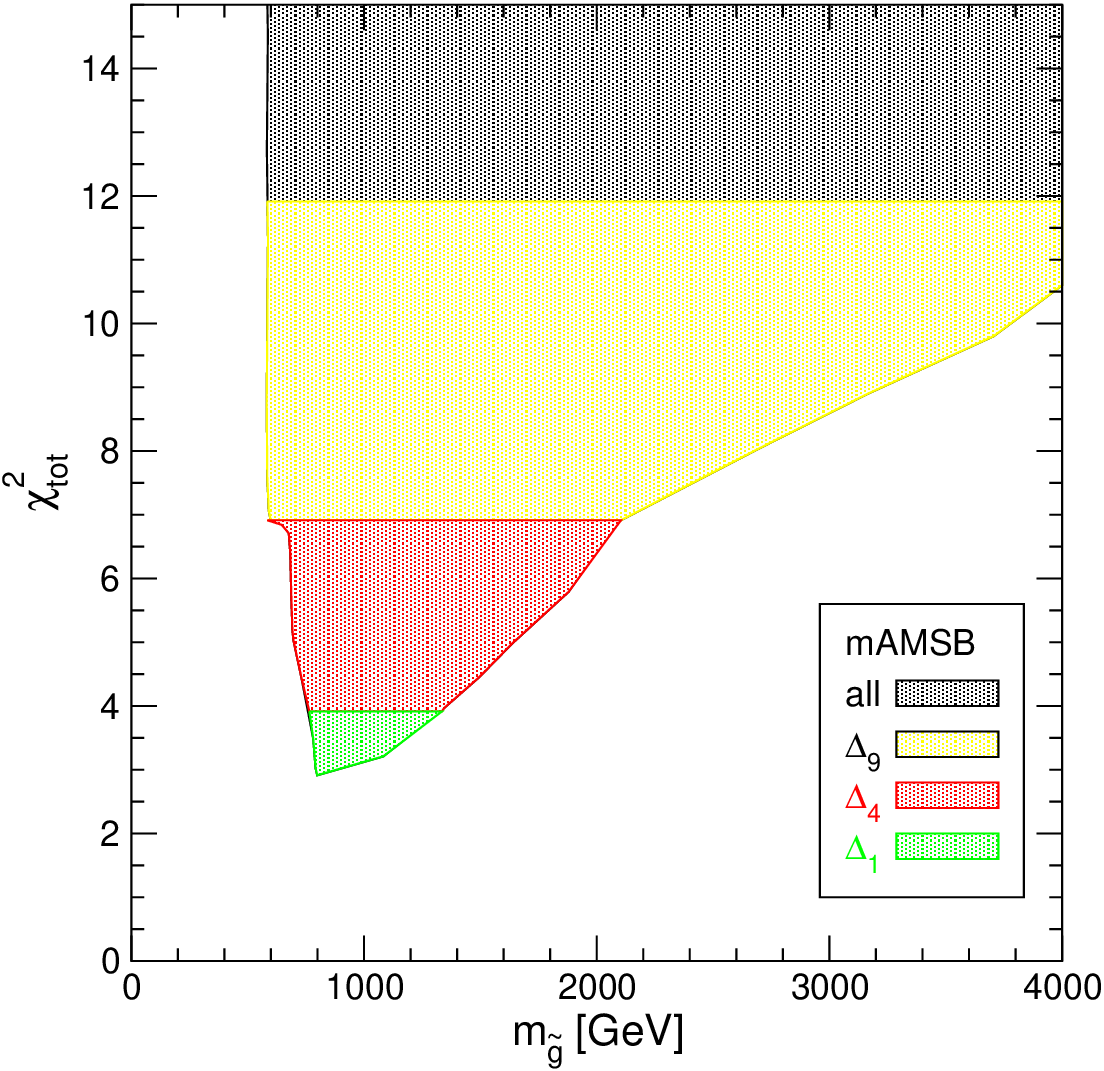}
\vspace{-1em}
\caption{%
$\mcha{1}$ (left) and $\mgl$ (right) are 
shown in the CMSSM (top), mGMSB (middle) and mAMSB (bottom)
scenarios for $\mu > 0$ with their respective total $\chi^2$.
The color coding is as in \reffi{fig:cmssmHigh}. 
}
\label{fig:asbs_cha}
\end{center} 
\vspace{-3em}
\end{figure}

\begin{figure}[htb!]
\begin{center}
\includegraphics[width=.45\textwidth,height=7.0cm]{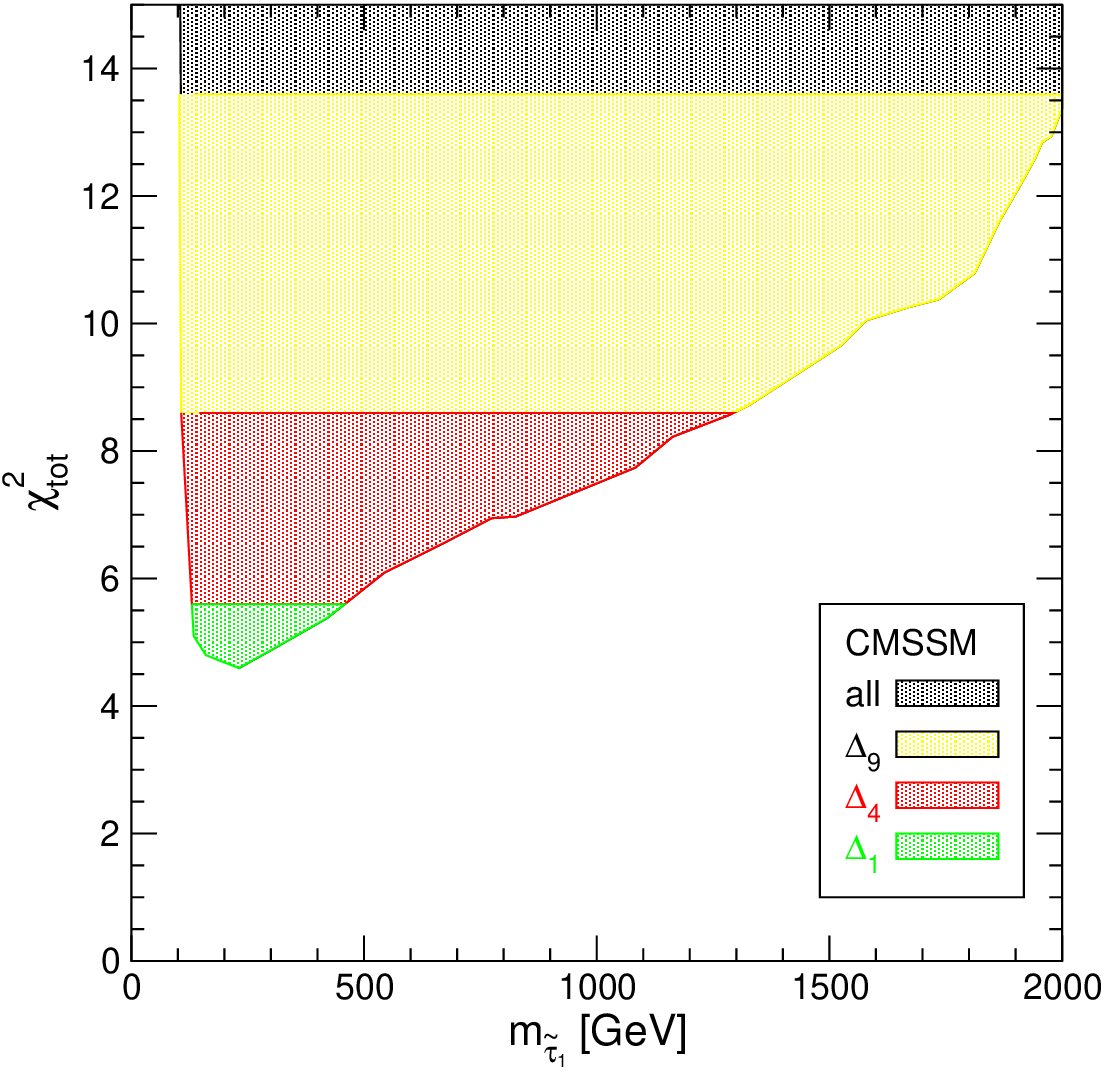}
\hspace{0.5cm}
\includegraphics[width=.45\textwidth,height=7.0cm]{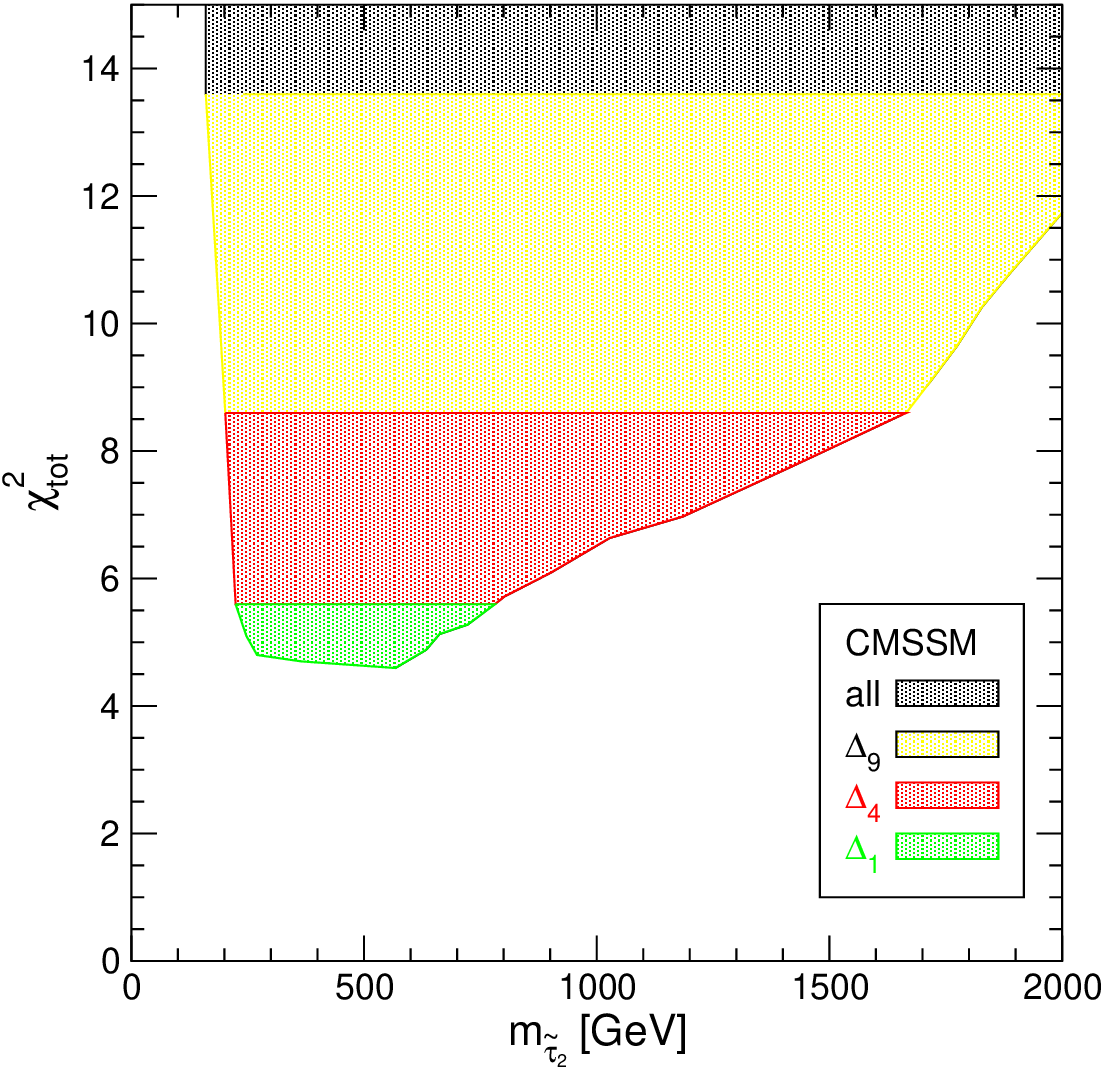}\\[.3em]
\includegraphics[width=.45\textwidth,height=7.0cm]{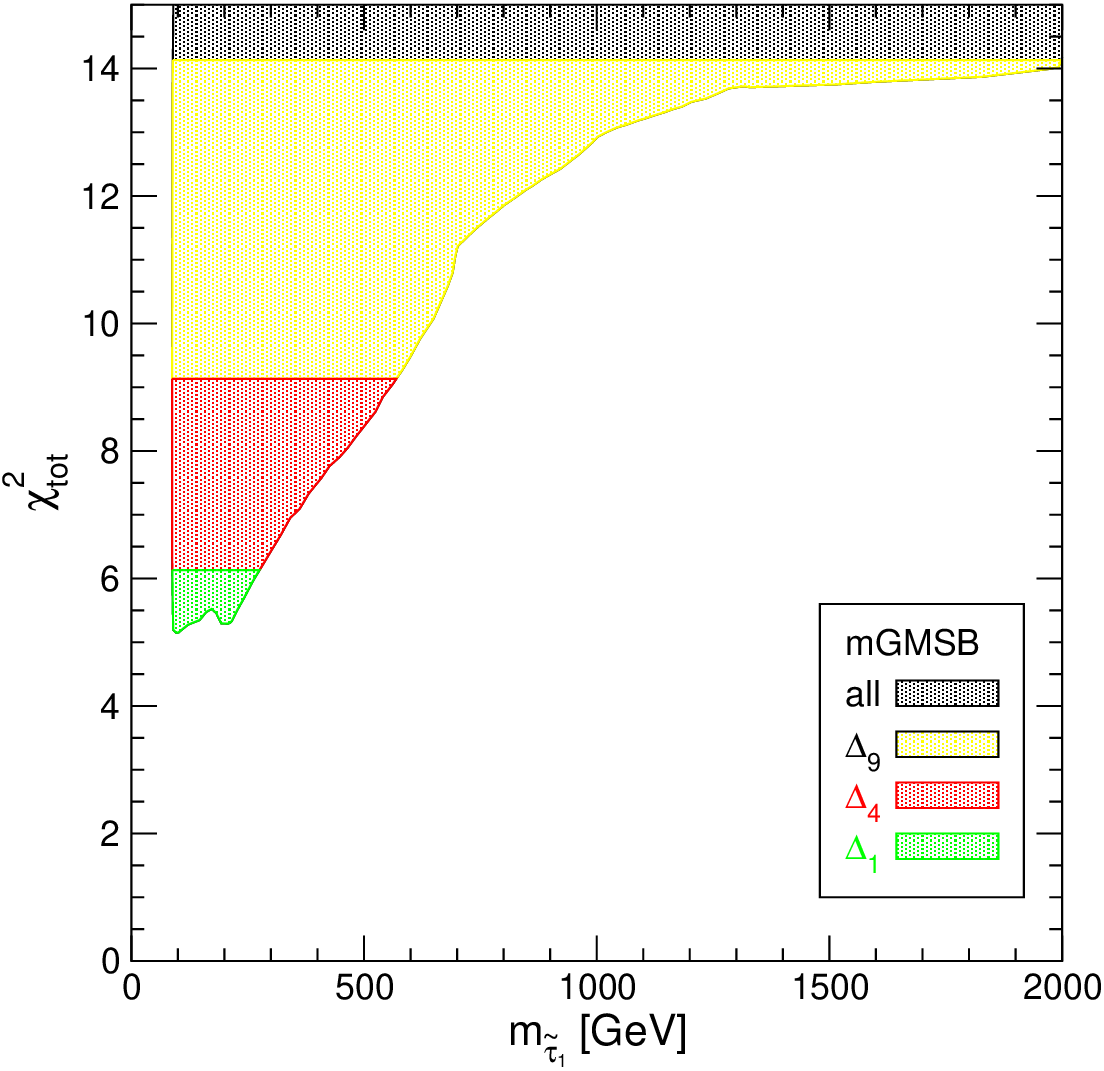}
\hspace{0.5cm}
\includegraphics[width=.45\textwidth,height=7.0cm]{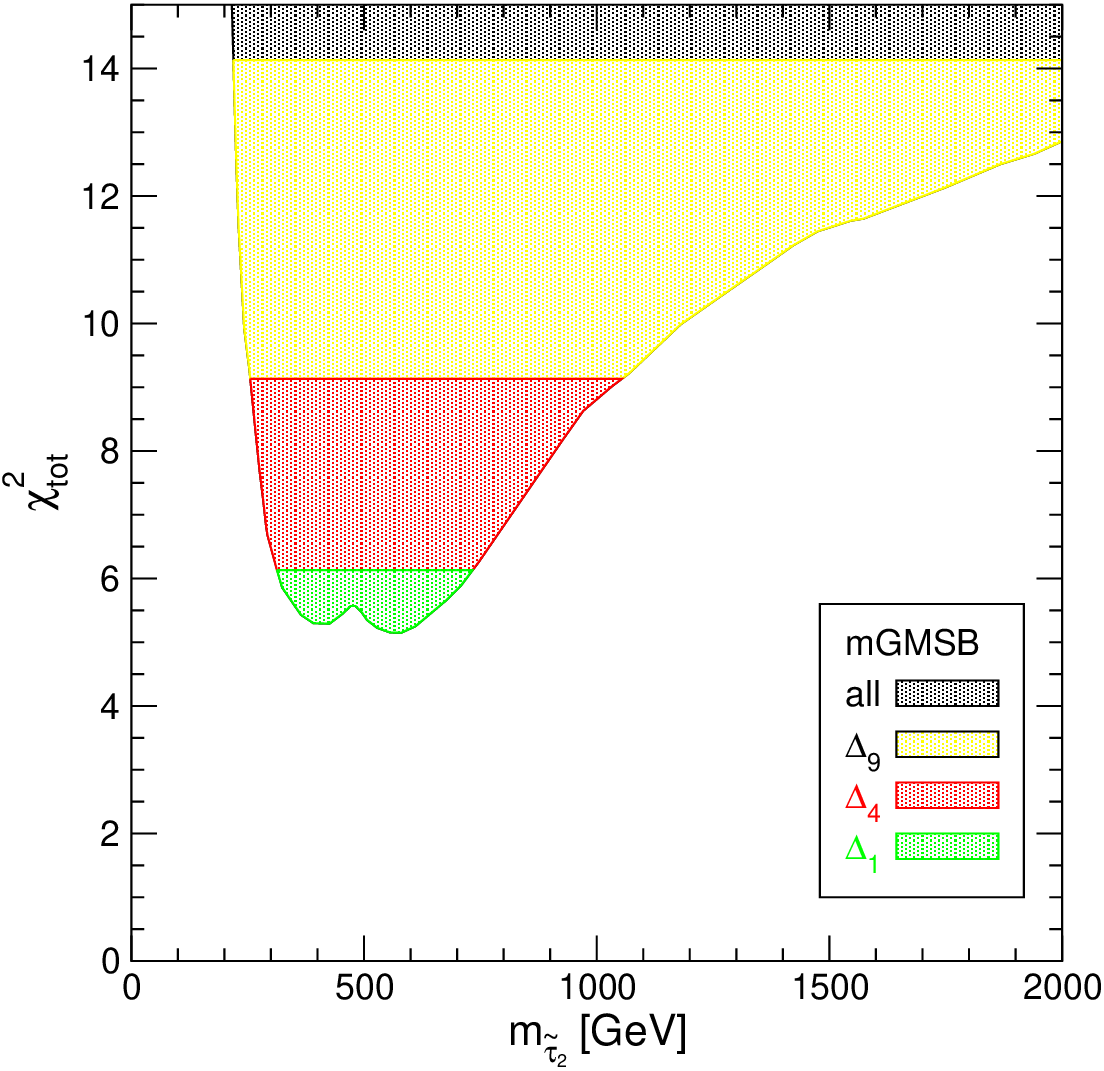}\\[.3em]
\includegraphics[width=.45\textwidth,height=7.0cm]{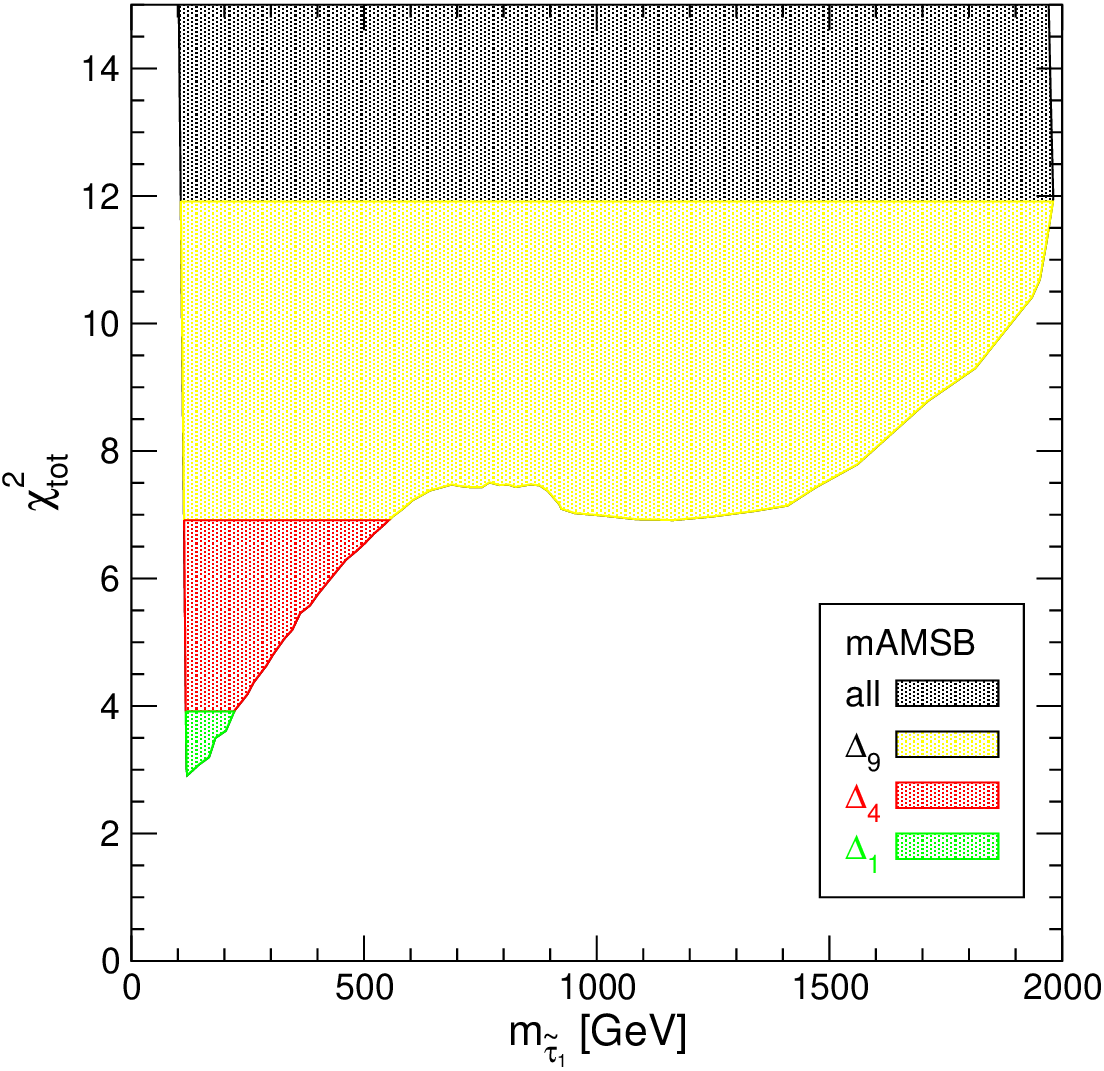}
\hspace{0.5cm}
\includegraphics[width=.45\textwidth,height=7.0cm]{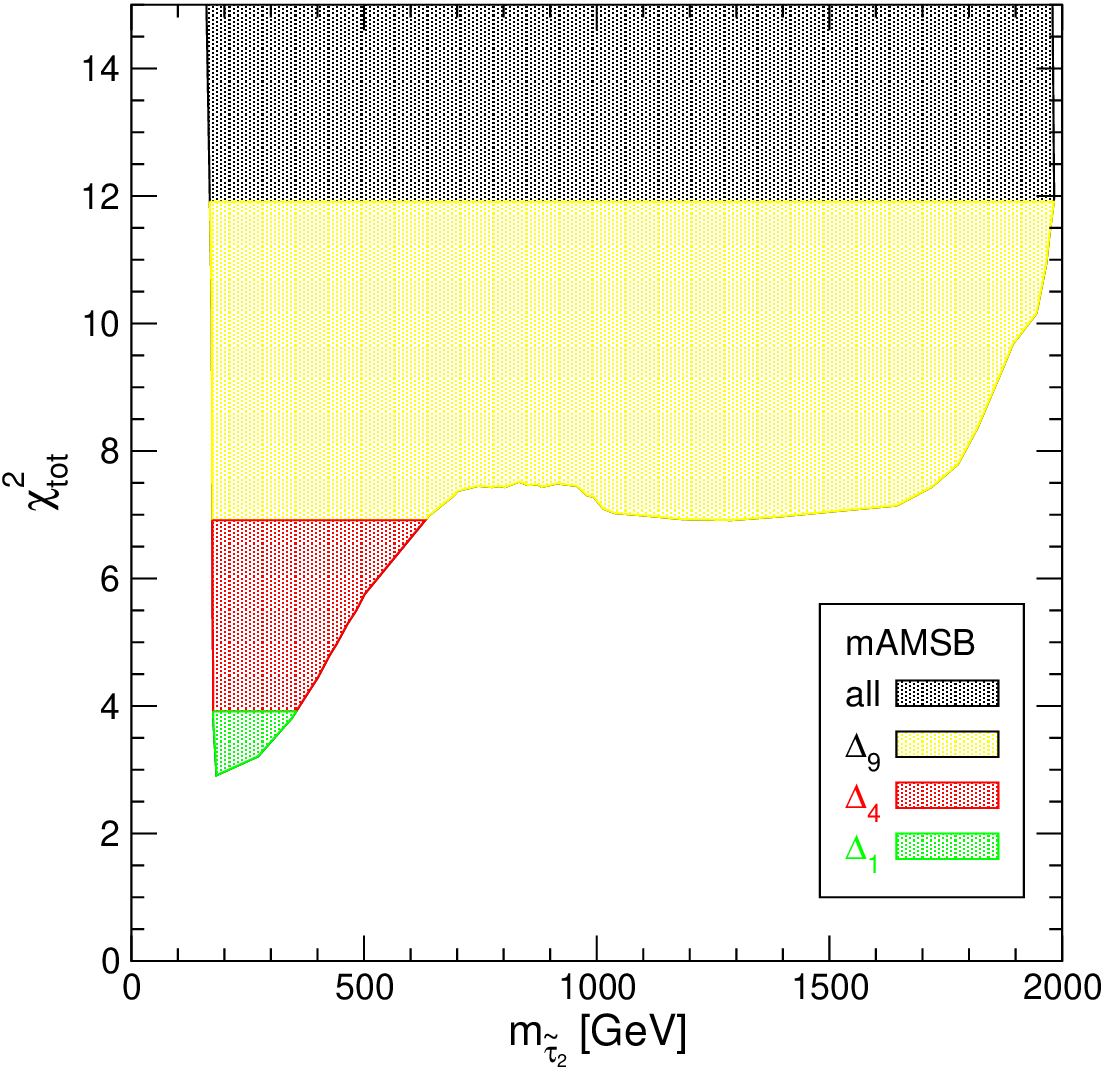}
\vspace{-1em}
\caption{%
$\mstaue$ (left) and $\mstauz$ (right) are 
shown in the CMSSM (top), mGMSB (middle) and mAMSB (bottom)
scenarios for $\mu > 0$ with their respective total $\chi^2$.
The color coding is as in \reffi{fig:cmssmHigh}. 
}
\label{fig:asbs_stau}
\end{center} 
\vspace{-3em}
\end{figure}

\begin{figure}[htb!]
\begin{center}
\includegraphics[width=.45\textwidth,height=7.0cm]{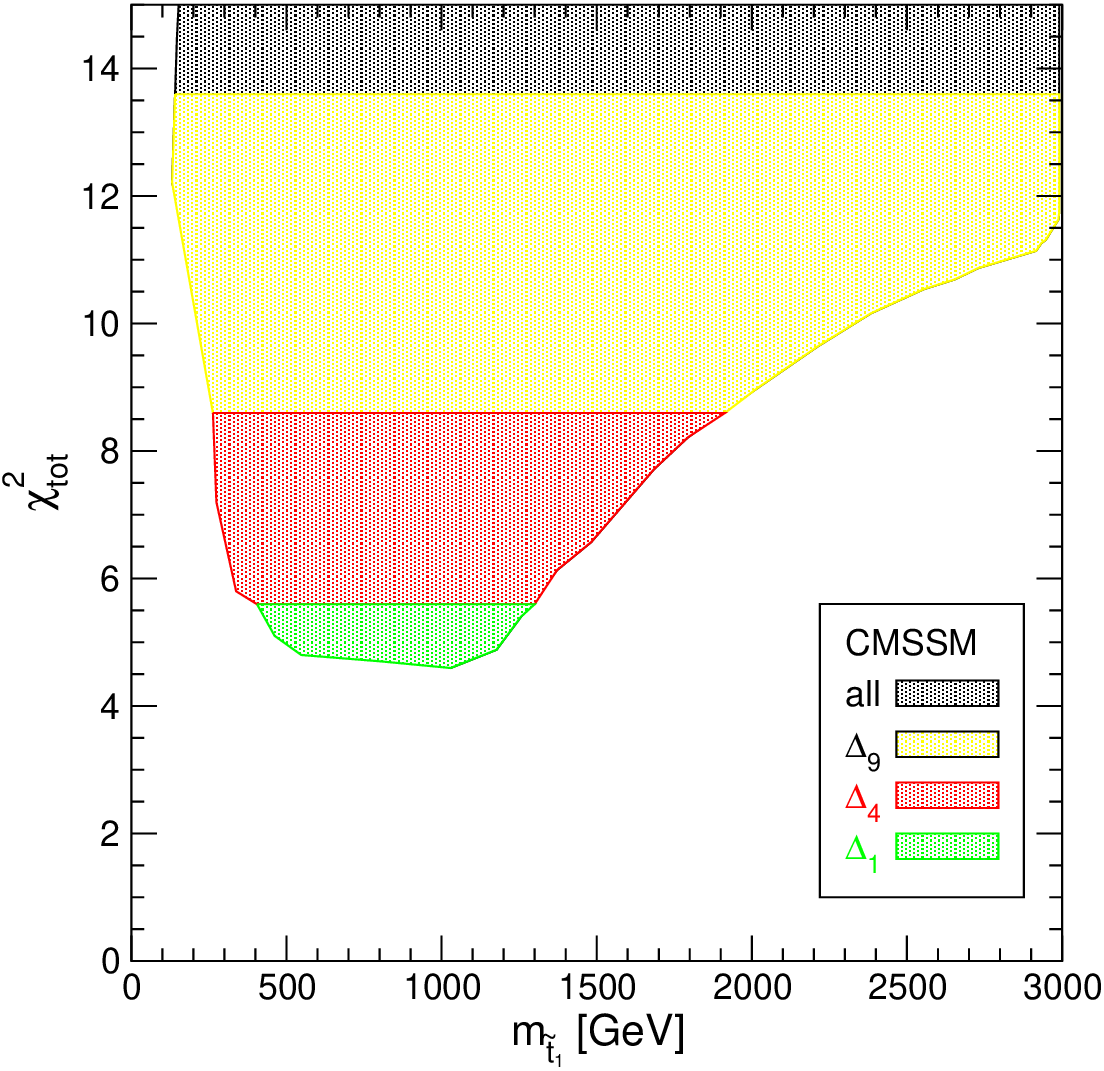}
\hspace{0.5cm}
\includegraphics[width=.45\textwidth,height=7.0cm]{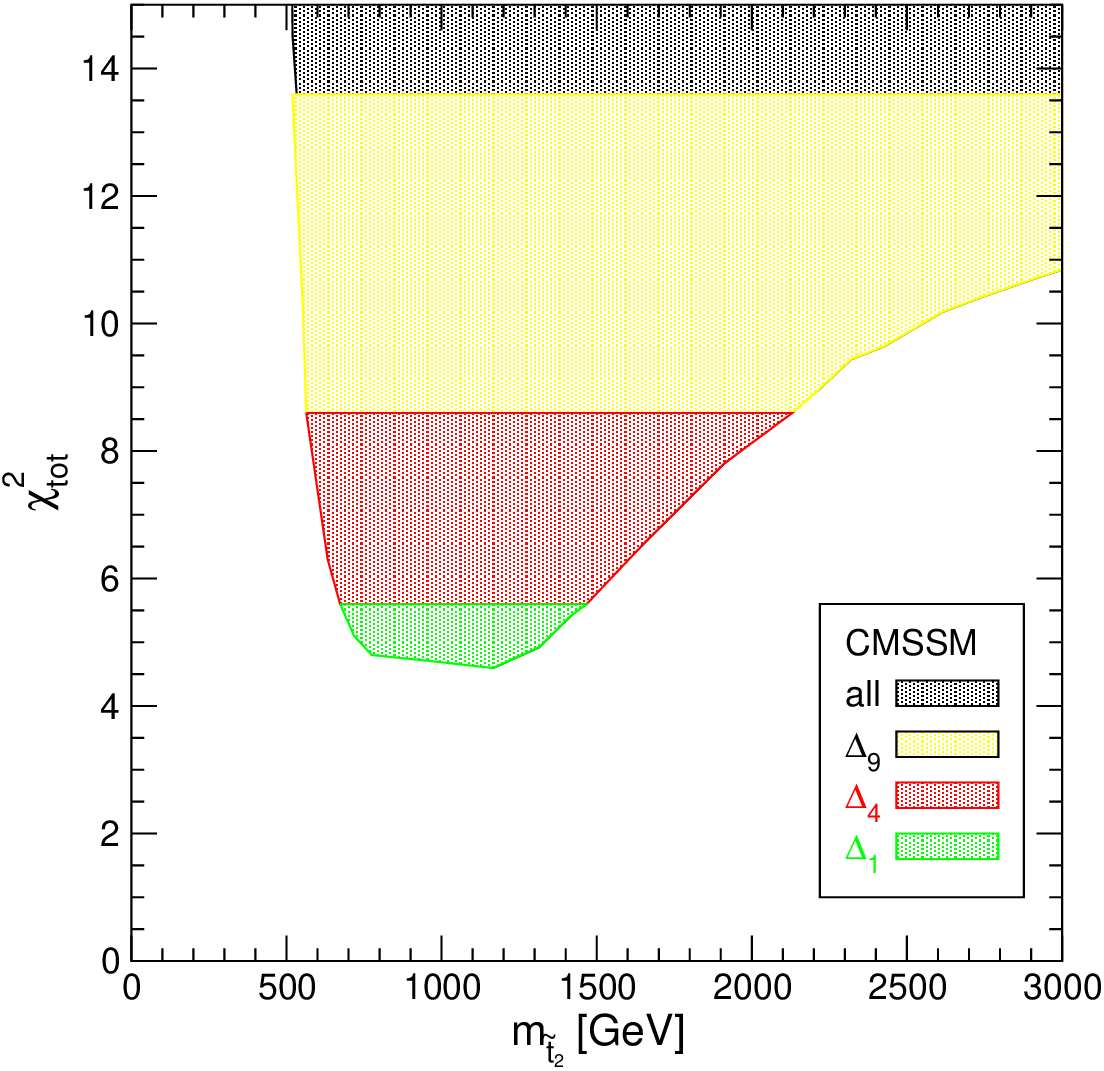}\\[.3em]
\includegraphics[width=.45\textwidth,height=7.0cm]{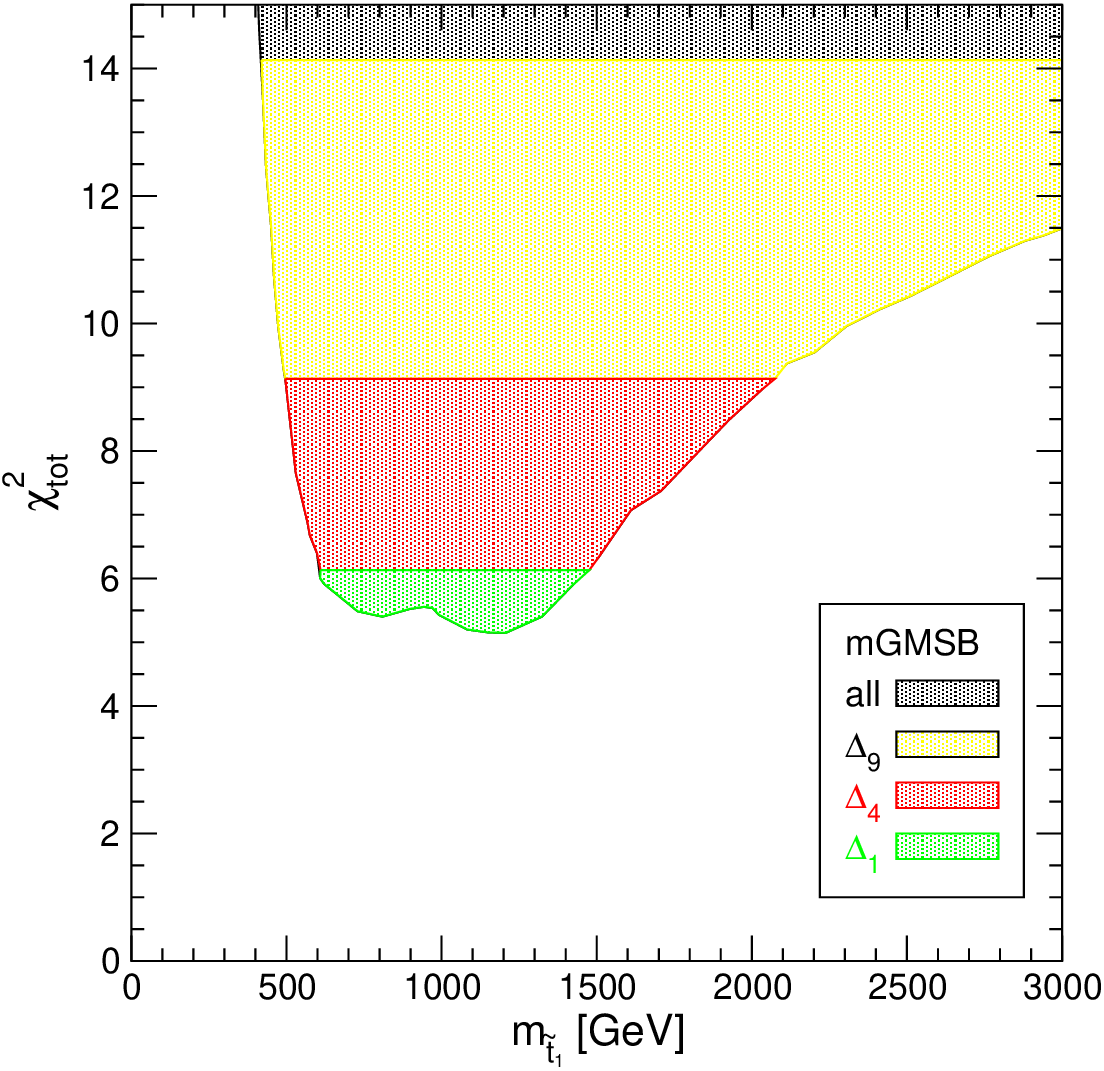}
\hspace{0.5cm}
\includegraphics[width=.45\textwidth,height=7.0cm]{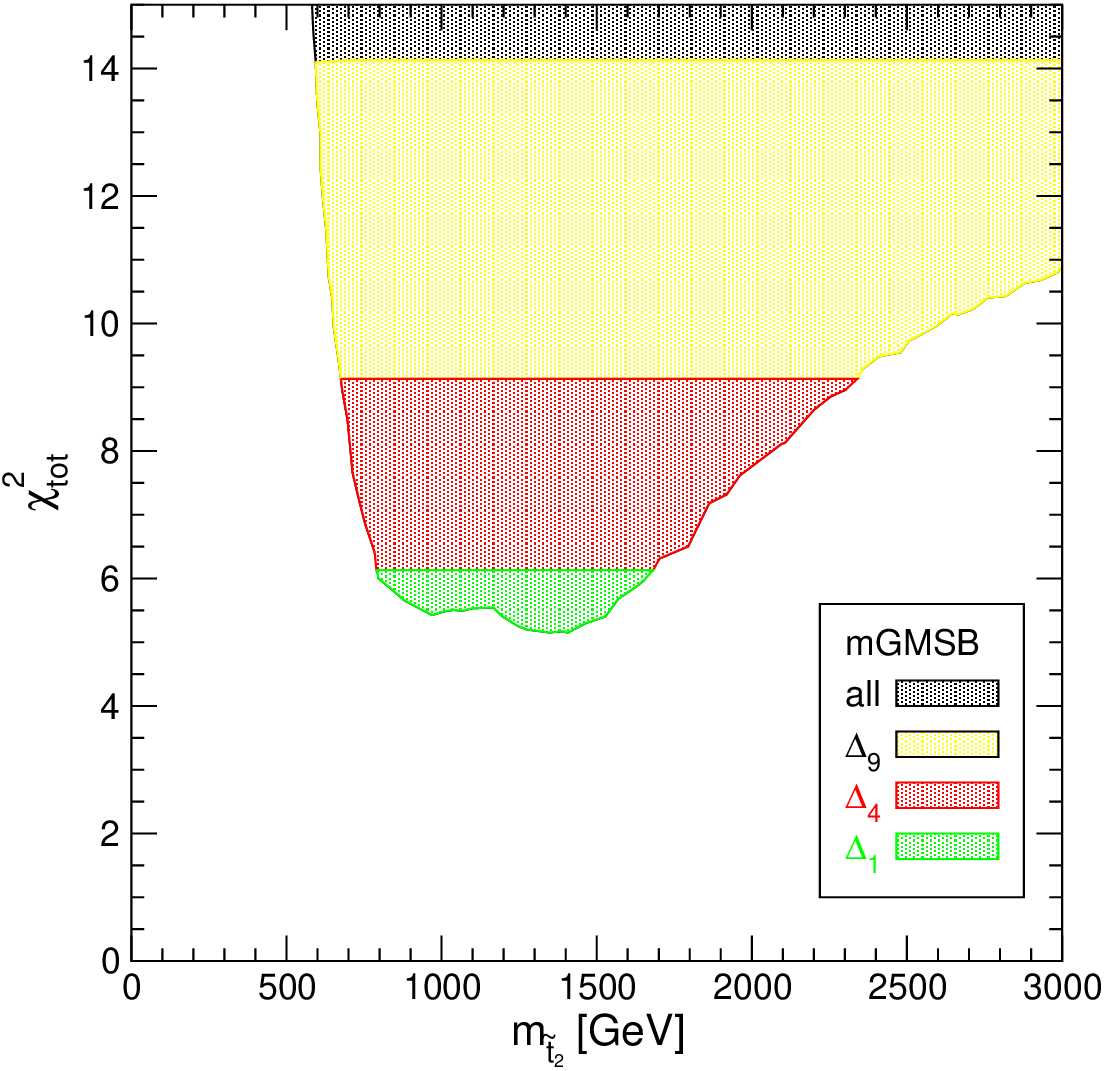}\\[.3em]
\includegraphics[width=.45\textwidth,height=7.0cm]{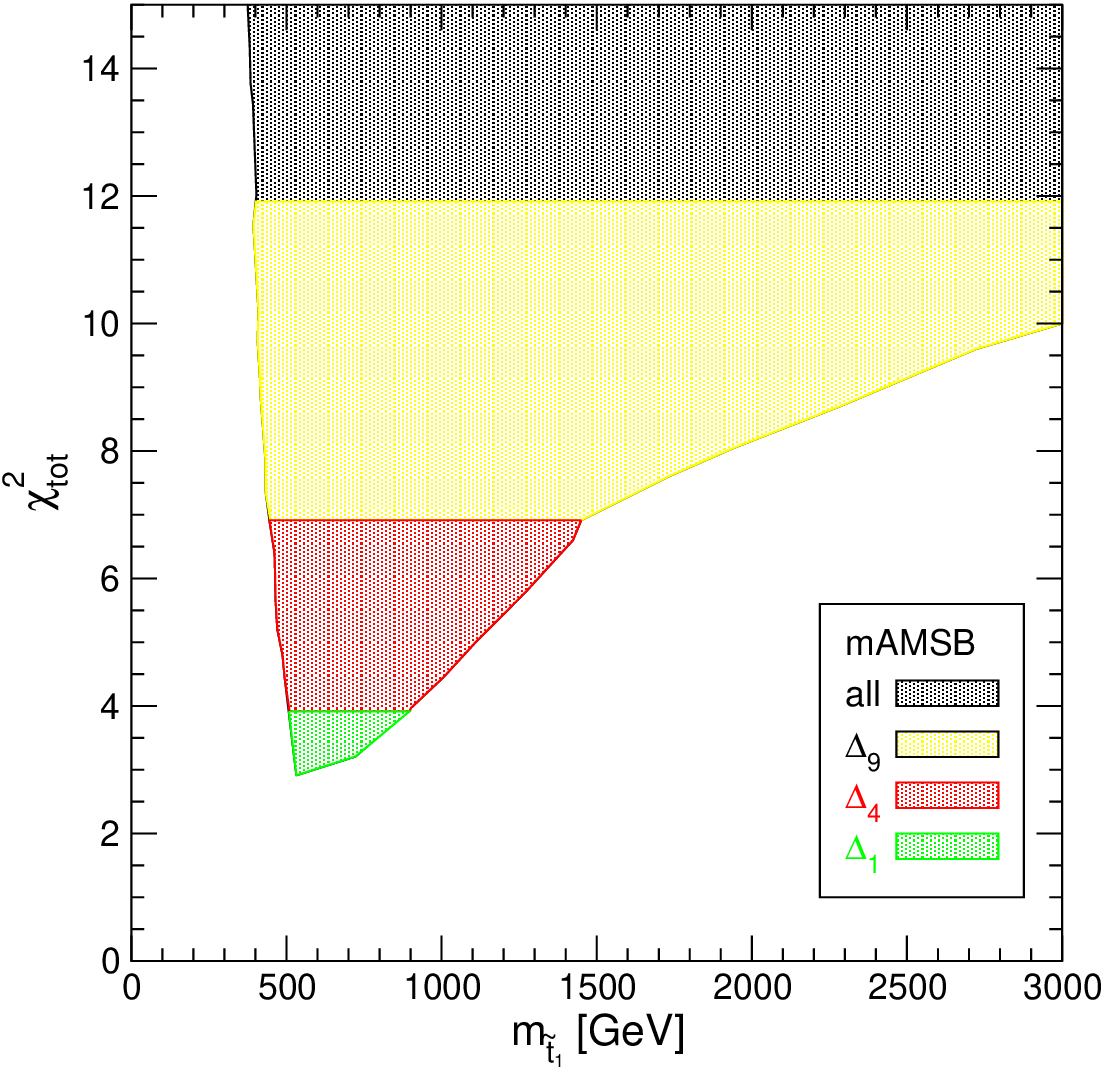}
\hspace{0.5cm}
\includegraphics[width=.45\textwidth,height=7.0cm]{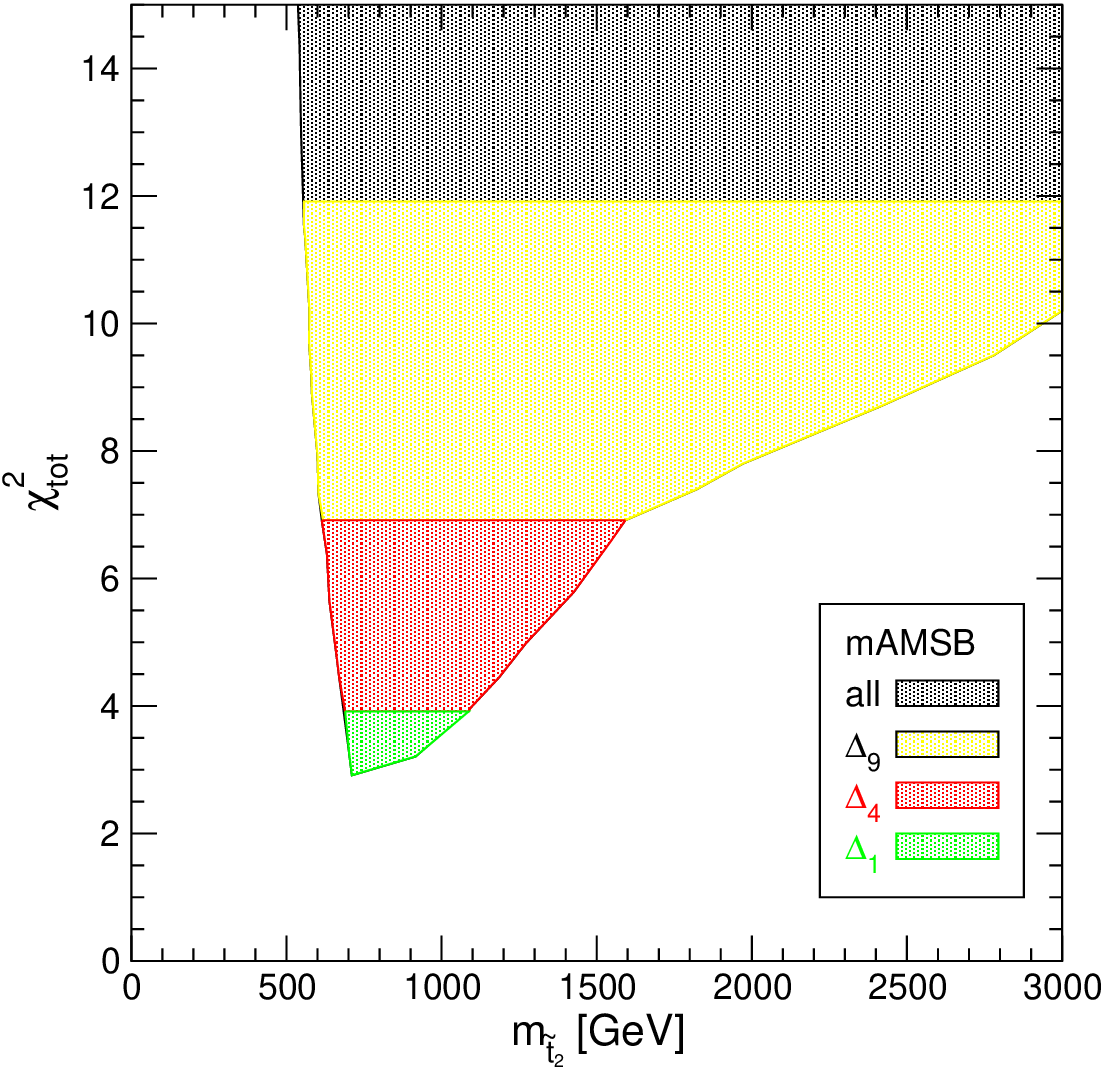}
\vspace{-1em}
\caption{%
$\mste$ (left) and $\mstz$ (right) are 
shown in the CMSSM (top), mGMSB (middle) and mAMSB (bottom)
scenarios for $\mu > 0$ with their respective total $\chi^2$.
The color coding is as in \reffi{fig:cmssmHigh}. 
}
\label{fig:asbs_stop}
\end{center} 
\vspace{-3em}
\end{figure}

\begin{figure}[htb!]
\begin{center}
\includegraphics[width=.45\textwidth,height=7.0cm]{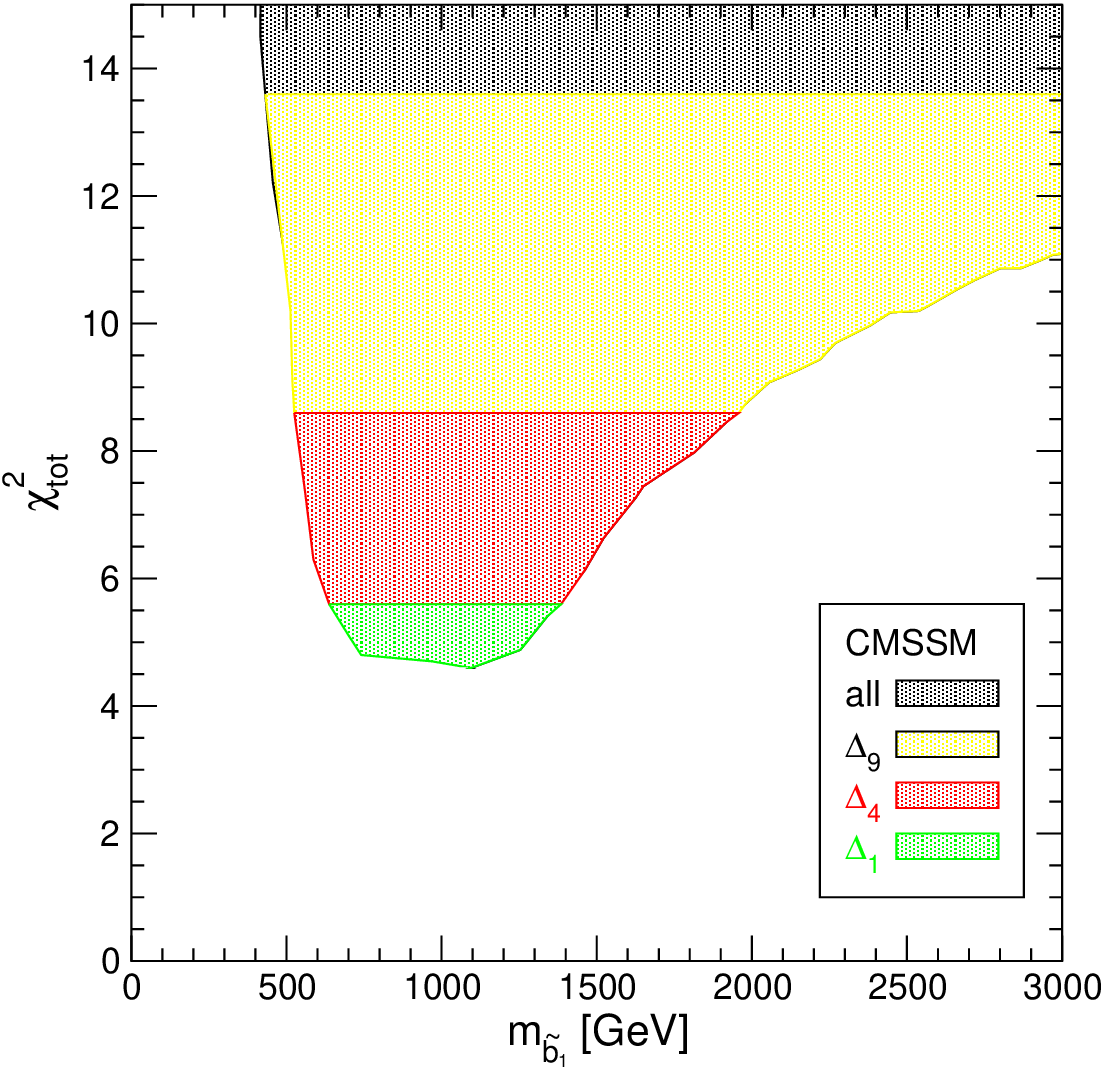}
\hspace{0.5cm}
\includegraphics[width=.45\textwidth,height=7.0cm]{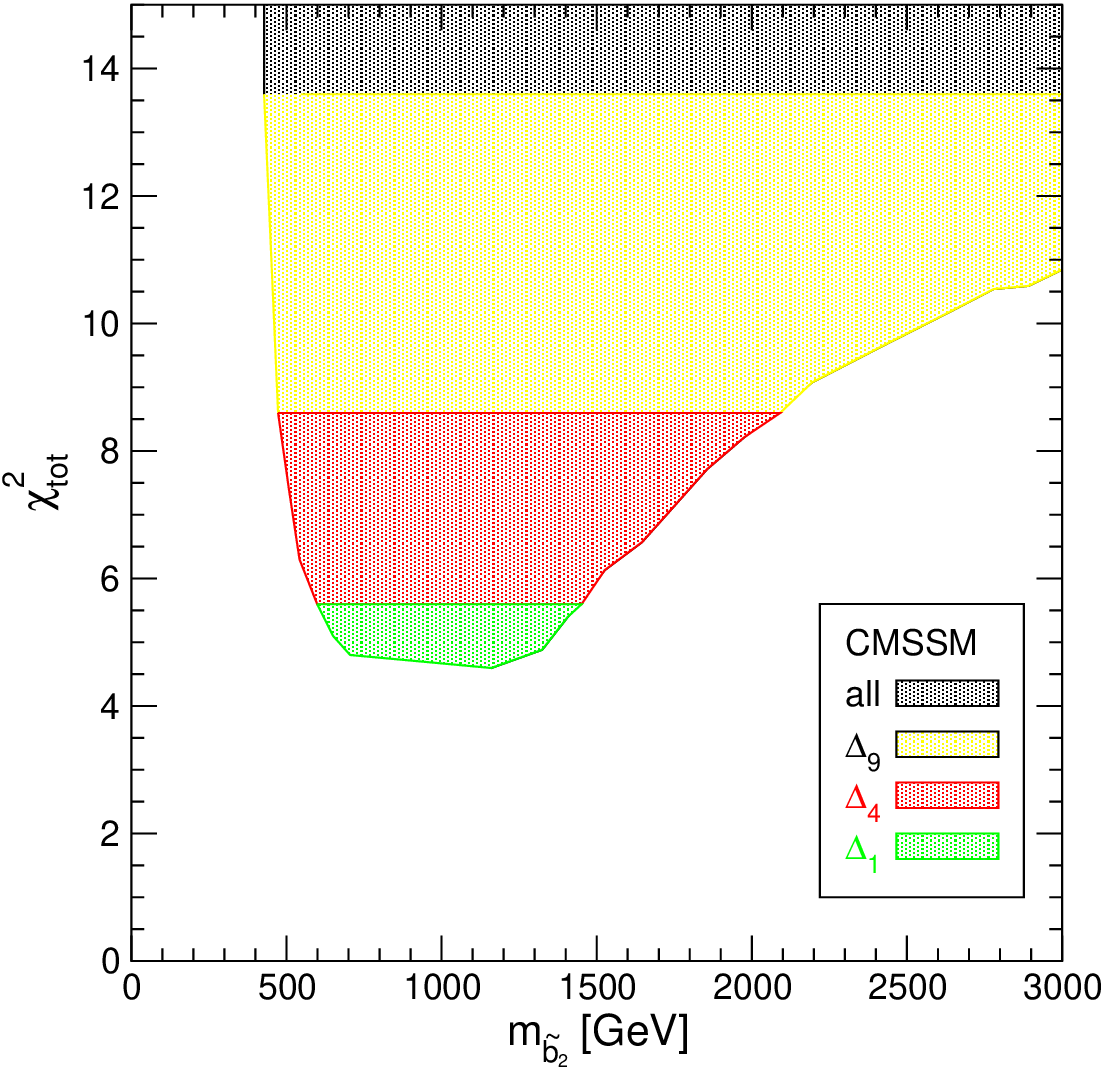}\\[.3em]
\includegraphics[width=.45\textwidth,height=7.0cm]{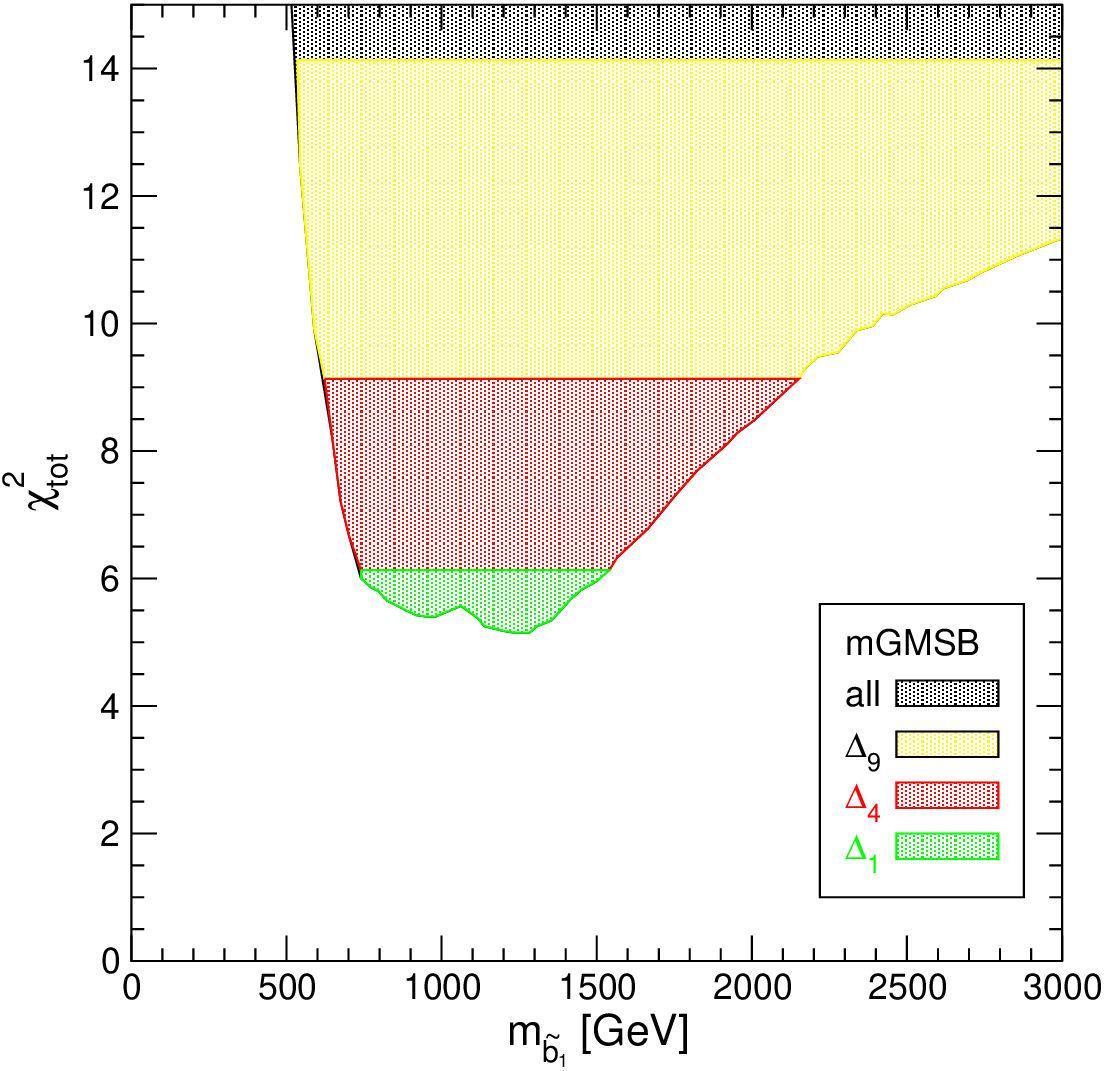}
\hspace{0.5cm}
\includegraphics[width=.45\textwidth,height=7.0cm]{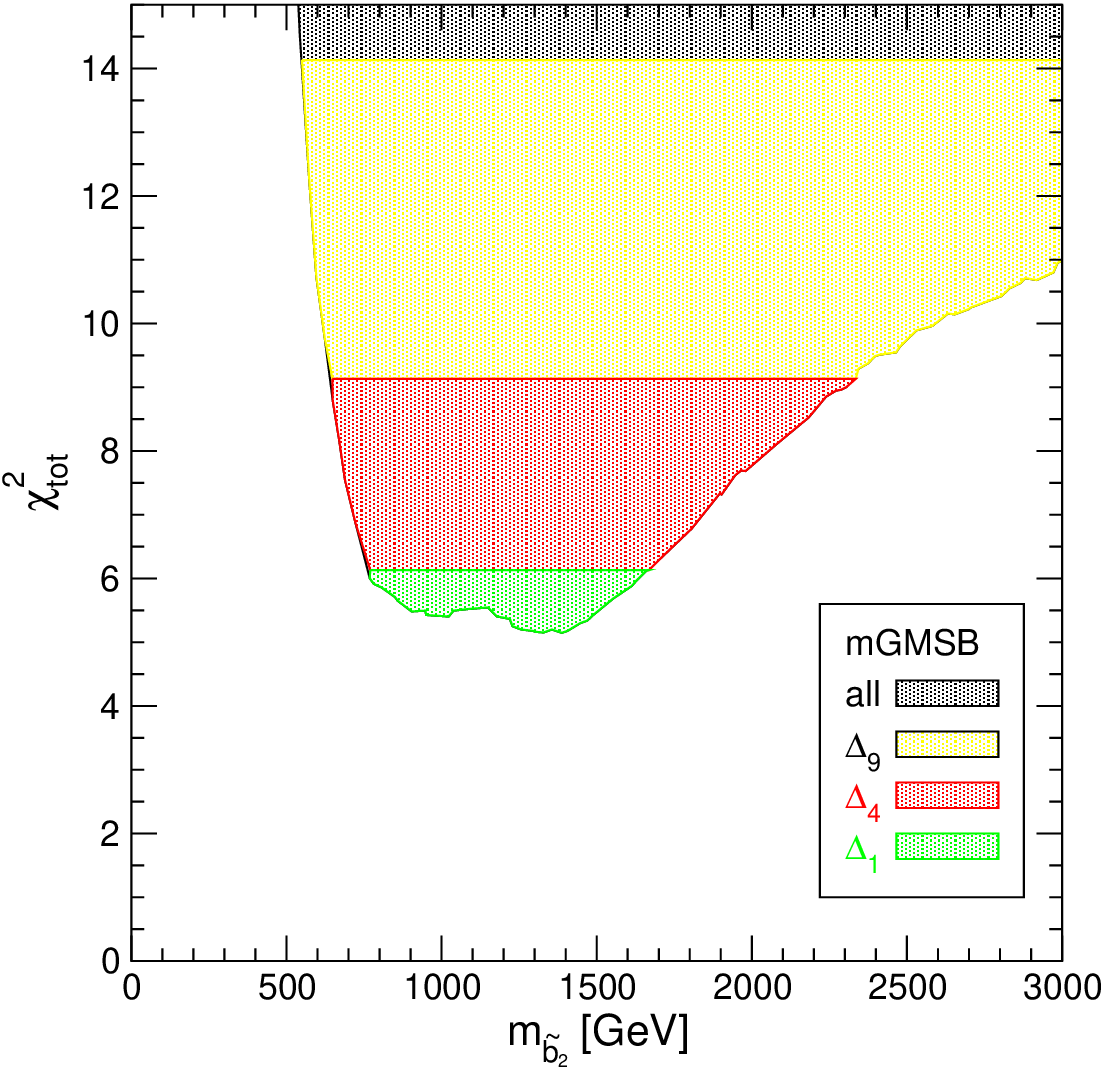}\\[.3em]
\includegraphics[width=.45\textwidth,height=7.0cm]{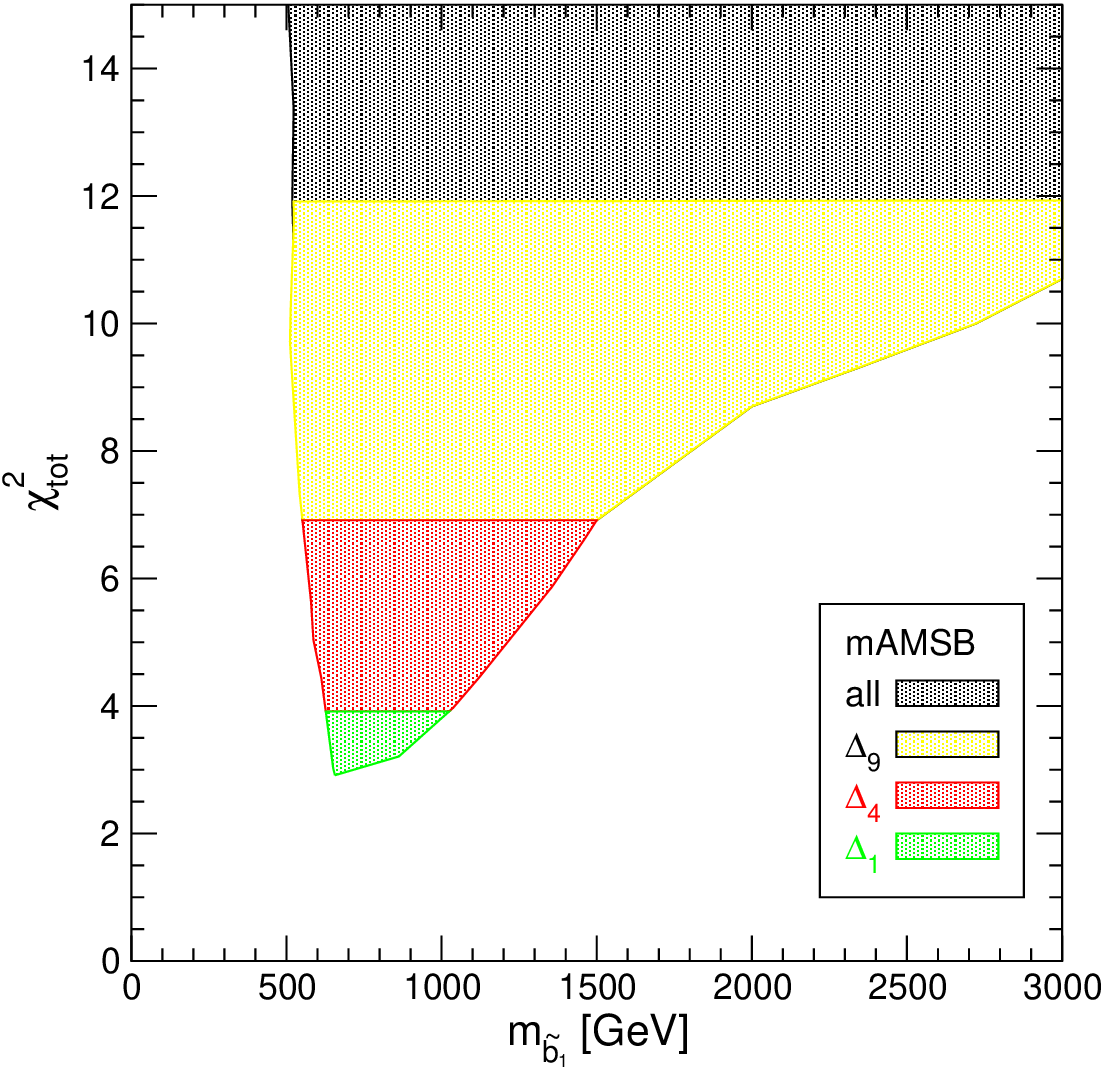}
\hspace{0.5cm}
\includegraphics[width=.45\textwidth,height=7.0cm]{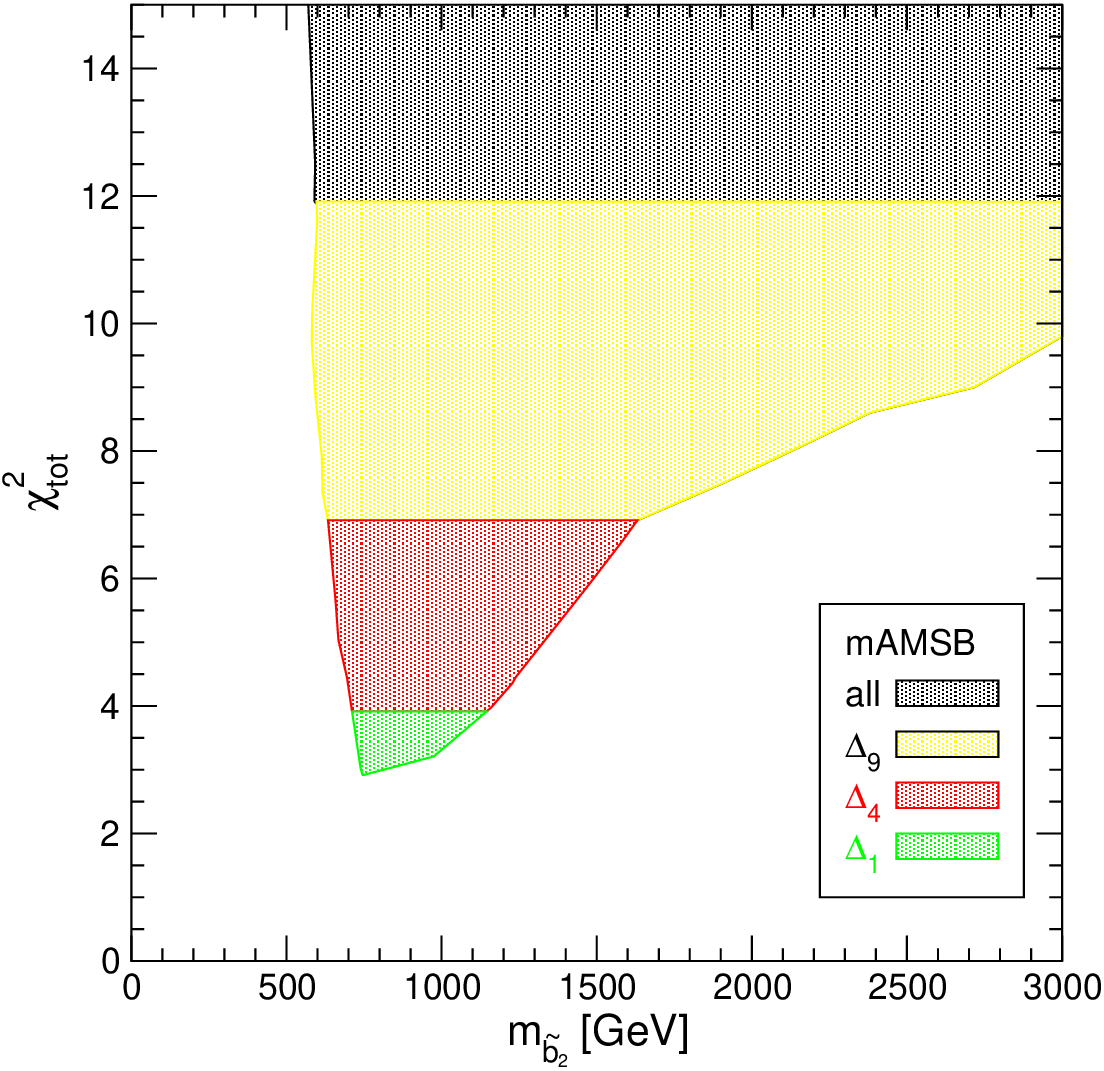}
\vspace{-1em}
\caption{%
$\msbe$ (left) and $\msbz$ (right) are 
shown in the CMSSM (top), mGMSB (middle) and mAMSB (bottom)
scenarios for $\mu > 0$ with their respective total $\chi^2$.
The color coding is as in \reffi{fig:cmssmHigh}. 
}
\label{fig:asbs_sbot}
\end{center} 
\vspace{-3em}
\end{figure}

In \reffi{fig:asbs_stop} we show 
the predictions for the two scalar top masses, $\mste$ (left) and $\mstz$
(right). 
As before, the masses are shown in the CMSSM (top), mGMSB (middle) and
mAMSB (bottom) scenarios for $\mu > 0$ with their respective total $\chi^2$.
The color coding is as in \reffi{fig:cmssmHigh}. 
The `preferred' mass ranges, i.e.\ $\De\chi^2 < 4$, range from about 
$300 \gev$ up to about $2300 \gev$, depending somewhat on the scenario.
Finally, the predictions for the sbottom masses are shown in
\reffi{fig:asbs_sbot}. 
The sbottom masses follow the same pattern as the stop masses.
Taking these values as representative scalar quark mass values, the LHC
should have no problem to discover the SUSY partners of the quarks,
whereas for the ILC(1000) only the lower part of the `preferred' values
could be in the kinematic reach.
However, it should be kept in mind that the $\De_9$ regions extend
beyond $\sim 3 \tev$, which could exceed even the discovery reach of the 
SLHC~\cite{slhc}. 

Apart from the values of the various SUSY and Higgs particle masses,
also the `preferred' values of $|\mu|$ and of $B$ (with $\mu\,B$ being
the prefactor of the Higgs mixing term in the potential) 
are of interest. In \refta{tab:bmu}
we list the current best fit points and the $\De_{1,4}$ ranges for $\mu$
(with $\mu > 0$, see \refse{sec:g-2}) and $B$. The `preferred' values
for $\mu$ range between $130 \gev$ and $1420 \gev$ in the mAMSB and
somewhat smaller intervals within in the two other scenarios. The
`preferred' values of $B$ are bounded from above by $\sim 540 \gev$ in
mAMSB, where also negative values down to $-275 \gev$ are reached in the
$\De_4$ area. In the other two scenarios the intervals are substantially
smaller, and only in the CMSSM negative values down to $-75 \gev$ are
reached.

\begin{table}[htb!]
\renewcommand{\arraystretch}{1.3}
\BC
\begin{tabular}{|c||c|c|c|} 
\cline{2-4} \multicolumn{1}{c||}{}
                       & CMSSM & mGMSB & mAMSB \\
\hline\hline
$\mu$ (best fit) & 588         & 810         & 604         \\ \hline
$\mu$ in $\De_1$ & 510 -- 730  & 460 -- 995  & 560 -- 980  \\ \hline
$\mu$ in $\De_4$ & 160 -- 1100 & 390 -- 1400 & 130 -- 1420 \\ \hline \hline
$B$ (best fit)   & 94         & 151       & 28          \\ \hline
$B$ in $\De_1$   & 65 -- 155  & 75 -- 210 & -105 -- 50  \\ \hline
$B$ in $\De_4$   & -75 -- 250 & 65 -- 330 & -275 -- 540 \\ \hline \hline
\end{tabular}
\EC
\vspace{-1em}
\caption{%
`Preferred' values of $\mu$ and $B$ (with $\mu\,B$ being the prefactor of the
Higgs mixing term in the potential). Shown are the best-fit points as
well as the intervals covered for $\De\chi^2 < 1, 4$. All values are in GeV.
}
\label{tab:bmu}
\renewcommand{\arraystretch}{1.0}
\end{table}

The results for the SUSY masses in the CMSSM can be compared with
previous analyses taking into account the CDM 
constraint~\cite{ehow3,ehow4,ehoww,other,masterfit}.
We focus here on \citeres{ehow3,ehow4,ehoww}, since similar sets of
precision observables and very similar $\chi^2$~analyses had been
used. 
Qualitative agreement can be found in the observed `preferred' mass
values. 
In our analysis the lower mass values in the $\De_1$ and $\De_4$ regions
are obtained for low $\tb$, where these masses are similar to 
to the ones in \citeres{ehow3,ehow4,ehoww} obtained for
$\tb = 10$. Higher mass values in the $\De_1$ and $\De_4$ regions, on
the other hand, are obtained for large $\tb$, where these masses are
similar to the ones in \citeres{ehow3,ehow4,ehoww} obtained for 
$\tb = 50$.
On the other hand, the following difference can be observed:
while the fit results obtained for the particle masses in
\citeres{ehow3,ehow4,ehoww} 
are `parabola shaped', whereas the mass plots presented in
\reffis{fig:asbs_neu} -- \ref{fig:asbs_sbot} show `full' areas.
This can easily be understood as an effect of taking the CDM constraint
into account in \citeres{ehow3,ehow4,ehoww}, while at the same time
$\tb$ had been restricted to the two discrete values $\tb = 10$ and~50.
The CDM constraint cuts out thin strips, for instance, in the
$m_0$--$m_{1/2}$ plane (for fixed $A_0$ and 
$\tb$)~\cite{WMAPstrips,wmapothers}. This yields naturally strips in the
mass vs.\ $\chi^2_{\rm tot}$ plots. 
Incorporating all $\tb$ values by scanning
over all allowed values simultaneously in our analysis
(where low (high) $\tb$ values yield lower (higher) best-fit masses),
broadens and fills automatically the $\De_1$ and $\De_4$ regions.
Another difference in our analysis compared to the ones in
\citeres{ehow3,ehow4,ehoww} is the lower value of $\mt$ that has been
used here.
Lowering the experimental value of $\mt$ in the $\chi^2$
analysis yields an increase in the minimum total $\chi^2$, as has been
analyzed for $\tb = 10$ in \citere{ehow4}. The minimum $\chi^2$ values
reached in \citeres{ehow3,ehow4,ehoww} and in our analysis roughly
follow the results presented in \citere{ehow4}.
However, it should be kept in mind that the latest value of $\mt$ that
has been published recently~\cite{mt1726} has moved upwards to
$\mt^{\rm exp} = 172.6 \pm 1.4 \gev$.


\section{Future sensitivities}
\label{sec:future}

We now turn to the analysis of the future sensitivities.
In a first step we take the current best-fit
point in each scenario and assume that the future measurements exactly
agree with this point. The experimental and theory uncertainties are set to
their `future' values as discussed in \refse{sec:expdata}. Also for $\Mh$ we
assume that its value is measured and include it into the $\chi^2$ fit
with the future uncertainties given in \refse{sec:mh}. 
In a second step, in order to compare the sensitivities in the three
scenarios, we have 
chosen one hypothetical best-fit point in each scenario, where the
low-energy spectrum is ``similar'' in all three scenarios.
In more detail, we have demanded that
\BE
\label{asbsfutbf}
\MA \approx 800 \gev, \quad
\tb \approx 40, \quad
\mste \approx 1225 \gev, \quad
\mstz \approx 1400 \gev, \quad
\mu > 0 ~.
\EE
These masses are somewhat higher than the current best-fit values and thus
illustrate a future scenario that is somewhat more in the decoupling
regime (i.e.\ where SUSY masses are heavy and loop corrections are
correspondingly smaller) than what is
currently favored. Furthermore the combination of $\MA$ and $\tb$, 
according to current
analyses~\cite{teslatdr,orangebook,acfarep,Snowmass05Higgs,atlastdr,cmstdr,jakobs,schumi,cmshiggsOrg,cmsHiggs,slhc}, 
is not in the discovery reach of the LHC or the ILC. In such a scenario
without experimental information on $\MA$ and $\tb$ from the observation
of the heavy Higgs bosons any sensitivity to
these parameters would constitute information {\em in addition} to the direct
collider data.
The three points are defined in terms of high-energy parameters as
\begin{align}
\label{cmssmfutbf}
{\rm CMSSM} ~:~ &m_0 = 640 \gev \\
                &m_{1/2} = 720 \gev \non \\
                &A_0 = 500 \gev \non \\
                &\tb = 41 \non \\[.5em]
\label{gmsbfutbf}
{\rm mGMSB} ~:~ &\La = 33200 \gev \\
                &M_{\rm mess} = 580000 \gev \non \\
                &N_{\rm mess} = 7 \non \\
                &\tb = 41 \non \\[.5em]
\label{amsbfutbf}
{\rm mAMSB} ~:~ &m_{\rm aux} = 50500 \gev \\
                &m_0 = 1600 \gev \non \\
                &\tb = 40 \non 
\end{align}
The choices in \refeq{asbsfutbf} ensure a ``similar'' behavior in the
Higgs and in the scalar top sector and their contributions to the EWPO
and BPO. This allows a comparison of the future sensitivities of the
EWPO and BPO in the three scenarios. The values for the lightest Higgs
boson mass at the three hypothecial best-fit points are 
$116.8 \gev$ (CMSSM), $117.5 \gev$ (mGMSB) and $119.1 \gev$ (mAMSB). 
The spread of $\sim 2.3 \gev$ has only a minor direct impact on the
predictions of the EWPO and BPO.


\subsection{Analysis of high-scale parameters}
\label{sec:anahighF}

We start by analyzing the CMSSM, mGMSB
and mAMSB in terms of their respective high-energy parameters, see
\refse{sec:asbs}.

\subsubsection{CMSSM}

In \reffi{fig:cmssmHighF} we show the results for the $\De_{1,4,9}$ areas
in terms of the high-energy parameters, using the $\chi^2$ result based
on the assumed future experimental
and theoretical precisions as described in \refse{sec:expdata}. 
As can been seen, the areas of the parameter space with 
$\De\chi^2 < 1, 4, 9$ shrink substantially in comparison with
\reffi{fig:cmssmHigh}.
At the $\De\chi^2 = 9$ level $m_{1/2}$ is
determined up to $\pm 200 \gev$ for the assumed best-fit point. For
$m_0$, on the other hand, still 
values up to $\sim 1500 \gev$ are permitted. The $\De_9$ interval for
$A_0$ shrinks to $\pm 1000 \gev$.

The reduction of the preferred parameter region with the assumed higher
precision in the future is so substantial because
the currently favored best-fit parameters are relatively small, where
smaller SUSY mass scales lead to larger loop effects in the precision
observables. This  
effect is less pronounced for larger GUT scale parameters. To illustrate
this effect we have chosen a CMSSM point as defined in
\refeq{cmssmfutbf}. 
We assume that the future experimental values agree exactly with the
low-energy parameters resulting from \refeq{cmssmfutbf}. The reduction of the
preferred parameter region as shown in \reffi{fig:cmssmHighFC} compared
to the present situation is still
visible, but much weaker than for the 
current best-fit point in \reffi{fig:cmssmHighF}.
Similar results (including the CDM constraint) had been found in
\citere{ehow3}. 


\subsubsection{mGMSB}

In \reffi{fig:mgmsbHighF} we show the results for the $\De_{1,4,9}$ areas
in terms of the high-energy parameters, using the future experimental
and theoretical precisions as described in \refse{sec:expdata}. 
The color coding is as in \reffi{fig:cmssmHigh}.
The plots in \reffi{fig:mgmsbHighF} show the $\La$--$M_{\rm mess}$
plane for $N_{\rm mess} = 1 \ldots 8$. For each $N_{\rm mess}$ a small
$\La$~interval is singled out, but hardly any limit on $M_{\rm mess}$ 
is obtained even with the future precisions.

The results look similar in \reffi{fig:mgmsbHighF2}, 
where we show the $N_{\rm mess}$--$\La$ plane. 
For each $N_{\rm mess}$ value a relatively small range of $\La$ is
favored, even at the $\De\chi^2 = 9$ level. If $N_{\rm mess}$ could be
determined in an independent way, the precision observables could give a
relatively precise determination of $\La$. On the other hand, if $\La$
could be determined, e.g.\ from the measurement of SUSY masses, the
precision observables would give a preference for certain $N_{\rm mess}$
values.

\begin{figure}[htb!]
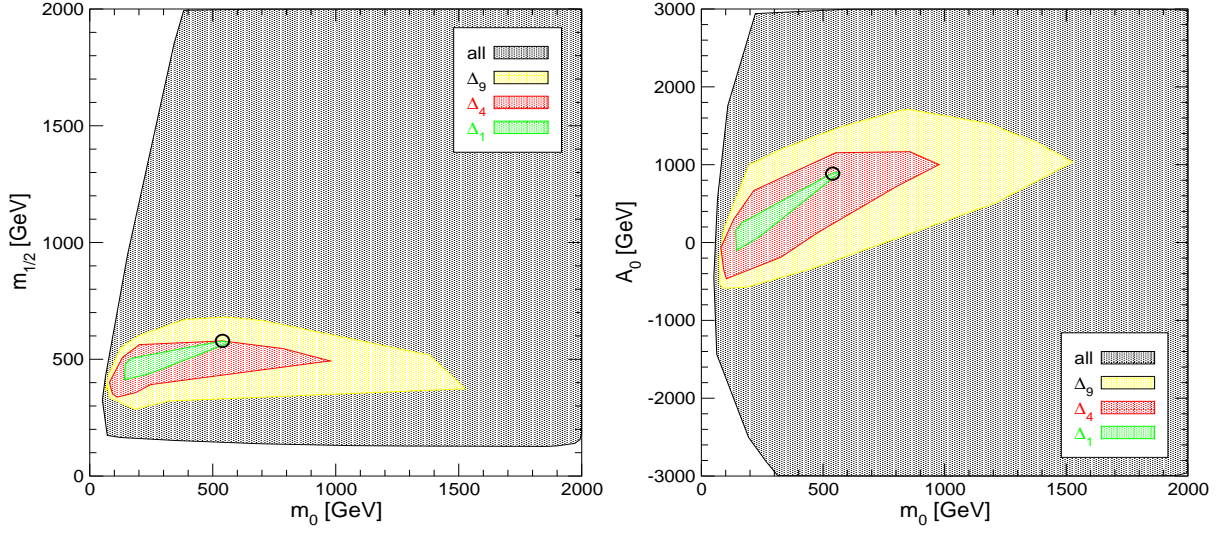

\vspace{4em}
\begin{center}
\includegraphics[width=.48\textwidth,height=7cm]{asbs3_s_para21_cl}
\includegraphics[width=.48\textwidth,height=7cm]{asbs3_s_para22_cl}
\caption{%
Future projection for 
the $\De_{1,4,9}$ regions in the $m_0$--$m_{1/2}$ plane (left) and in
the $m_0$--$A_0$ plane (right) in the CMSSM assuming that the future
experimental data agree exactly with the current best-fit point.
The color code is as in \reffi{fig:cmssmHigh}.
}
\label{fig:cmssmHighF}
\end{center}
\end{figure}

\begin{figure}[htb!]
\vspace{2em}
\begin{center}
\includegraphics[width=.48\textwidth,height=7cm]{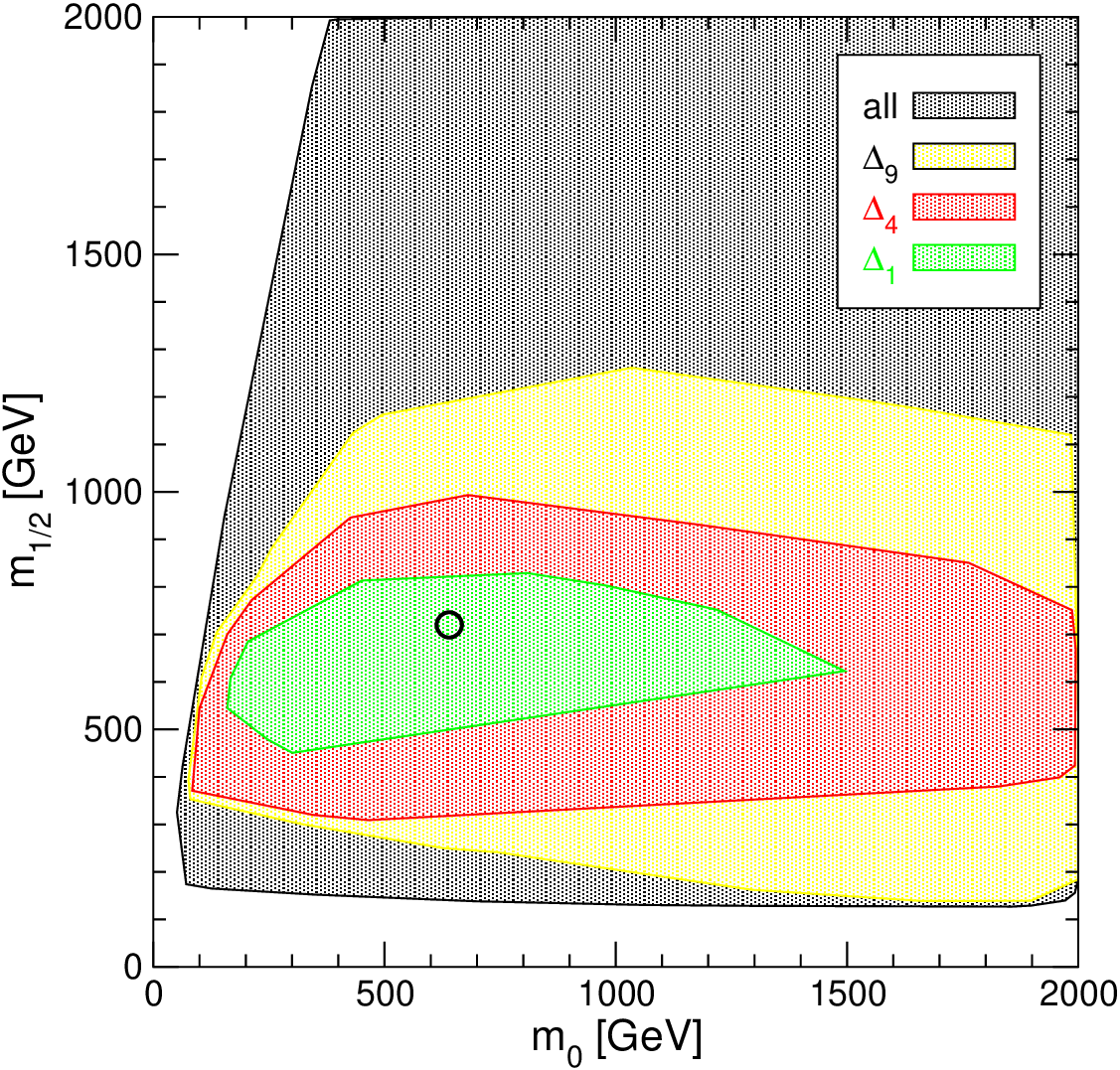}
\includegraphics[width=.48\textwidth,height=7cm]{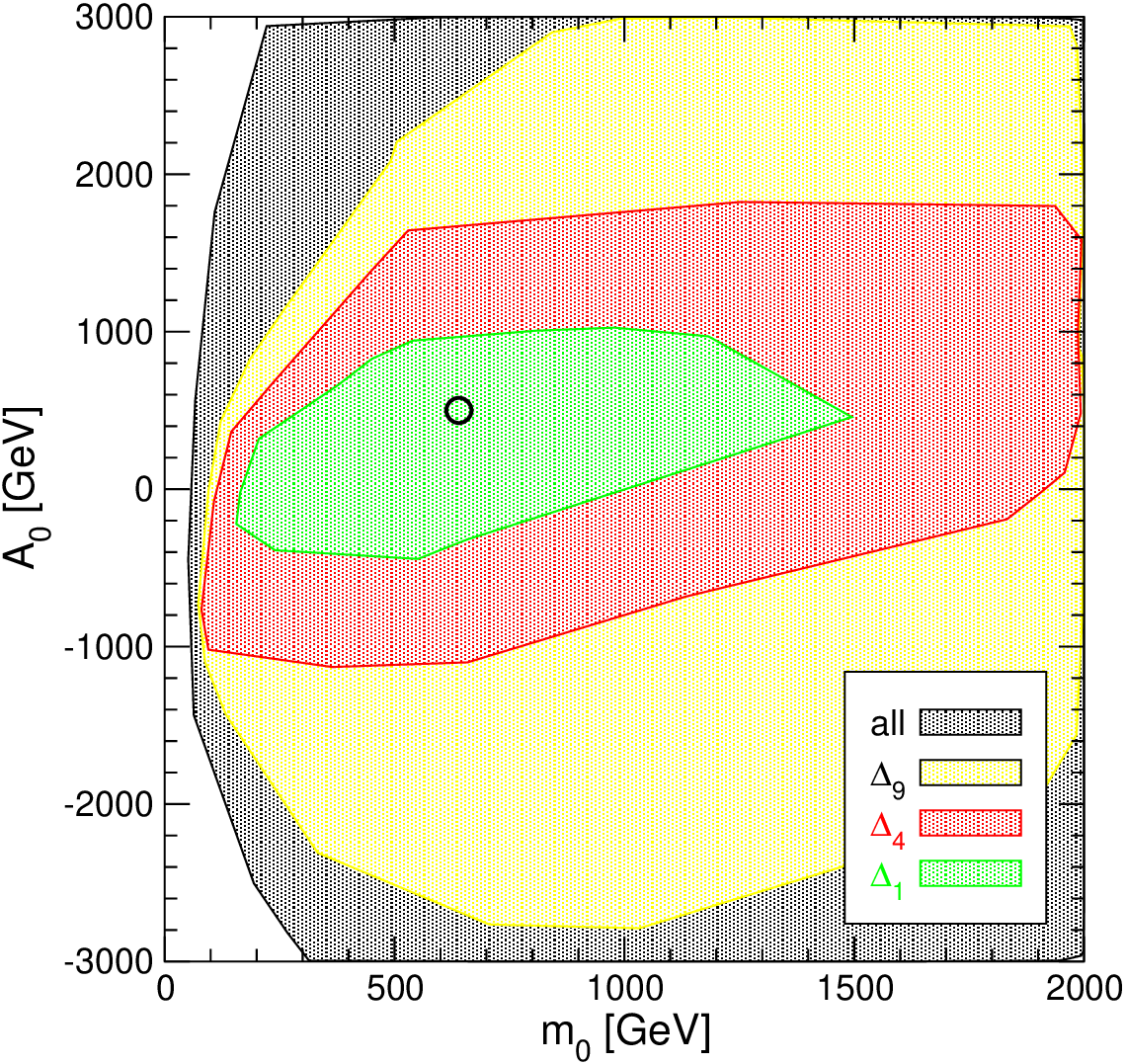}
\caption{%
Future projection for 
the $\De_{1,4,9}$ regions in the $m_0$--$m_{1/2}$ plane (left) and in
the $m_0$--$A_0$ plane (right) in the CMSSM assuming that the future 
experimental data agree exactly with a hypothetical best-fit point as
specified in \refeq{cmssmfutbf}. 
The color code is as in \reffi{fig:cmssmHigh}.
}
\label{fig:cmssmHighFC}
\end{center}
\end{figure}

\begin{figure}[htb!]
\begin{center}
\includegraphics[width=.4\textwidth,height=4.9cm]{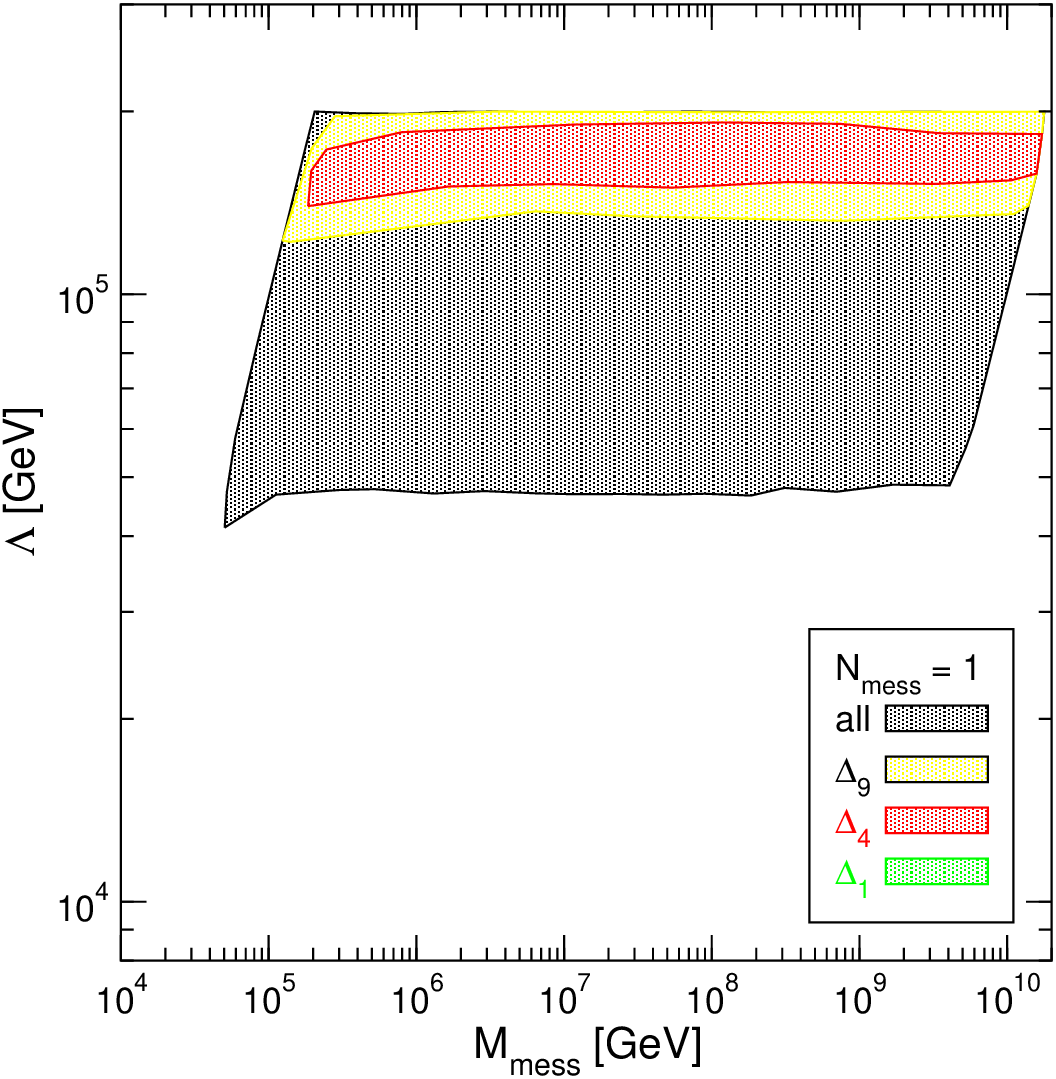}\hspace{1cm}
\includegraphics[width=.4\textwidth,height=4.9cm]{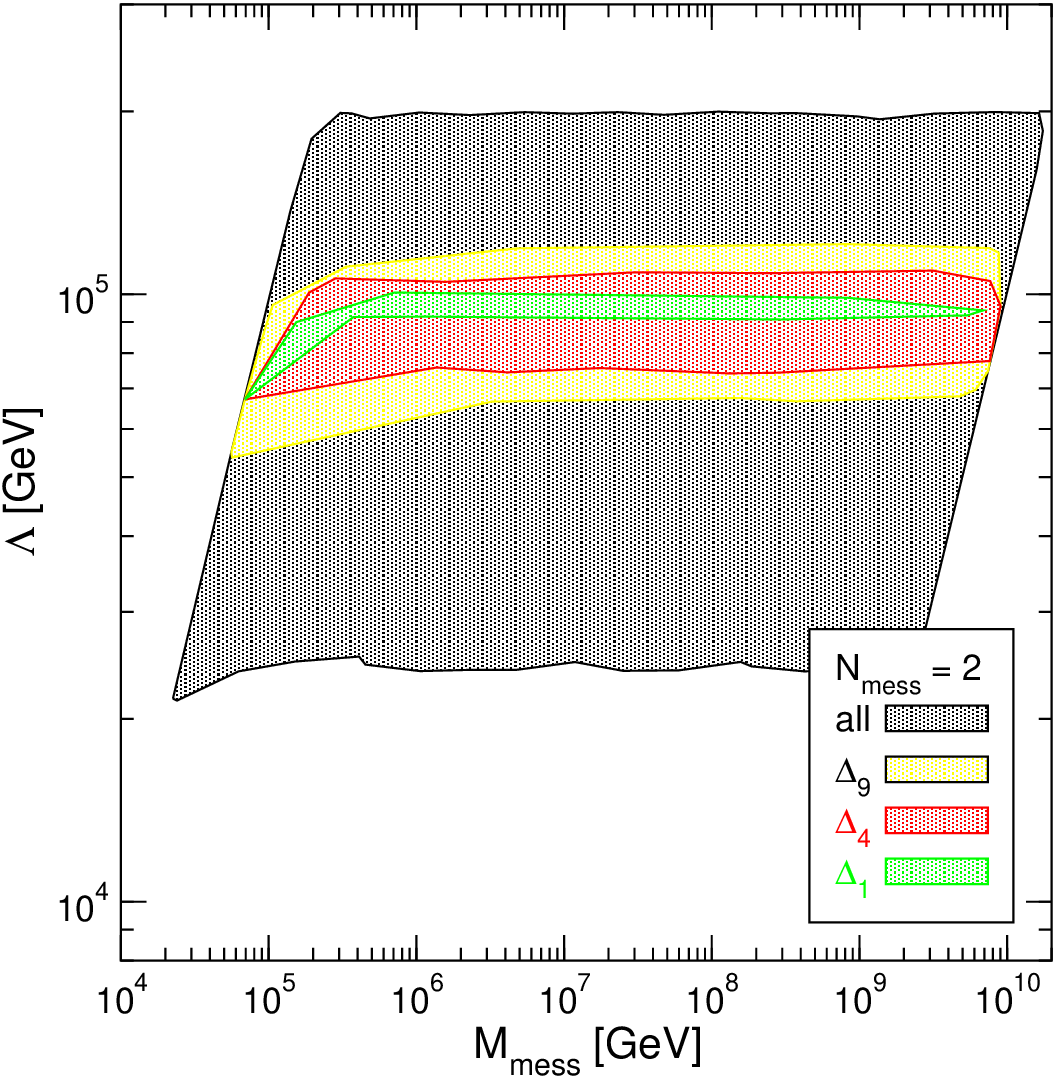}\\[0.3cm]
\includegraphics[width=.4\textwidth,height=4.9cm]{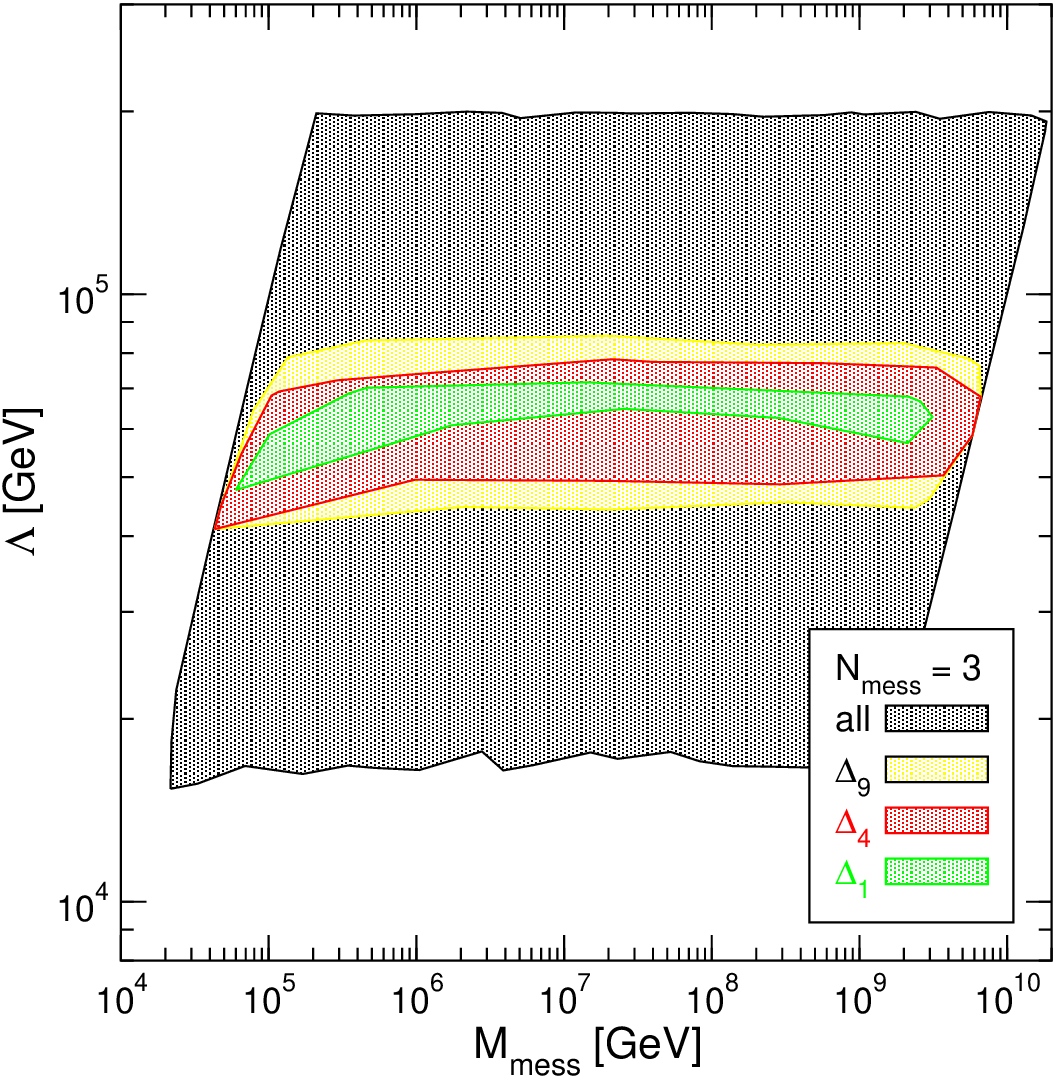}\hspace{1cm}
\includegraphics[width=.4\textwidth,height=4.9cm]{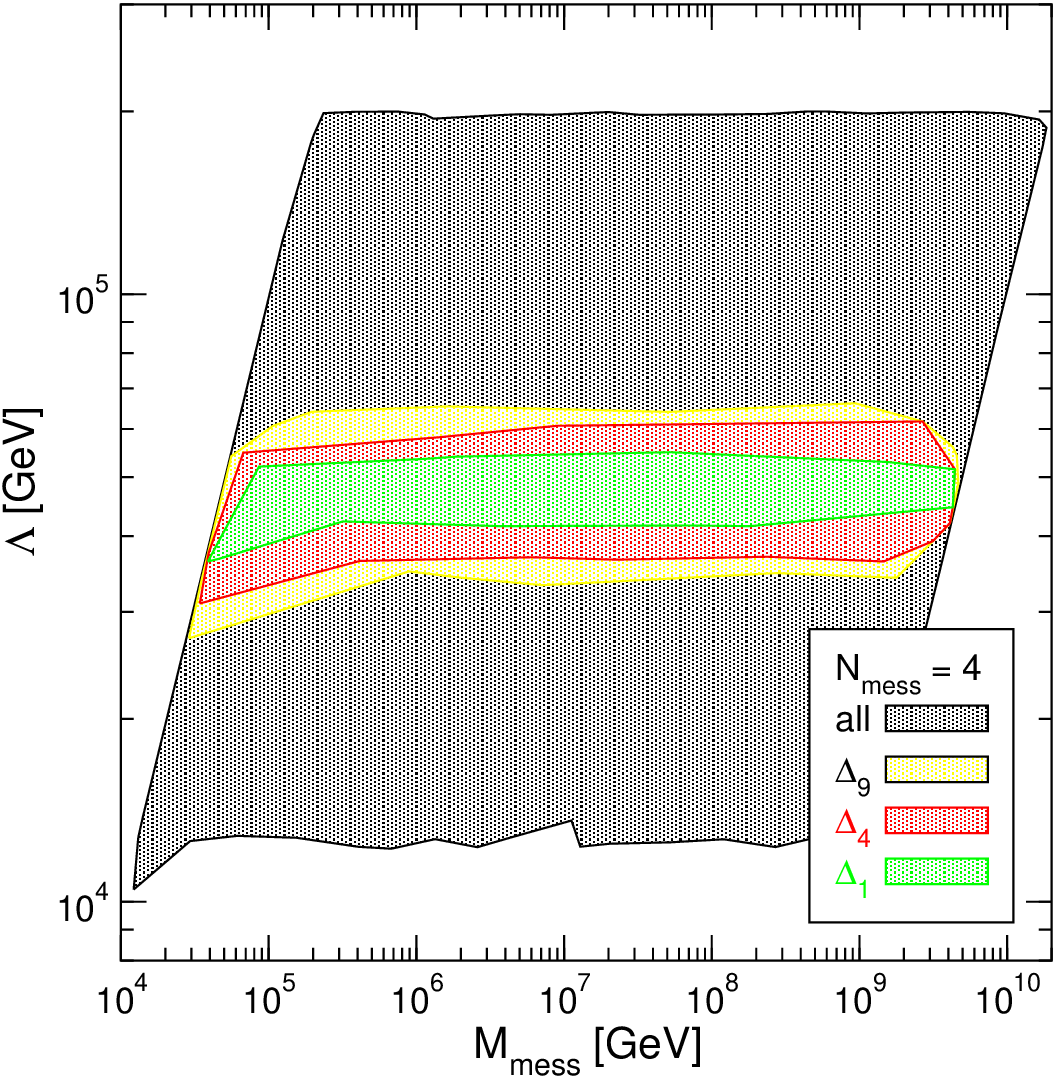}\\[0.3cm]
\includegraphics[width=.4\textwidth,height=4.9cm]{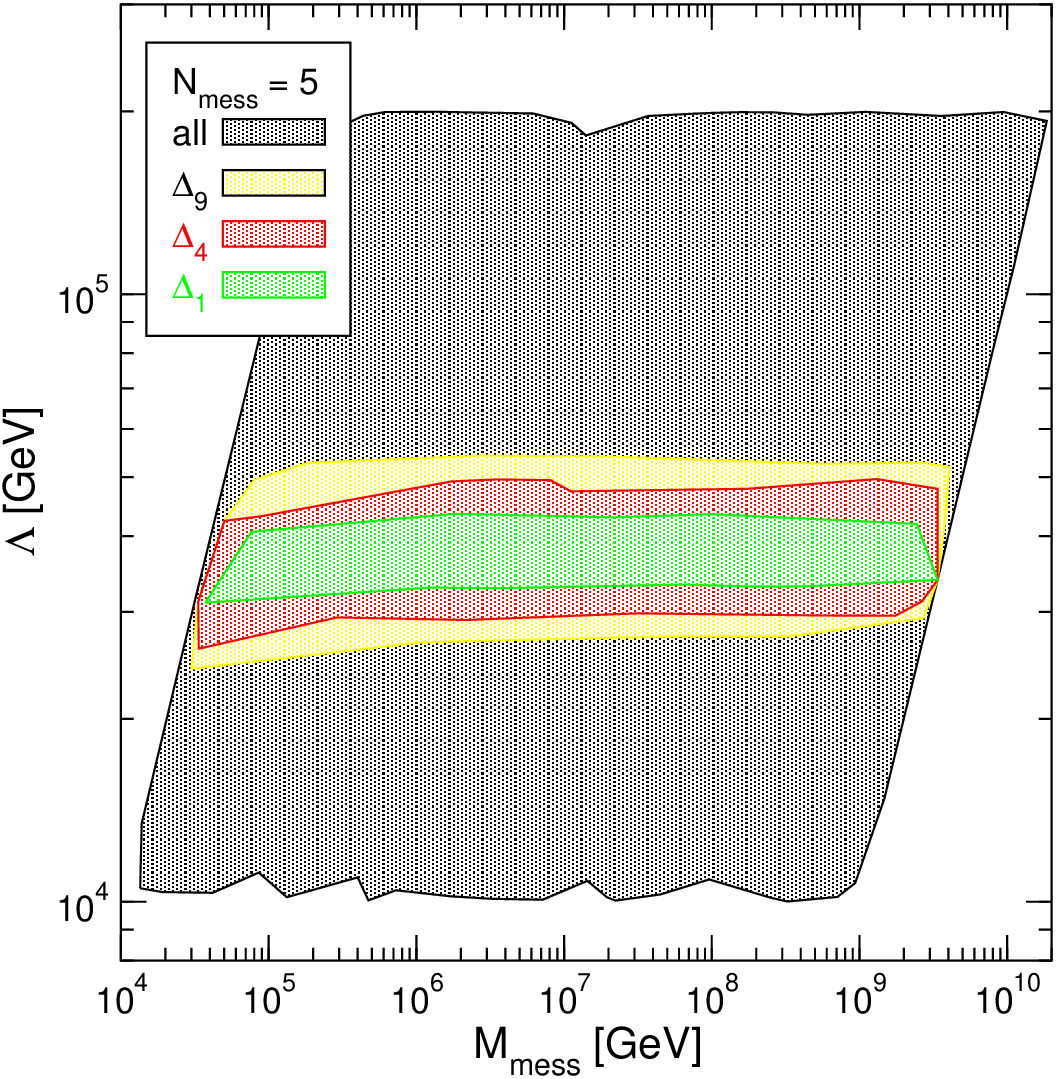}\hspace{1cm}
\includegraphics[width=.4\textwidth,height=4.9cm]{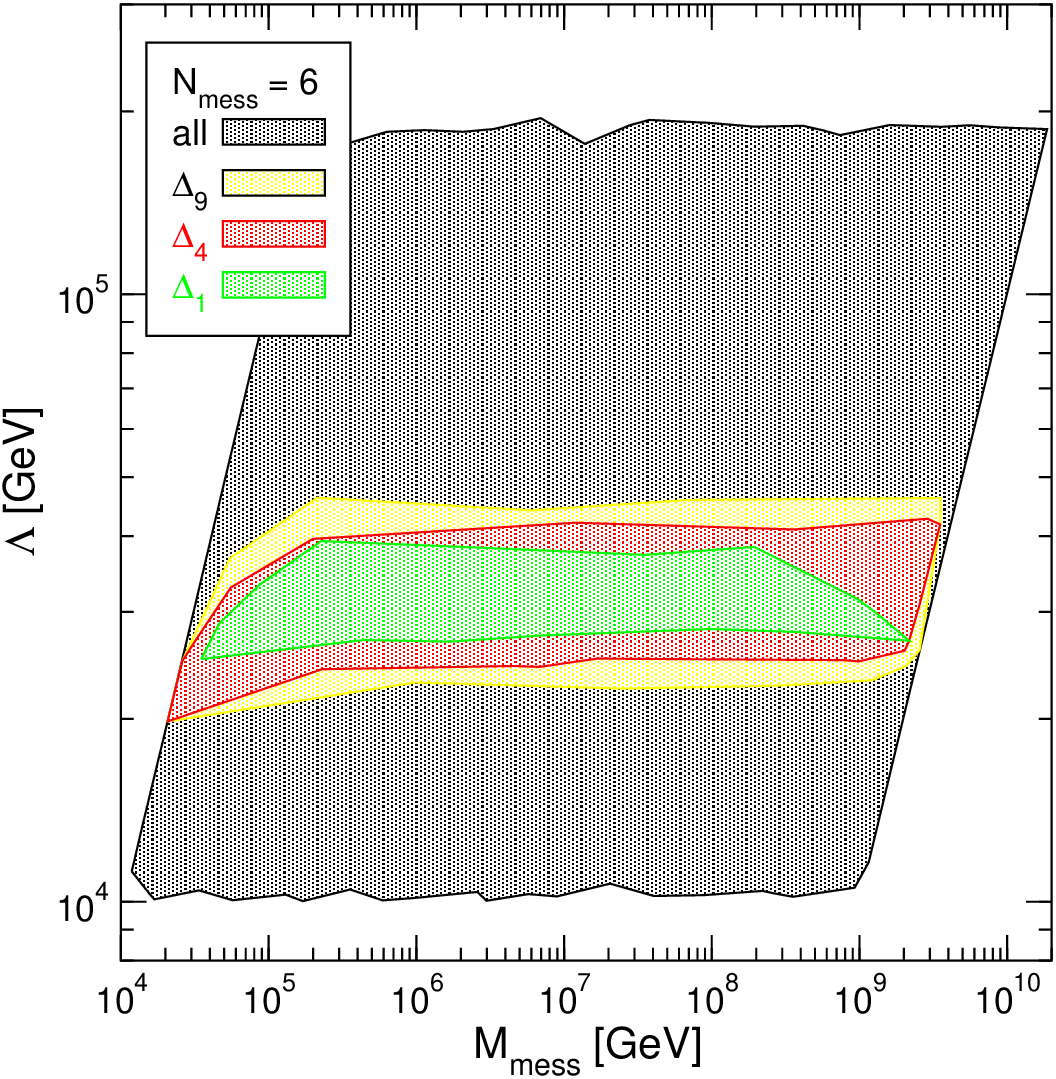}\\[0.3cm]
\includegraphics[width=.4\textwidth,height=4.9cm]{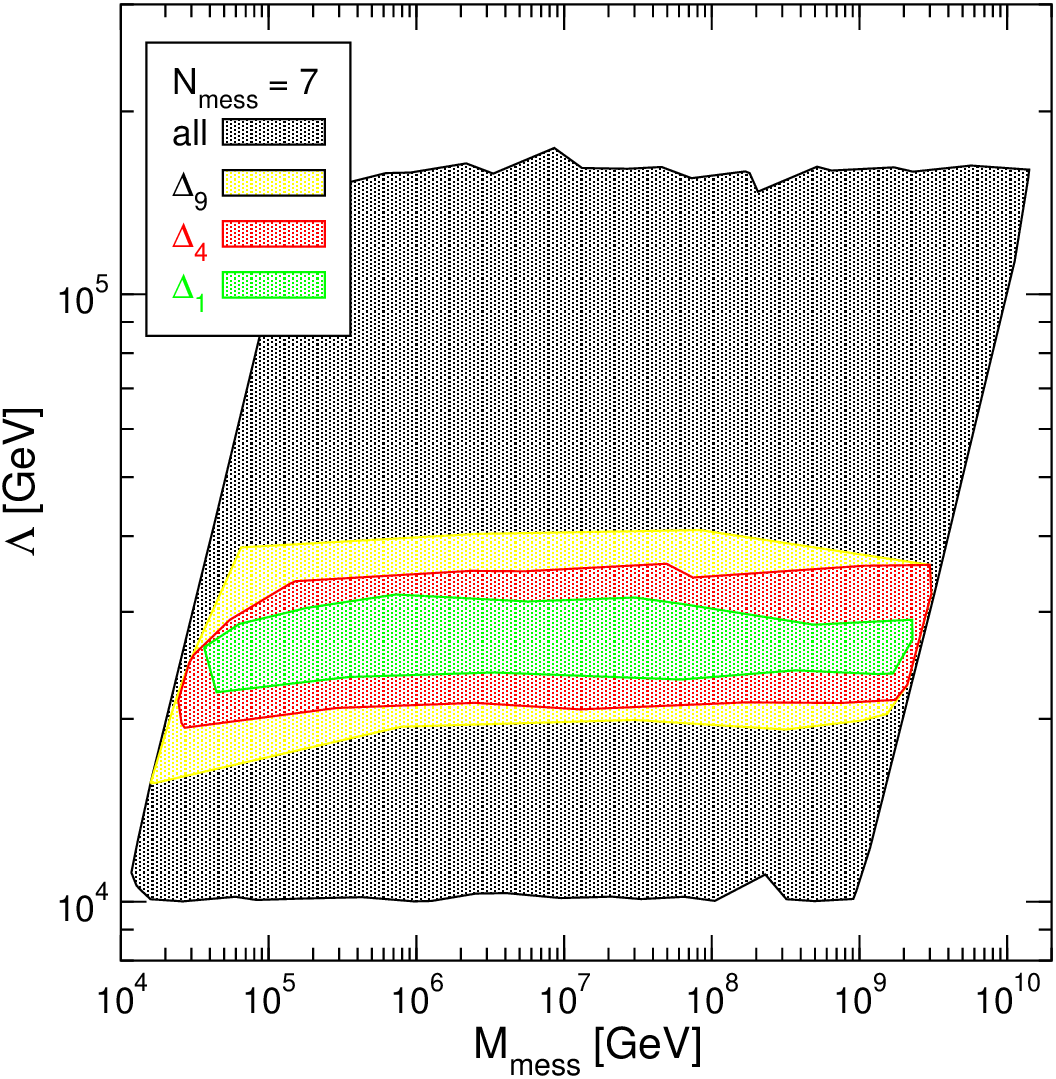}\hspace{1cm}
\includegraphics[width=.4\textwidth,height=4.9cm]{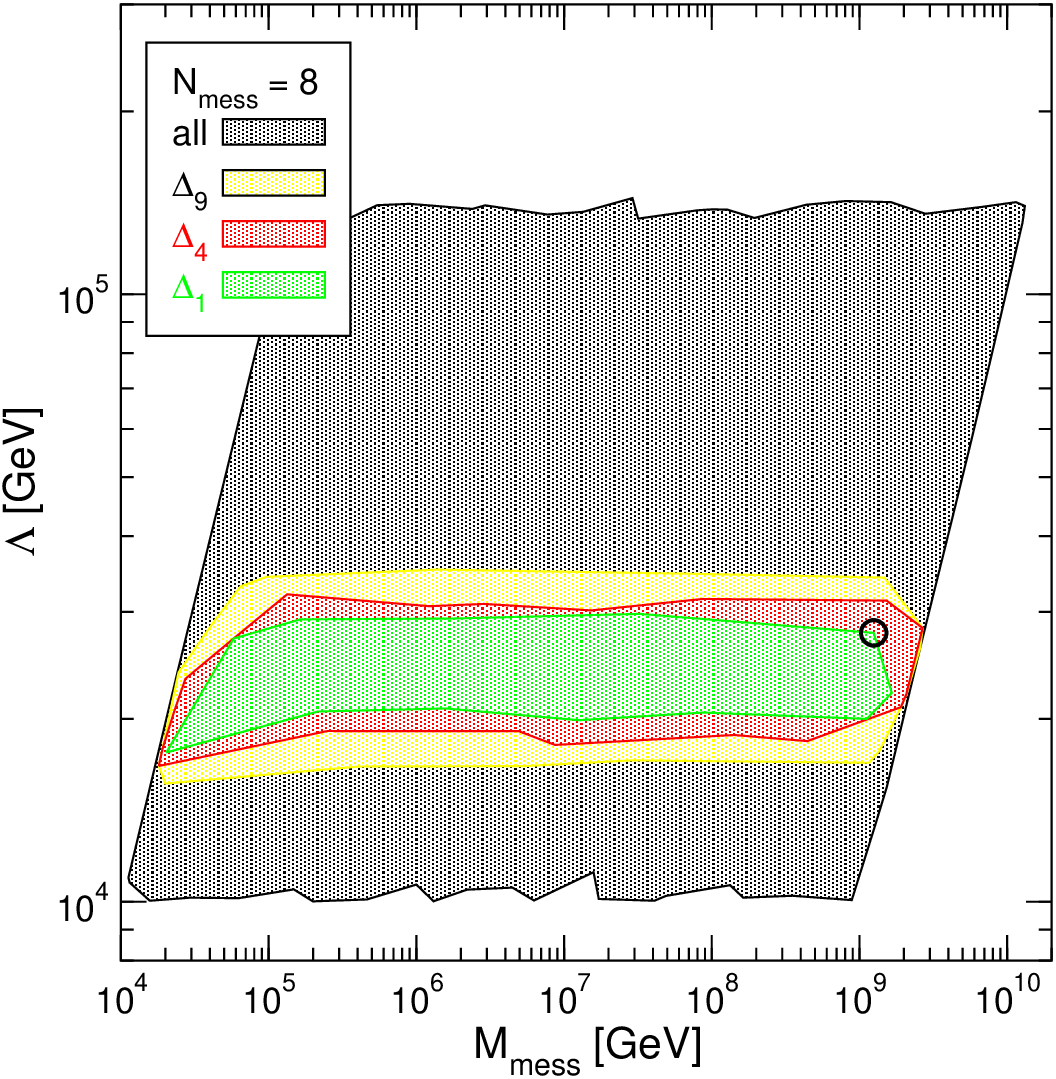}
\caption{%
Future projection for 
the $\De_{1,4,9}$ regions in the $\La$--$M_{\rm mess}$ plane for the 
$N_{\rm mess} = 1 \ldots 8$ in the mGMSB assuming that the
future experimental data agree exactly with the current best-fit point (marked
by a circle). 
The color coding is as in \reffi{fig:cmssmHigh}.
}
\label{fig:mgmsbHighF}
\end{center}
\vspace{-2em}
\end{figure}

\begin{figure}[htb!]
\begin{center}
\includegraphics[width=.4\textwidth,height=4.9cm]{asbs3_g_para21F_cl}\hspace{1cm}
\includegraphics[width=.4\textwidth,height=4.9cm]{asbs3_g_para22F_cl}\\[0.3cm]
\includegraphics[width=.4\textwidth,height=4.9cm]{asbs3_g_para23F_cl}\hspace{1cm}
\includegraphics[width=.4\textwidth,height=4.9cm]{asbs3_g_para24F_cl}\\[0.3cm]
\includegraphics[width=.4\textwidth,height=4.9cm]{asbs3_g_para25F_cl}\hspace{1cm}
\includegraphics[width=.4\textwidth,height=4.9cm]{asbs3_g_para26F_cl}\\[0.3cm]
\includegraphics[width=.4\textwidth,height=4.9cm]{asbs3_g_para27F_cl}\hspace{1cm}
\includegraphics[width=.4\textwidth,height=4.9cm]{asbs3_g_para28F_cl}
\caption{%
Future projection for 
the $\De_{1,4,9}$ regions in the $\La$--$M_{\rm mess}$ plane for the 
$N_{\rm mess} = 1 \ldots 8$ in the mGMSB assuming that the
future experimental data agree exactly with the hypothetical best-fit point as
defined in \refeq{gmsbfutbf} (marked by a circle).
The color coding is as in \reffi{fig:cmssmHigh}.
}
\label{fig:mgmsbHighFX}
\end{center}
\vspace{-3em}
\end{figure}

\begin{figure}[htb!]
\vspace{2em}
\begin{center}
\includegraphics[width=.60\textwidth]{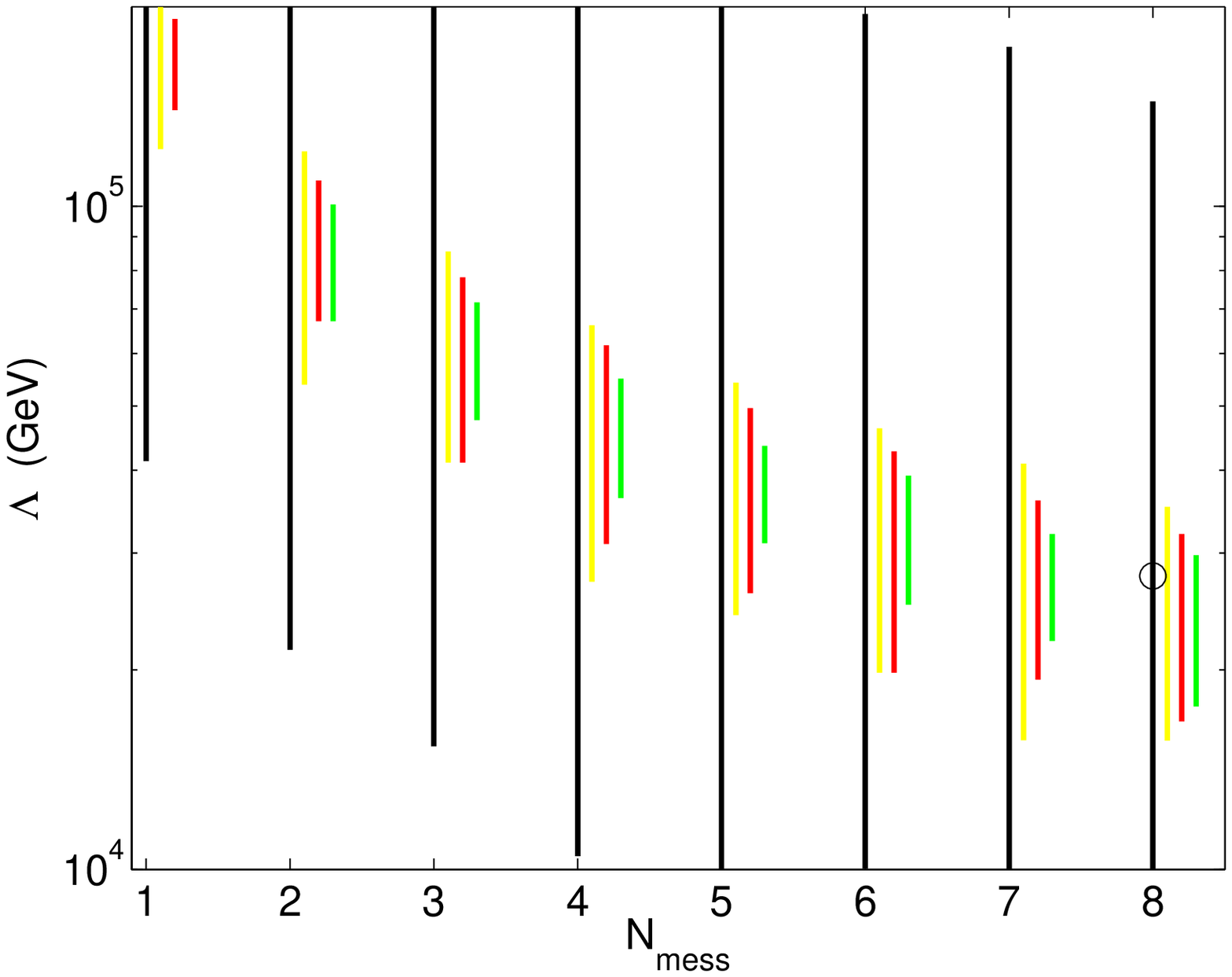}
\caption{%
Future projection for 
the $\De_{1,4,9}$ regions in 
the $N_{\rm mess}$--$\La$ plane in the mGMSB assuming that the future
experimental data agree exactly with the current best-fit point (marked
by a circle).
The color coding is as in \reffi{fig:cmssmHigh}.
}
\label{fig:mgmsbHighF2}
\end{center}
\end{figure}

\begin{figure}[htb!]
\vspace{2em}
\begin{center}
\includegraphics[width=.60\textwidth]{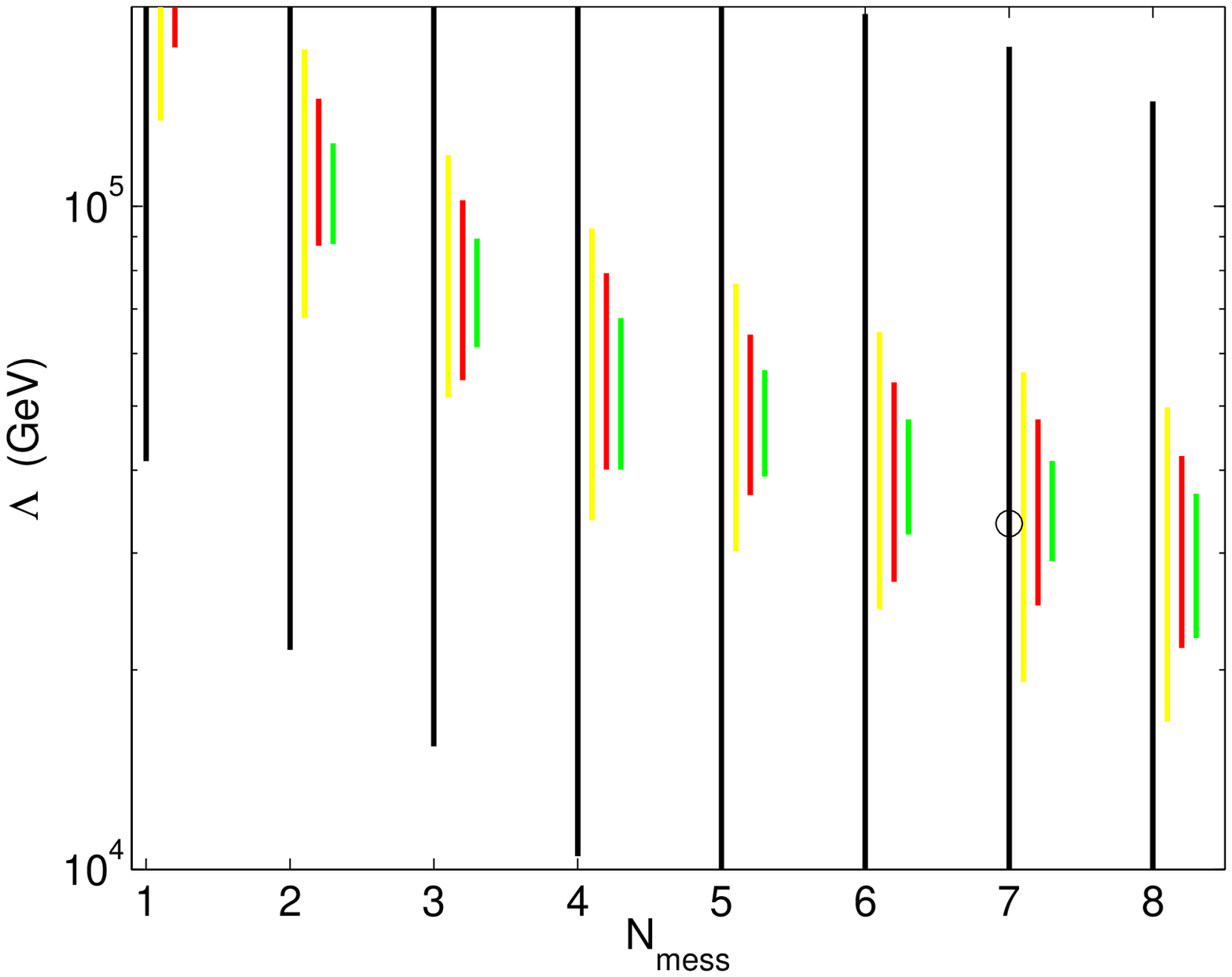}
\caption{%
Future projection for 
the $\De_{1,4,9}$ regions in 
the $N_{\rm mess}$--$\La$ plane in the mGMSB assuming that the future
experimental data agree exactly with the hypothetical best-fit point as
defined in \refeq{gmsbfutbf} (marked by a circle).
The color coding is as in \reffi{fig:cmssmHigh}.
}
\label{fig:mgmsbHighFX2}
\end{center}
\end{figure}

As for the CMSSM scenario also in mGMSB we have chosen a
hypothetical future best-fit point with higher mass scales, defined by
\refeq{gmsbfutbf}. 
As for the CMSSM, we assume that the future experimental values agree
exactly with the low-energy parameters corresponding to \refeq{gmsbfutbf}. 
The reduction of the $\De_{1,4,9}$ regions can be observed in
\reffi{fig:mgmsbHighFX}. It is at the same level as for the
current best-fit point in \reffi{fig:mgmsbHighF}.
These results are also shown in the $N_{\rm mess}$--$\La$ plane in
\reffi{fig:mgmsbHighFX2}, where the same sensitivity is found as for the
current best-fit point displayed in \reffi{fig:mgmsbHighF2}.


\subsubsection{mAMSB}
\label{sec:analhighmamsb}

In \reffi{fig:mamsbHighF} we show the only high-energy parameter plane in
the mAMSB, $m_{\rm aux}$ vs.\ $m_0$, with the same color coding as in 
\reffi{fig:cmssmHigh}. 
Within this scenario the precision observables will allow an extremely
precise determination of the high-energy parameters. For the case that
the current best-fit point agrees exactly with the future measurements, 
at the $\De\chi^2 = 9$ level $m_{\rm aux}$ is determined to 
$\pm 3 \times 10^3 \gev$, i.e.\ to $\sim 10\%$.
The absolute precision for $m_0$ is $\pm 100 \gev$,
whereas the relative precision reaches only $\sim 30\%$.
(The $\De_4$ and $\De_1$ regions are very small and nearly invisible
inside (by definition) the $\De_9$ region.)
This result is to a large extent due to the fact that the $\tb$ value
for the current best-fit point is relatively low (see also the
discussion of the hypothetical best-fit point below).

As for the other two scenarios, also in mAMSB we have chosen a
hypothetical future best-fit point with higher mass scales, defined by
\refeq{amsbfutbf}. It should be noted that for mAMSB the increase in
$\MA$ from the current best-fit point to the hypothetical best-fit point
is a bit smaller than in the other two scenarios, while the shift in
$\tb$ is substantially larger. 
Again we assume that the future experimental values agree
exactly with the low-energy parameters corresponding to \refeq{amsbfutbf}. 
We show the preferred parameter space for this hypothetical point in
\reffi{fig:mamsbHighFX}. The reduction in the size of the 
$\De_{1,4,9}$ regions compared to the present situation is much weaker
than for the 
current best-fit point in \reffi{fig:mamsbHighF}. At the $\De_9$ level
no limit on $m_0$ can be set. This shows that the very high precision
obtainable 
with the current best-fit point is not generally valid in the mAMSB
scenario.


\subsection{Low-energy analysis}
\label{sec:analowF}

We now turn to the comparison of the three soft SUSY-breaking scenarios
in terms of $\MA$ and $\tb$, assuming the future experimental and theory
precisions as discussed in \refse{sec:expdata}.  
In \reffi{fig:asbsLowF} we show the $\MA$--$\tb$ plane for the CMSSM
(top), mGMSB (middle) and mAMSB (bottom) with the same color coding as
in \reffi{fig:cmssmHigh}. In each scenario we assume that the future
measurements will agree exactly with the current best-fit point.

\begin{figure}[htb!]
\vspace{4em}
\begin{center}
\includegraphics[width=.6\textwidth,height=7cm]{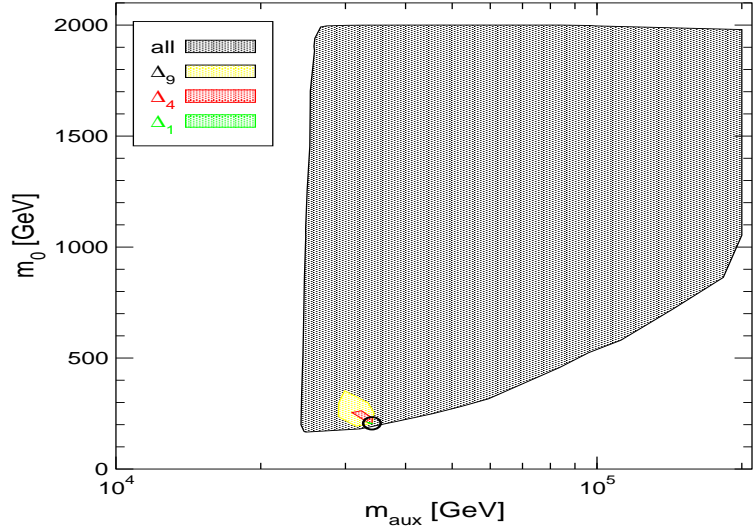}
\caption{%
Future projection for 
the $\De_{1,4,9}$ regions in the $m_{\rm aux}$--$m_0$ plane
in the mAMSB assuming that the future
experimental data agree exactly with the current best-fit point.
The color coding is as in \reffi{fig:cmssmHigh}.
}
\label{fig:mamsbHighF}
\end{center}
\end{figure}

\begin{figure}[htb!]
\vspace{2em}
\begin{center}
\includegraphics[width=.6\textwidth,height=7cm]{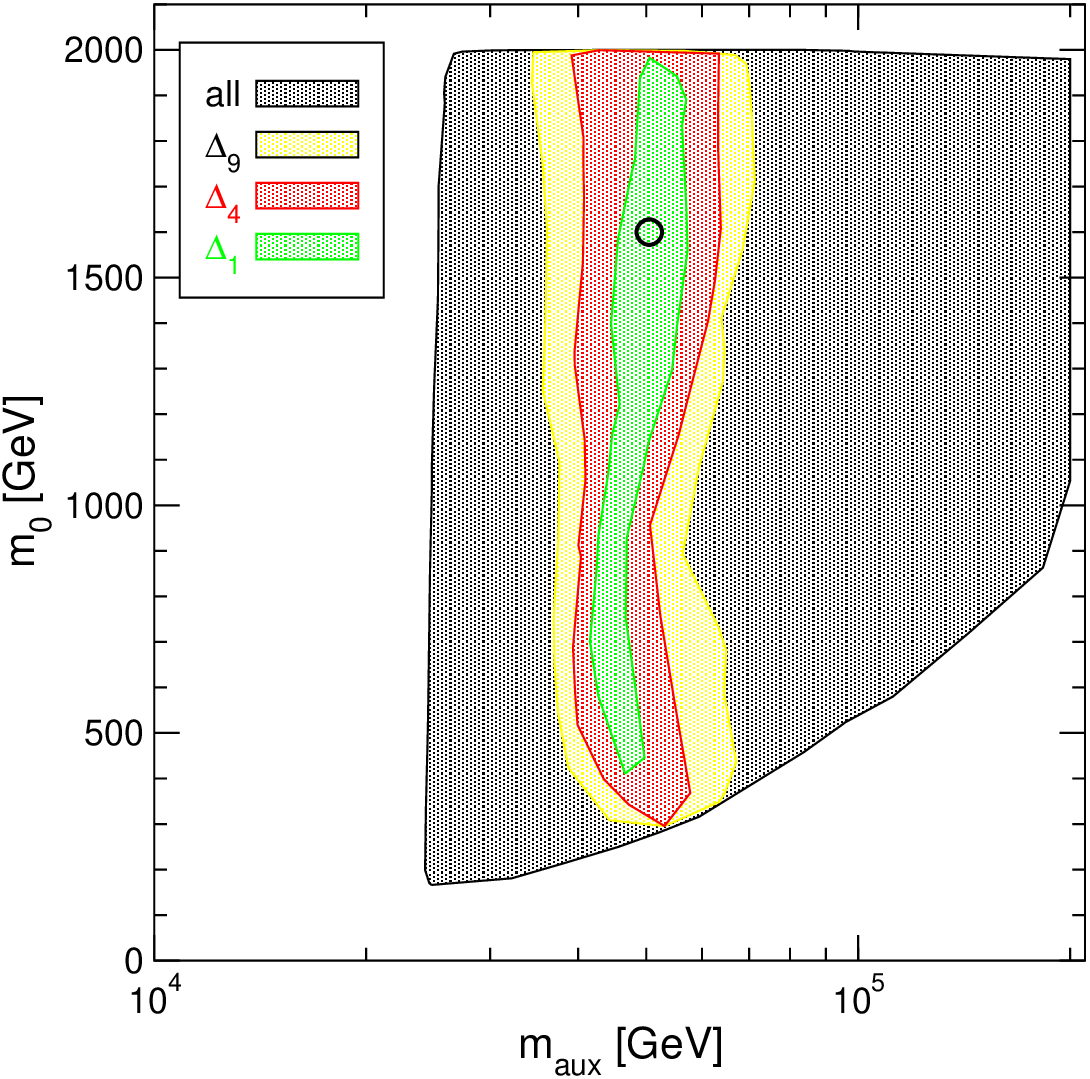}
\caption{%
Future projection for 
the $\De_{1,4,9}$ regions in the $m_{\rm aux}$--$m_0$ plane
in the mAMSB assuming that the future
experimental data agree exactly with the hypothetical best-fit point as
defined in \refeq{amsbfutbf}
The color coding is as in \reffi{fig:cmssmHigh}.
}
\label{fig:mamsbHighFX}
\end{center}
\end{figure}

A drastic improvement compared to the present situation can be observed
in all three scenarios.  
However, also for the low-energy parameters the quality of the
improvement going to the future sensitivities depends on the fact that
currently relatively low mass scales are favored, see below. 
The results look quite different in mAMSB as compared to
the CMSSM and mGMSB.
Within the latter two the $\De_9$ region is confined to 
$\MA \lsim 1000 \gev$ with a width of $300 (400) \gev$ for the CMSSM
(mGMSB), whereas $\tb$ is only weakly restricted, 
$10 (20) \lsim \tb \lsim 60$.
Within mAMSB, as for the high-energy parameters, a very precise
indirect determination of $\MA$ and $\tb$ can be performed. At the
$\De\chi^2 = 9$ level $\MA$ is confined to $\pm 50 \gev$, i.e.\ to
about~6\%.  $\tb$ is determined to $\pm 3$, corresponding to a precision
of $\sim 8\%$. However, as discussed in \refse{sec:analhighmamsb}, this is
largely due to the relatively small value of $\tb$ within the mAMSB
scenario at the current best-fit point.

\begin{figure}[htb!]
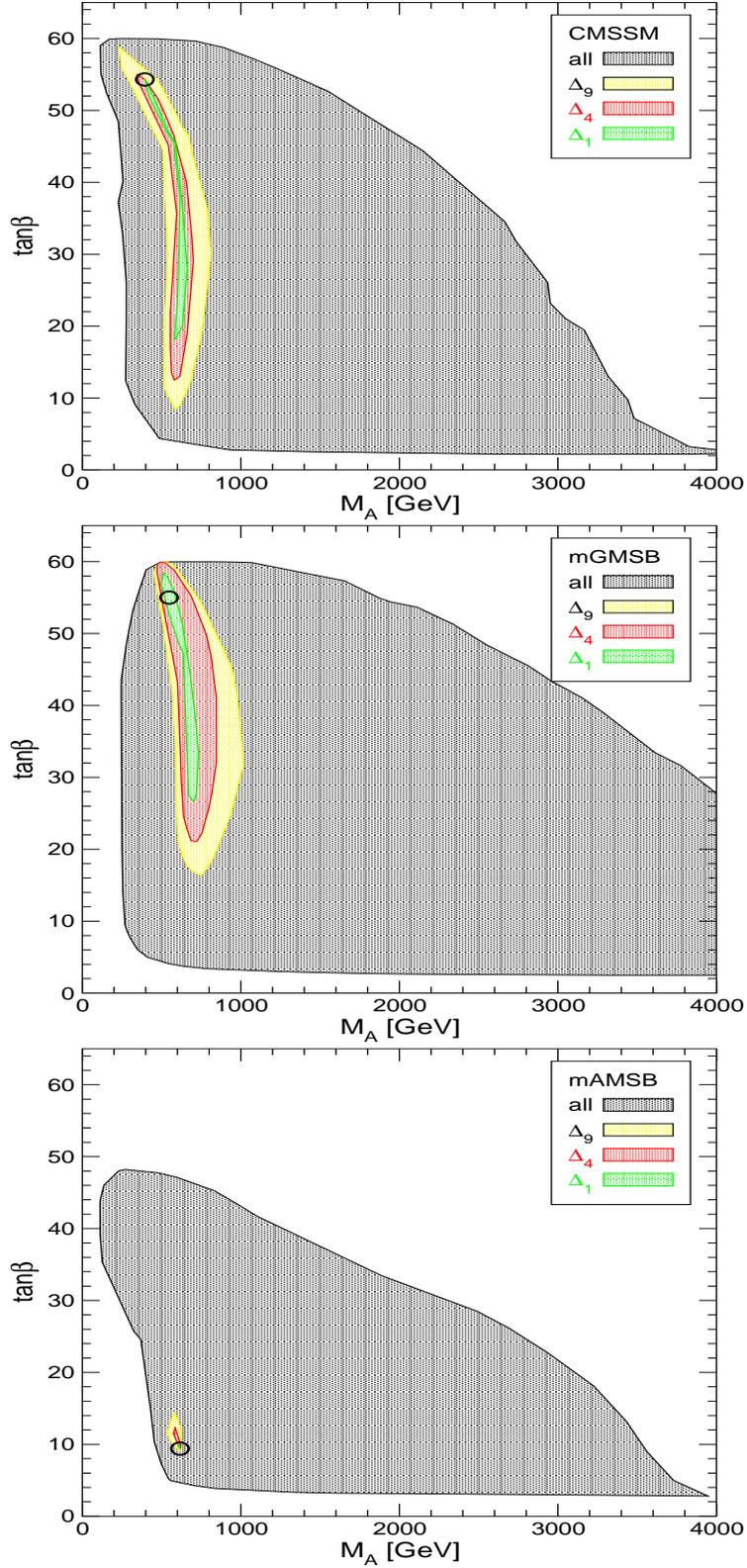

\begin{center}
\includegraphics[width=.6\textwidth,height=7.0cm]{asbs3_s_mass21_cl}
\includegraphics[width=.6\textwidth,height=7.0cm]{asbs3_g_mass21_cl}
\includegraphics[width=.6\textwidth,height=7.0cm]{asbs3_a_mass21_cl}
\caption{%
Future projection for 
the $\De_{1,4,9}$ regions in the $\MA$--$\tb$ planes
in the CMSSM (top), mGMSB (middle) and mAMSB (bottom) assuming that the
future measurements will agree exactly with the current best-fit point.
The color coding is as in \reffi{fig:cmssmHigh}.
}
\label{fig:asbsLowF}
\end{center}
\vspace{-3em}
\end{figure}

\begin{figure}[htb!]
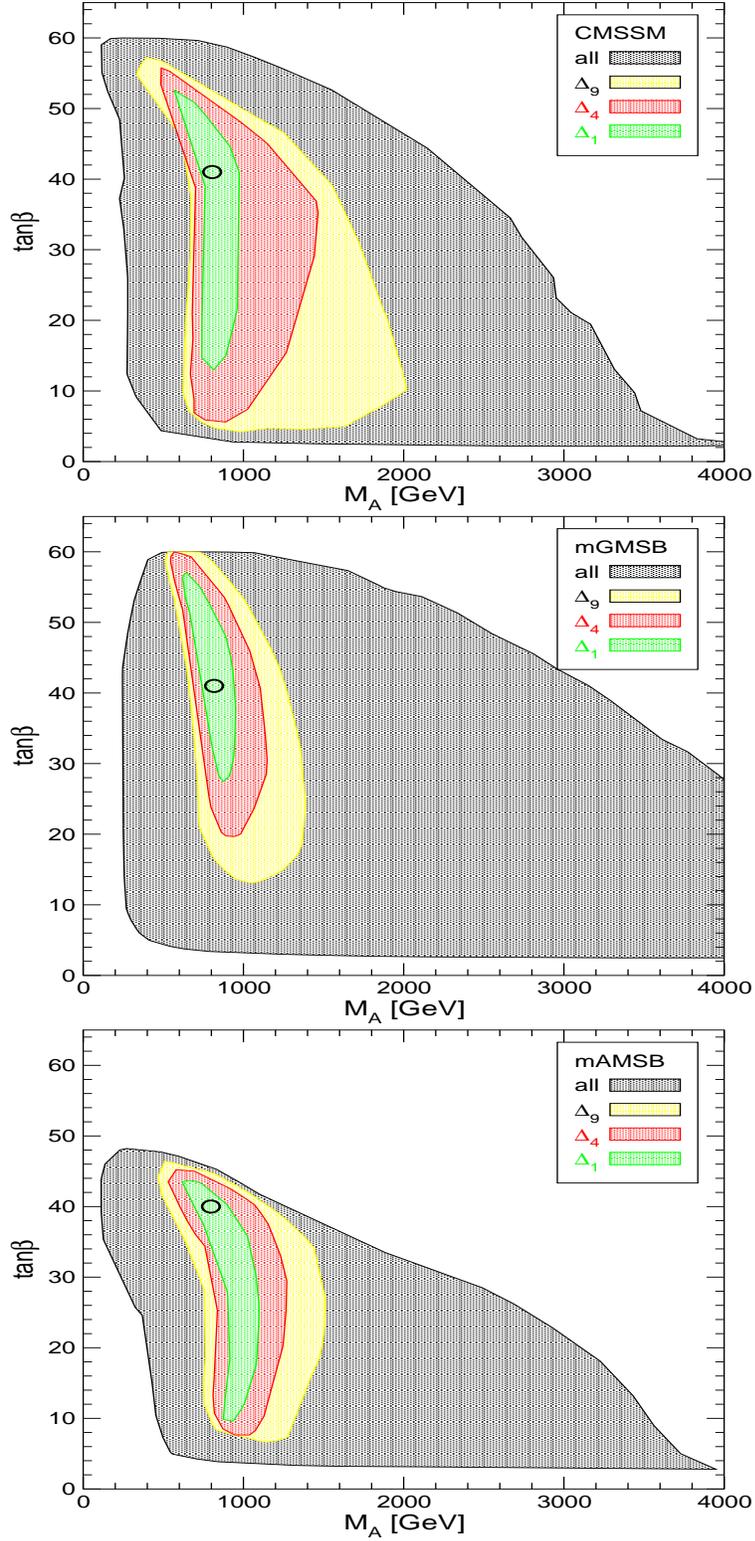

\begin{center}
\includegraphics[width=.6\textwidth,height=6.8cm]{asbs3_s_mass21F_cl}
\includegraphics[width=.6\textwidth,height=6.8cm]{asbs3_g_mass21F_cl}
\includegraphics[width=.6\textwidth,height=6.8cm]{asbs3_a_mass21F_cl}
\caption{%
Future projection for 
the $\De_{1,4,9}$ regions in the $\MA$--$\tb$ planes
in the CMSSM (top), mGMSB (middle) and mAMSB (bottom) assuming the
hypothetical future best fit points defined \refeqs{cmssmfutbf},
(\ref{gmsbfutbf}), (\ref{amsbfutbf}) for the CMSSM, mGMSB and mAMSB,
respectively. 
The color coding is as in \reffi{fig:cmssmHigh}.
}
\label{fig:asbsLowFX}
\end{center}
\vspace{-4em}
\end{figure}

We finally investigate the future sensitivity of the three soft
SUSY-breaking scenarios for the hypothetical best-fit point. 
In \reffi{fig:asbsLowFX} we show the results for the hypothetical best-fit
points as defined in \refeqs{cmssmfutbf}, (\ref{gmsbfutbf}), 
(\ref{amsbfutbf}) for the CMSSM, mGMSB and mAMSB, respectively.
By definition, see \refeq{asbsfutbf}, the hypothetical best-fit values for
$\MA$ and $\tb$ are very similar in the three scenarios, 
$\MA \approx 800 \gev$ and $\tb \approx 40$. 
These $\MA$ values are somewhat larger than the current best-fit values,
see \refta{tab:chi2min}. 
In combination with $\tb \approx 40$ such heavy MSSM Higgs bosons could
not be detected at the
LHC~\cite{atlastdr,cmstdr,jakobs,schumi,cmshiggsOrg,cmsHiggs,slhc} or
the ILC~\cite{teslatdr,orangebook,acfarep,Snowmass05Higgs}. 
Despite the fact that these
values are already in the decoupling regime (i.e.\ where SUSY masses are
large and loop effects are correspondingly small), the precision
observables are still able to provide upper (and lower) limits on $\MA$
and $\tb$ with 
similar results in the three soft SUSY-breaking scenarios. The upper
limit at the $\De\chi^2 = 9$ level on $\MA$ varies between 
$\sim 2000 \gev$ in the CMSSM and 
$\sim 1400 \gev$ in mGMSB. This means that the limits obtainable for
$\MA$ and $\tb$ depend only to a small extent on the details of
the underlying physics scenario and can thus be viewed as a more general
result for scenarios resulting from a high-scale theory.
In conclusion, the precision observables could allow one to
set an indirect bound on $\MA$ (and mildly also on $\tb$) beyond the direct 
collider reach. 
This sensitivity would improve even more if the future collider data
(SUSY masses etc.) would be included (see e.g.\ \citere{deschi}). 
Such an analysis, however, would at the present state be highly
speculative and is beyond the scope of our paper.


\section{Conclusions}
\label{sec:conclusions}

We investigated the constraints arising from electroweak precision
observables (EWPO) and $B$-physics 
observables (BPO) providing a comparison of the CMSSM, the
mGMSB and the mAMSB. 
We performed a $\chi^2$ analysis based on the mass of the $W$~boson,
$\MW$, the effective weak leptonic mixing angle, $\sweff$, the anomalous
magnetic moment of the muon $(g-2)_\mu$, the mass of the lightest $\cp$-even
MSSM Higgs boson, $\Mh$, as well as on $\br(b \to s \ga)$ and 
$\br(B_s \to \mu^+\mu^-)$. 
Our analysis should be viewed as an exploratory study for the comparison
of the scenarios, providing a starting point for 
a more refined investigation using more precision data and an elaborate
$\chi^2$ analysis~\cite{masterfit}.

Our results are analyzed separately in terms
of the high-scale parameters of the respective model as well as in terms of
low-energy parameters such as $\MA$, $\tb$ and SUSY particle masses.
Using todays measurements, uncertainties and exclusion bounds, 
we find that relatively low mass scales in all three scenarios are
favored at the level of $\De\chi^2 < 1$~or~4. 
However, the current data of EWPO and BPO can hardly set any upper bound
on the SUSY mass scales at the level of $\De\chi^2 = 9$. 
The best fit-values for $\MA$ range from $\sim 400 \gev$ in the CMSSM up
to $\sim 600 \gev$ in mAMSB, whereas the $\tb$ values are only weakly
constrained. 
Remarkably the mAMSB scenario, despite
having one free GUT scale parameter less than the other two scenarios,
has a somewhat lower total minimum $\chi^2$. This can be traced back to
a better agreement with the combination of the $\br(b \to s \ga)$ and
$(g-2)_\mu$ measurements (with some help from $\MW$) for a heavier
scalar quark spectrum and a corresponding slightly larger value of $\Mh$.

We presented predictions for the lightest $\cp$-even Higgs boson mass,
based on the 
current $\chi^2$ data, but without imposing the current LEP bound from Higgs
boson searches and its corresponding $\chi^2$ contribution. 
Best-fit values of $\Mh \sim 105 \gev$ are found for
the CMSSM and mGMSB, and $\Mh \sim 113 \gev$ for mAMSB. In all three
scenarios a relatively good compatibility with the direct bounds from
the Higgs searches at LEP is found. 
Within mAMSB the $\De\chi^2 < 1$ region extends up to $\Mh \lsim 122 \gev$. 

We also presented the predictions for the masses of various SUSY
particles such as $\mste$, $\mstz$, $\msbe$, $\msbz$, $\mstaue$,
$\mstauz$, $\mneu{1}$, $\mneu{2}$, $\mcha{1}$ and $\mgl$ in the three
soft SUSY-breaking scenarios. As a general feature lowest masses are
found in the mAMSB and heaviest in mGMSB. 
All three scenarios offer good prospects for the discovery of some
color-neutral particles at the ILC (with a center-of-mass energy up to
$\sqrt{s} = 1 \tev$) and for colored particles at the LHC. 
There are also good prospects for the discovery of uncolored particles
such as charginos, neutralinos and light sleptons,
especially if they are produced in cascade decays.
Some part of the preferred parameter space in the three scenarios 
is currently probed at the Tevatron.
Within the CMSSM qualitative agreement in the preferred mass ranges with
previous analyses \cite{ehow3,ehow4,ehoww} has been found.

Finally, we explored the projection for 
the future sensitivities of the EWPO and BPO in the three soft
SUSY-breaking scenarios. Here we also assumed a measurement of the
lightest MSSM Higgs boson mass. In a first step we analyzed the future
sensitivities assuming that the future measurements agree with the
current best-fit results. We found a strong 
improvement with respect to the current sensitivity. Within the
mAMSB scenario $\MA$ and $\tb$ can be determined indirectly with very
high precision, largely due to the fact that the current best-fit point
has a relatively low $\tb$ value. 
On the other hand, in the CMSSM and mGMSB the $\tb$ determination
remains relatively weak, where the current best-fit points have very
large $\tb$ values.
In a second step we assumed that the future measurements will agree 
in each scenario with a certain hypothetical point. These three points were
defined for each scenario 
such that they result in a similar Higgs and SUSY spectrum with 
$\MA \approx 800 \gev$ and $\tb \approx 40$. In general the Higgs and
SUSY mass scales are somewhat higher than for the current best-fit
points, i.e.\ loop corrections are correspondingly somewhat smaller.
These points would not permit
a direct determination of the heavy Higgs-boson mass scale. We find that 
the EWPO and BPO exhibit a similar future sensitivity in the 
CMSSM, mGMSB and mAMSB giving rise to an upper limit on the high-scale
parameters at the $\De\chi^2 = 9$ level.
The future EWPO and BPO sensitivities depend only mildly
on the underlying physics scenario. 
The precision observables could allow one to constrain the
Higgs sector parameters even beyond the direct reach of the LHC or the ILC.

Once LHC (and ILC) data on SUSY masses will be available, the assumption
about the underlying scenario itself will be investigated.
While information from the direct production of SUSY particles will
obviously be crucial for disentangling the underlying scenario of
SUSY-breaking, also the EWPO and BPO will certainly play an
important role in this context.


\subsubsection*{Acknowledgements}

\vspace{-0.5em}
We thank B.~Allanach for help with in the use of {\tt SoftSUSY}.
We thank A. Dedes and C.E.M.~Wagner for helpful discussions.
The work of S.H.\ was partially supported by CICYT (grant FPA~2007--66387).
X.M.\ and S.S.\ are supported
under U.S. Department of Energy contract \# DE-FG02-04ER-41298.
Work supported in part by the European Community's Marie-Curie Research
Training Network under contract MRTN-CT-2006-035505
`Tools and Precision Calculations for Physics Discoveries at Colliders'.





\begin{thebibliography}{999}


\bibitem{susy} H.~Nilles, 
               {\em Phys.\ Rept.} {\bf 110} (1984) 1.

\bibitem{susy2} H.~Haber and G.~Kane, 
               {\em Phys.\ Rept.} {\bf 117} (1985) 75; \\
               R.~Barbieri, 
               {\em Riv.\ Nuovo Cim.} {\bf 11} (1988) 1. 

\bibitem{hfag} Heavy Flavor Analysis group, 
               see: {\tt www.slac.stanford.edu/xorg/hfag/} .

\bibitem{GUTs} J.~Ellis, S.~Kelley and D.~Nanopoulos, 
               {\em Phys.\ Lett.} {\bf B 260} (1991) 131;\\
               U.~Amaldi, W.~de Boer and H.~Furstenau,
               {\em Phys.\ Lett.} {\bf B 260} (1991) 447;\\
               C.~Giunti, C.~Kim and U.~Lee,
               {\em Mod.\ Phys.\ Lett.} {\bf A 6} (1991) 1745.

\bibitem{oldGMSB} M.~Dine, W.~Fischler and M.~Srednicki, 
                  {\em Nucl. Phys.} {\bf B 189} (1981) 575;\\
                  S.~Dimopoulos and S.~Raby, 
                  {\em Nucl. Phys.} {\bf B 192} (1981) 353;\\
                  M.~Dine and W.~Fischler, 
                  {\em Phys. Lett.} {\bf B 110} (1982) 227;\\
                  M.~Dine and M.~Srednicki, 
                  {\em Nucl. Phys.} {\bf B 202} (1982) 238;\\
                  M.~Dine and W.~Fischler, 
                  {\em Nucl. Phys.} {\bf B 204} (1982) 346;\\
                  L.~Alvarez-Gaum\'e, M.~Claudson and M.~Wise, 
                  {\em Nucl. Phys.} {\bf B 207} (1982) 96;\\
                  C.~Nappi and B.~Ovrut, 
                  {\em Phys. Lett.} {\bf B 113} (1982) 175;\\
                  S.~Dimopoulos and S.~Raby, 
                  {\em Nucl. Phys.} {\bf B 219} (1983) 479.

\bibitem{newGMSB} M.~Dine and A.~Nelson, 
                  {\em Phys. Rev.} {\bf D 48} (1993) 1277
                  [arXiv:hep-ph/9303230];\\
                  M.~Dine, A.~Nelson and Y.~Shirman, 
                  {\em Phys. Rev.} {\bf D 51} (1995) 1362
                  [arXiv:hep-ph/9408384];\\
                  M.~Dine, A.~Nelson, Y.~Nir and Y.~Shirman, 
                  {\em Phys. Rev.} {\bf D 53} (1996) 2658
                  [arXiv:hep-ph/9507378].

\bibitem{GR-GMSB} For a review, see: G.~Giudice and R.~Rattazzi, 
                  {\em Phys. Rept.} {\bf 322} (1999) 419
                  [arXiv:hep-ph/9801271].

\bibitem{lr} L.~Randall and R.~Sundrum, 
             {\em Nucl. Phys.} {\bf B 557} (1999) 79
             [arXiv:hep-th/9810155].

\bibitem{giudice} G.~Giudice, M.~Luty, H.~Murayama and
                  R.~Rattazzi, 
                  {\em JHEP} {\bf 9812} (1998) 027
                  [arXiv:hep-ph/9810442].

\bibitem{wells} T.~Gherghetta, G.~Giudice and J.~Wells,
                {\em Nucl. Phys.} {\bf B 559} (1999) 27
                [arXiv:hep-ph/9904378].

\bibitem{ehow3} J.~Ellis, S.~Heinemeyer, K.~Olive and G.~Weiglein,
                {\em JHEP} {\bf 0502} (2005) 013
                [arXiv:hep-ph/0411216].

\bibitem{ehow4} J.~Ellis, S.~Heinemeyer, K.~Olive and G.~Weiglein,
                {\em JHEP} {\bf 0605} (2006) 005
                [arXiv:hep-ph/0602220].

\bibitem{ehoww} J.~Ellis, S.~Heinemeyer, K.~Olive, A.M.~Weber and 
                G.~Weiglein, 
                {\em JHEP} {\bf 0708} (2007) 083
                [arXiv:0706.0652 [hep-ph]].

\bibitem{ehhow} J.~Ellis, T.~Hahn, S.~Heinemeyer, K.~Olive and G.~Weiglein,
                {\em JHEP} {\bf 0710} (2007) 092
                [arXiv:0709.0098 [hep-ph]].

\bibitem{ehow5} J.~Ellis, S.~Heinemeyer, K.~Olive and G.~Weiglein,
                {\em Phys.\ Lett.} {\bf B 653} (2007) 292
                [arXiv:0706.0977 [hep-ph]].

\bibitem{other} J.~Ellis, K.~Olive, Y.~Santoso and V.~Spanos,
                {\em Phys.\ Rev.} {\bf D 69} (2004) 095004
                [arXiv:hep-ph/0310356];\\
                B.~Allanach and C.~Lester,
                {\em Phys.\ Rev.} {\bf D 73} (2006) 015013
                [arXiv:hep-ph/0507283];\\
                B.~Allanach,
                {\em Phys.\ Lett.} {\bf B 635} (2006) 123
                [arXiv:hep-ph/0601089];\\
                R.~de Austri, R.~Trotta and L.~Roszkowski,
                {\em JHEP} {\bf 0605} (2006) 002
                [arXiv:hep-ph/0602028];
                {\em JHEP} {\bf 0704} (2007) 084
                [arXiv:hep-ph/0611173];
                {\em JHEP} {\bf 0707} (2007) 075
                [arXiv:0705.2012 [hep-ph]];\\
                B.~Allanach, C.~Lester and A.~M.~Weber,
                {\em JHEP} {\bf 0612} (2006) 065
                [arXiv:hep-ph/0609295];
                {\em JHEP} {\bf 0708} (2007) 023
                [arXiv:0705.0487 [hep-ph]];\\
                S.~Heinemeyer,
                arXiv:hep-ph/0611372.

\bibitem{LSPlargeTB} G.~Isidori, F.~Mescia, P.~Paradisi and D.~Temes,
                     {\em Phys.\ Rev.} {\bf D 75} (2007) 115019
                     [arXiv:hep-ph/0703035];\\
                     M.~Carena, A.~Menon and C.~Wagner,
                     {\em Phys.\ Rev.} {\bf D 76} (2007) 035004
                     [arXiv:0704.1143 [hep-ph]].

\bibitem{masterfit} O.~Buchmueller et al.,
                    {\em Phys.\ Lett.} {\bf B 657} (2007) 87
                    [arXiv:0707.3447 [hep-ph]].

\bibitem{fut} S.~Heinemeyer, M.~Mondrag\'on and G.~Zoupanos, 
              arXiv:0712.3630 [hep-ph];\\
              A.~Djouadi, S.~Heinemeyer, M.~Mondrag\'on and G.~Zoupanos, 
              {\em Springer Proc.\ Phys.} {\bf 98} (2005) 273
              [arXiv:hep-ph/0404208].

\bibitem{nazilla2} F.~Mahmoudi,
                   {\em JHEP} {\bf 0712} (2007) 026
                   [arXiv:0710.3791 [hep-ph]].

\bibitem{nmssmBPO} F.~Domingo and U.~Ellwanger,
                   {\em JHEP} {\bf 0712} (2007) 090
                   [arXiv:0710.3714 [hep-ph]].

\bibitem{MSSM11fit} J.~Kasahara, K.~Freese and P.~Gondolo,
                    arXiv:0805.0999 [hep-ph].

\bibitem{WMAP} C.~Bennett et al.,
               {\em Astrophys. J. Suppl.} {\bf 148} (2003) 1
               [arXiv:astro-ph/0302207];\\
               D.~Spergel et al.\ [WMAP Collaboration],
               {\em Astrophys.\ J.\ Suppl.} {\bf 148} (2003) 175
               [arXiv:astro-ph/0302209];\\
               D.~Spergel et al.\ [WMAP Collaboration],
               {\em Astrophys.\ J.\ Suppl.} {\bf 170} (2007) 377
               [arXiv:astro-ph/0603449].

\bibitem{nazilla} A.~Arbey and F.~Mahmoudi,
                  arXiv:0803.0741 [hep-ph].

\bibitem{herbi} H.~Dreiner,
                arXiv:hep-ph/9707435;\\
                G.~Bhattacharyya,
                arXiv:hep-ph/9709395;\\
                B.~Allanach, A.~Dedes and H.~Dreiner,
                {\em Phys.\ Rev.} {\bf D 60} (1999) 075014
                [arXiv:hep-ph/9906209].

\bibitem{thermalinf} D.~Lyth and E.~Stewart, 
                     {\em Phys. Rev.} {\bf D 53} (1996) 1784
                     [arXiv:hep-ph/9510204].

\bibitem{latetimeentropy} G.~Gelmini and P.~Gondolo,
                          {\em Phys.\ Rev.} {\bf D 74} (2006) 023510
                          [arXiv:hep-ph/0602230].

\bibitem{nmfv} G.~Degrassi, P.~Gambino and P.~Slavich,
               {\em Phys.\ Lett.} {\bf B 635} (2006) 335
               [arXiv:hep-ph/0601135];\\
               E.~Lunghi, W.~Porod and O.~Vives,
               {\em Phys.\ Rev.} {\bf D 74} (2006) 075003
               [arXiv:hep-ph/0605177].

\bibitem{Hall} H.~Nilles,
               {\em Phys. Lett.} {\bf B 115} (1982) 193;
               {\em Nucl. Phys.} {\bf B 217} (1983) 366;\\
               A.~Chamseddine, R.~Arnowitt and P.~Nath,
               {\em Phys. Rev. Lett.}  {\bf 49} (1982) 970;\\
               R.~Barbieri, S.~Ferrara and C.~Savoy,
               {\em Phys. Lett.} {\bf B 119} (1982) 343;\\
               H.~Nilles, M.~Srednicki and D.~Wyler,
               {\em Phys. Lett.} {\bf B 120} (1983) 346;\\
               E.~Cremmer, P.~Fayet and L.~Girardello,
               {\em Phys. Lett.} {\bf B 122} (1983) 41;\\
               S.~Ferrara, L.~Girardello and H.~Nilles,
               {\em Phys. Lett.} {\bf B 125} (1983) 457;\\
               L.~Hall, J.~Lykken and S.~Weinberg,
               {\em Phys. Rev.} {\bf D 27} (1983) 2359;\\
               S.~Soni and H.~Weldon,
               {\em Phys. Lett.} {\bf B 126} (1983) 215;\\
               R.~Arnowitt, A.~Chamseddine and P.~Nath,
               {\em Nucl. Phys.} {\bf B 227} (1983) 121.\\
For more details see, S.~Weinberg,
``The quantum theory of fields.  Vol. 3: Supersymmetry,''
{\it  Cambridge University Press (2000)}. 

\bibitem{mSUGRArev} For reviews see also: \\
                    \citere{susy} and the first article in \citere{susy2};\\
                    A.~Lahanas and D.~Nanopoulos,
                    {\em Phys. Rept.}  {\bf 145} (1987) 1; \\
                    S.~Martin, 
                    in {\em ``Perspectives on supersymmetry''}, ed. G.~Kane,
                    arXiv:hep-ph/9709356,
                    see: {\tt zippy.physics.niu.edu/primer.shtml}.

\bibitem{GMSBrecent} H.~Murayama and Y.~Nomura,
                     {\em Phys.\ Rev.\ Lett.} {\bf 98} (2007) 151803
                     [arXiv:hep-ph/0612186];
                     {\em Phys.\ Rev.} {\bf D 75} (2007) 095011
                     [arXiv:hep-ph/0701231].

\bibitem{GMSBrecent2} S.~Abel, C.~Durnford, J.~Jaeckel and V.~V.~Khoze,
                      {\em JHEP} {\bf 0802} (2008) 074
                      [arXiv:0712.1812 [hep-ph]].

\bibitem{pdg} W.~Yao et al.\  [Particle Data Group Collaboration],
              {\em J.\ Phys.} {\bf G 33} (2006) 1.

\bibitem{Fayet} P.~Fayet, 
                {\em Phys. Lett.} {\bf B 70} (1977) 461;
                {\em Phys. Lett.} {\bf B 86} (1979) 272;
                {\em Phys. Lett.} {\bf B 175} (1986) 471;
    and in {\em ``Unification of the fundamental 
    particle interactions"}, eds.~S.~Ferrara, J.~Ellis,   
    P.~van Nieuwenhuizen (Plenum, New York, 1980) p.~587.

\bibitem{negative} A.~Pomarol and R.~Rattazzi,
                   {\em JHEP} {\bf 9905} (1999) 013
                   [arXiv:hep-ph/9903448].

\bibitem{clm} Z.~Chacko, M.~Luty and E.~Ponton, 
              {\em JHEP} {\bf 0004} (2000) 001
              [arXiv:hep-ph/9905390]. 

\bibitem{kss} E.~Katz, Y.~Shadmi and Y.~Shirman,
              {\em JHEP} {\bf 9908} (1999) 015
              [arXiv:hep-ph/9906296].

\bibitem{jjw} I.~Jack, D.~Jones and R.~Wild,
              {\em Phys. Lett.} {\bf B 535}, 193 (2002)
              [arXiv:hep-ph/0202101].

\bibitem{softsusy} B.~Allanach,
                   {\em Comput.\ Phys.\ Commun.} {\bf 143} (2002) 305
                   [arXiv:hep-ph/0104145].

\bibitem{PomssmRep} S.~Heinemeyer, W.~Hollik and G.~Weiglein, 
                    {\em Phys.\ Rept.} {\bf 425} (2006) 265
                    [arXiv:hep-ph/0412214].

\bibitem{ZOpope} S.~Heinemeyer, W.~Hollik, A.M.~Weber and G.~Weiglein,
                 {\em JHEP} {\bf 0804} (2008) 039
                 [arXiv:0710.2972 [hep-ph]].

\bibitem{mt1714} E.~Brubaker et al. [Tevatron Electroweak Working Group],
                 arXiv:hep-ex/0608032,\\
                 see: {\tt tevewwg.fnal.gov/top/}~.

\bibitem{mt1726} Tevatron Electroweak Working Group,
                 axXiv:0803.1683 [hep-ex];\\
                 see: {\tt tevewwg.fnal.gov/top/}~.

\bibitem{sirlin} A.~Sirlin, 
                 {\em Phys. Rev.} {\bf D 22} (1980) 971;\\
                 W.~Marciano and A.~Sirlin, 
                 {\em Phys. Rev.} {\bf D 22} (1980) 2695.

\bibitem{deltarMSSM1lA} P.~Chankowski, A.~Dabelstein, W.~Hollik,
                        W.~M\"osle, S.~Pokorski and J.~Rosiek,
                        {\em Nucl. Phys.} {\bf B 417} (1994) 101.

\bibitem{deltarMSSM1lB} D.~Garcia and J.~Sol\`a,
                        {\em Mod. Phys. Lett.} {\bf A 9} (1994) 211.

\bibitem{drSMgfals} A.~Djouadi and C.~Verzegnassi,
                    {\em Phys. Lett.} {\bf B 195} (1987) 265;\\
                    A.~Djouadi,
                    {\em Nuovo Cim.} {\bf A 100} (1988) 357.

\bibitem{deltarSMgfals} B.~Kniehl, 
                        {\em Nucl. Phys.} {\bf B 347} (1990) 89;\\
                        F.~Halzen and B.~Kniehl, 
                        {\em Nucl. Phys.} {\bf B 353} (1991) 567;\\
                        B.~Kniehl and A.~Sirlin, 
                        {\em Nucl. Phys.} {\bf B 371} (1992) 141;
                        {\em Phys. Rev.} {\bf D 47} (1993) 883.

\bibitem{drSMgfals2} K.~Chetyrkin, J.~K\"uhn and M.~Steinhauser,
                     {\em Phys. Rev. Lett.} {\bf 75} (1995) 3394
                     [arXiv:hep-ph/9504413];\\
                     L.~Avdeev et al.,
                     {\em Phys. Lett.} {\bf B 336} (1994) 560,
                     [Erratum-ibid.\ {\bf B 349} (1995) 597]
                     [arXiv:hep-ph/9406363].

\bibitem{drSMgfals2LF} K.~Chetyrkin, J.~K\"uhn and M.~Steinhauser,
                       {\em Nucl. Phys.} {\bf B 482} (1996) 213
                       [arXiv:hep-ph/9606230].

\bibitem{dr2lA} A.~Djouadi, P.~Gambino, S.~Heinemeyer, W.~Hollik,
                C.~J\"unger and G.~Weiglein, 
                {\em Phys. Rev. Lett.} {\bf 78} (1997) 3626
                [arXiv:hep-ph/9612363];
                {\em Phys. Rev.} {\bf D 57} (1998) 4179
                [arXiv:hep-ph/9710438].

\bibitem{drMSSMal2B} S.~Heinemeyer and G.~Weiglein, 
                     {\em JHEP} {\bf 0210} (2002) 072
                     [arXiv:hep-ph/0209305];
                     arXiv:hep-ph/0301062.

\bibitem{drMSSMal2} J.~Haestier, S.~Heinemeyer, D.~St\"ockinger and
                     G.~Weiglein,
                     {\em JHEP} {\bf 0512} (2005) 027
                     [arXiv:hep-ph/0508139];
                     arXiv:hep-ph/0506259.

\bibitem{rho} M.~Veltman, 
              {\em Nucl. Phys.} {\bf B 123} (1977) 89. 

\bibitem{MWpope} S.~Heinemeyer, W.~Hollik, D.~St\"ockinger, 
                 A.M.~Weber and G.~Weiglein,
                 {\em JHEP} {\bf 0608} (2006) 052
                 [arXiv:hep-ph/0604147];
                 arXiv:hep-ph/0611371.

\bibitem{mtdet1} A.~Hoang et al., 
                 {\em Eur. Phys. J. direct} {\bf C 2} (2000) 1
                 [arXiv:hep-ph/0001286].

\bibitem{mtdet2} M.~Martinez and R.~Miquel,
                 {\em Eur. Phys. J.} {\bf C 27} (2003) 49
                 [arXiv:hep-ph/0207315].

\bibitem{fredl} F.~Jegerlehner,
                Talk presented at the LNF Spring School,
                Frascati, Italy, 1999,\\ see:
           {\tt www-com.physik.hu-berlin.de/$\sim$fjeger/Frascati99.ps.gz} ;\\
           arXiv:hep-ph/0105283.

\bibitem{deltamt} S.~Heinemeyer, S.~Kraml, W.~Porod and G.~Weiglein,
                  {\em JHEP} {\bf 0309} (2003) 075
                  [arXiv:hep-ph/0306181].

\bibitem{lepewwg} The ALEPH, DELPHI, L3, OPAL, SLD Collaborations,
                  the LEP Electroweak Working Group,
                  the SLD Electroweak and Heavy Flavour Groups,
                  arXiv:hep-ex/0509008;\\ 
                  {}[The ALEPH, DELPHI, L3 and OPAL Collaborations, the LEP
                  Electroweak Working Group], 
                  arXiv:hep-ex/0612034.

\bibitem{LEPEWWG} LEP Electroweak Working Group,\\
                  see: {\tt lepewwg.web.cern.ch/LEPEWWG/Welcome.html}.

\bibitem{TEVEWWG} Tevatron Electroweak Working Group,
                  see: {\tt tevewwg.fnal.gov}~.

\bibitem{MWcdf} CDF collaboration,
                arXiv:0707.0085 [hep-ex];\\
                see: {\tt www-cdf.fnal.gov/physics/ewk/}~.

\bibitem{MWworld} M.~Gr\"unewald, 
                  private communication. 

\bibitem{ssdd} S.~Brensing, S.~Dittmaier, M.~Kr\"amer and A.~M\"uck,
               {\em Phys.\ Rev.} {\bf D 77} (2008) 073006
               [arXiv:0710.3309 [hep-ph]].

\bibitem{mwgigaz} G.~Wilson, LC-PHSM-2001-009, see:
                  {\tt www.desy.de/$\sim$lcnotes/notes.html} .

\bibitem{blueband} U.~Baur, R.~Clare, J.~Erler, S.~Heinemeyer,
                   D.~Wackeroth, G.~Weiglein and D.~Wood,
                   arXiv:hep-ph/0111314.

\bibitem{sw2eff2l} M.~Awramik, M.~Czakon, A.~Freitas and G.~Weiglein,
                   {\em Phys.\ Rev.\ Lett.} {\bf 93} (2004) 201805
                   [arXiv:hep-ph/0407317].

\bibitem{ewpo:gigaz2} R.~Hawkings and K.~M\"onig,
                     {\em Eur. Phys. J. direct} {\bf C 8} (1999) 1
                     [arXiv:hep-ex/9910022].

\bibitem{gigaz} J.~Erler, S.~Heinemeyer, W.~Hollik, G.~Weiglein 
                and P.~Zerwas,
                {\em Phys. Lett.} {\bf B 486} (2000) 125
                [arXiv:hep-ph/0005024].

\bibitem{g-2review} A.~Czarnecki and W.~Marciano,
                    {\em Phys. Rev.} {\bf D 64} (2001) 013014
                    [arXiv:hep-ph/0102122].

\bibitem{g-2review2} M.~Knecht,
                     {\em Lect.\ Notes Phys.} {\bf 629} (2004) 37
                     [arXiv:hep-ph/0307239];\\
                     M.~Passera,
                     {\em Nucl.\ Phys.\ Proc.\ Suppl.} {\bf 155} (2006) 365
                     [arXiv:hep-ph/0509372].

\bibitem{g-2reviewDS} D.~St\"ockinger,
                      {\em J.\ Phys.} {\bf G 34} (2007) R45
                      [arXiv:hep-ph/0609168].

\bibitem{g-2reviewMRR} J.~Miller, E.~de Rafael and B.~Roberts,
                       {\em Rept.\ Prog.\ Phys.} {\bf 70} (2007) 795
                       [arXiv:hep-ph/0703049].

\bibitem{g-2reviewFJ} F.~Jegerlehner,
                      {\em Acta Phys.\ Polon.} {\bf B 38} (2007) 3021
                      [arXiv:hep-ph/0703125].

\bibitem{g-2reviewPMS} M.~Passera, W.~Marciano and A.~Sirlin,
                       arXiv:0804.1142 [hep-ph].

\bibitem{Kinoshita} T.~Kinoshita and M.~Nio,
                    {\em Phys.\ Rev.} {\bf D 70} (2004) 113001
                    [arXiv:hep-ph/0402206];
                    {\em Phys.\ Rev.} {\bf 73} (2006) 053007
                    [arXiv:hep-ph/0512330].

\bibitem{g-2QEDmassdep} M.~Passera,
                        {\em Phys.\ Rev.} {\bf D 75} (2007) 013002
                        [arXiv:hep-ph/0606174].

\bibitem{g-2QEDrecent} T.~Aoyama, M.~Hayakawa, T.~Kinoshita and M.~Nio,
                       {\em Phys.\ Rev.\ Lett.} {\bf 99} (2007) 110406
                       [arXiv:0706.3496 [hep-ph]];
                       {\em Nucl.\ Phys.} {\bf B 796} (2008) 184
                       [arXiv:0709.1568 [hep-ph]];
                       arXiv:0712.2607 [hep-ph].

\bibitem{DEHZ} M.~Davier, S.~Eidelman, A.~H\"ocker and Z.~Zhang,
               {\em Eur.\ Phys.\ J.}\  {\bf C 31} (2003) 503
               [arXiv:hep-ph/0308213].

\bibitem{g-2HMNT} K.~Hagiwara, A.~Martin, D.~Nomura and T.~Teubner,
                  {\em Phys. Rev.} {\bf D 69} (2004) 093003
                  [arXiv:hep-ph/0312250].

\bibitem{g-2HMNT2} K.~Hagiwara, A.~Martin, D.~Nomura and T.~Teubner,
                   {\em Phys.\ Lett.} {\bf B 649} (2007) 173
                   [arXiv:hep-ph/0611102].

\bibitem{Jegerlehner} S.~Ghozzi and F.~Jegerlehner,
                      {\em Phys. Lett.} {\bf B 583} (2004) 222
                      [arXiv:hep-ph/0310181].

\bibitem{Yndurain} J.~de Troconiz and F.~Yndurain,
                   {\em Phys.\ Rev.} {\bf D 71} (2005) 073008
                   [arXiv:hep-ph/0402285].

\bibitem{DDDD} M.~Davier,
               {\em Nucl.\ Phys.\ Proc.\ Suppl.} {\bf 169} (2007) 288
               [arXiv:hep-ph/0701163].

\bibitem{LBLrev} J.~Bijnens and J.~Prades,
                 {\em Mod.\ Phys.\ Lett.} {\bf A 22} (2007) 767
                 [arXiv:hep-ph/0702170].

\bibitem{LBL} M.~Knecht and A.~Nyffeler,
              {\em Phys. Rev.} {\bf D 65} (2002) 073034
              [arXiv:hep-ph/0111058];\\
              M.~Knecht, A.~Nyffeler, M.~Perrottet and E.~De Rafael,
              {\em Phys. Rev. Lett.} {\bf 88} (2002) 071802
              [arXiv:hep-ph/0111059];\\
              I.~Blokland, A.~Czarnecki and K.~Melnikov,
              {\em Phys. Rev. Lett.} {\bf 88} (2002) 071803
              [arXiv:hep-ph/0112117];\\
              M.~Ramsey-Musolf and M.~Wise,
              {\em Phys. Rev. Lett.} {\bf 89} (2002) 041601
              [arXiv:hep-ph/0201297];\\
              J.~K\"uhn, A.~Onishchenko, A.~Pivovarov and O.~Veretin,
              {\em Phys. Rev.} {\bf D 68} (2003) 033018
              [arXiv:hep-ph/0301151].

\bibitem{LBLnew} K.~Melnikov and A.~Vainshtein,
                 {\em Phys. Rev.} {\bf D 70} (2004) 113006
                 [arXiv:hep-ph/0312226].

\bibitem{LBLnew2} M.~Davier and W.~Marciano,
                  {\em Ann.\ Rev.\ Nucl.\ Part.\ Sci.} {\bf 54} (2004) 115.

\bibitem{belle} H.~Hayashii  [Belle Collaboration],
                PoS {\bf HEP2005} (2006) 291.

\bibitem{g-2tauiso} M.~Benayoun, P.~David, L.~DelBuono, 
                    O.~Leitner and H.~O'Connell,
                    arXiv:0711.4482 [hep-ph].

\bibitem{KLOE} A.~Aloisio et al.  [KLOE Collaboration],
               {\em Phys.\ Lett.} {\bf 606} (2005) 12
               [arXiv:hep-ex/0407048];\\
               D.~Leone  [KLOE Collaboration],
               {\em Nucl.\ Phys.\ Proc.\ Suppl.} {\bf 162} (2006) 95.

\bibitem{CMD2} R.~Akhmetshin et al.\ [CMD-2 Collaboration],
               {\em Phys.\ Lett.} {\bf B 578} (2004) 285
               [arXiv:hep-ex/0308008];
               {\em Phys.\ Lett.} {\bf B 648} (2007) 28
               [arXiv:hep-ex/0610021].

\bibitem{SND} M.~Achasov et al.\ [SND Collaboration],
              {\em J.\ Exp.\ Theor.\ Phys.} {\bf 101} (2005) 1053
              [arXiv:hep-ex/0506076].

\bibitem{KLOEeps} S.~M\"uller, 
                  talk given at the EPS07, Manchester, July 2007, see:\\
                  {\tt agenda.hep.man.ac.uk/contribution} \\
                  {\tt Display.py?contribId=36\&sessionId=26\&confId=70} .

\bibitem{g-2exp} G.~Bennett et al.\ [The Muon g-2 Collaboration],
                 {\em Phys. Rev. Lett.} {\bf 92} (2004) 161802
                 [arXiv:hep-ex/0401008].

\bibitem{g-2exp2} G.~Bennett et al.\ [The Muon g-2 Collaboration],
                  {\em Phys.\ Rev.} {\bf D 73} (2006) 072003
                  [arXiv:hep-ex/0602035].

\bibitem{g-2SEtalk} S.~Eidelman, 
            talk given at the ICHEP06, Moscow, July 2006,\\ see:
            {\tt ichep06.jinr.ru/reports/333\_6s1\_9p30\_Eidelman.pdf}~;\\
            V.~Druzhinin,
            talk given at {\em Lepton Photon 07}, 
            August 2007, Daegu, Korea,\\ 
            see: {\tt chep.knu.ac.kr/lp07/htm/S5/S05-15.pdf}~.


\bibitem{g-2MSSMf1l} T.~Moroi,
                     {\em Phys. Rev.} {\bf D 53} (1996) 6565
                     [Erratum-ibid.\ {\bf D 56} (1997) 4424]
                     [arXiv:hep-ph/9512396].

\bibitem{correlation} J.~Lopez, D.~Nanopoulos and X.~Wang,
                      {\em Phys.\ Rev.} {\bf D 49} (1994) 366
                      [arXiv:hep-ph/9308336];\\
                      U.~Chattopadhyay and P.~Nath,
                      {\em Phys.\ Rev.} {\bf D 53} (1996) 1648
                      [arXiv:hep-ph/9507386].

\bibitem{g-2MSSMlog2l} G.~Degrassi and G.~Giudice,
                       {\em Phys. Rev.} {\bf D 58} (1998) 053007
                       [arXiv:hep-ph/9803384].

\bibitem{g-2FSf} S.~Heinemeyer, D.~St\"ockinger and G.~Weiglein,
                 {\em Nucl. Phys.} {\bf B 690} (2004) 62
                 [arXiv:hep-ph/0312264].

\bibitem{g-2CNH} S.~Heinemeyer, D.~St\"ockinger and G.~Weiglein,
                 {\em Nucl. Phys.} {\bf B 699} (2004) 103
                 [arXiv:hep-ph/0405255].

\bibitem{mhiggsf1lC} A.~Dabelstein,
                     {\em Nucl. Phys.} {\bf B 456} (1995) 25
                     [arXiv:hep-ph/9503443];
                     {\em Z. Phys.} {\bf C 67} (1995) 495
                     [arXiv:hep-ph/9409375].

\bibitem{mhiggsletter} S.~Heinemeyer, W.~Hollik and G.~Weiglein, 
                       {\em Phys. Rev.} {\bf D 58} (1998) 091701
                       [arXiv:hep-ph/9803277]; 
                       {\em Phys. Lett.} {\bf B 440} (1998) 296
                       [arXiv:hep-ph/9807423].

\bibitem{feynhiggs} S.~Heinemeyer, W.~Hollik and G.~Weiglein, 
                    {\em Comput. Phys. Commun.} {\bf 124} 2000 76
                    [arXiv:hep-ph/9812320].
                    The code is accessible via
                    {\tt www.feynhiggs.de} .

\bibitem{mhiggslong} S.~Heinemeyer, W.~Hollik and G.~Weiglein,
                     {\em Eur. Phys. J.} {\bf C 9} (1999) 343
                     [arXiv:hep-ph/9812472].

\bibitem{mhiggsAEC} G.~Degrassi, S.~Heinemeyer, W.~Hollik,
                    P.~Slavich, G.~Weiglein, 
                    {\em Eur. Phys. J.} {\bf C 28} (2003) 133
                    [arXiv:hep-ph/0212020].

\bibitem{mhcMSSMlong} M.~Frank, T.~Hahn, S.~Heinemeyer, W.~Hollik,  
                      H.~Rzehak and G.~Weiglein,
                      {\em JHEP} {\bf 0702} (2007) 047
                      [arXiv:hep-ph/0611326].

\bibitem{ERZ} Y.~Okada, M.~Yamaguchi, T.~Yanagida,
              {\em Prog.\ Theor.\ Phys. } {\bf 85} (1991) 1;\\
              J.~Ellis, G.~Ridolfi, F.~Zwirner,
              {\em Phys.\ Lett.} {\bf B 257} (1991) 83;\\
              H.~Haber, R.~Hempfling,
              {\em Phys.\ Rev.\ Lett.}  {\bf 66} (1991) 1815.

\bibitem{mhiggsf1lB} P.~Chankowski, S.~Pokorski, J.~Rosiek,
                     {\em Phys. Lett.} {\bf B 286} (1992) 307;
                     {\em Nucl. Phys.} {\bf B 423} (1994) 437
                     [arXiv:hep-ph/9303309].

\bibitem{deltamb} M.~Carena, D.~Garcia, U.~Nierste and C.~Wagner,
                  {\em Nucl. Phys.} {\bf B 577} (2000) 577
                  [arXiv:hep-ph/9912516];\\
                  H.~Eberl, K.~Hidaka, S.~Kraml, W.~Majerotto and
                  Y.~Yamada,
                  {\em Phys. Rev.} {\bf D 62} (2000) 055006
                  [arXiv:hep-ph/9912463].

\bibitem{deltamb1} T.~Banks, 
                   {\em Nucl.\ Phys.} {\bf B 303} (1988) 172;\\
                   L.~Hall, R.~Rattazzi and U.~Sarid,
                   {\em Phys.\ Rev.} {\bf D 50} (1994) 7048
                   [arXiv:hep-ph/9306309];\\
                   R.~Hempfling, 
                   {\em Phys.\ Rev.} {\bf D 49} (1994) 6168;\\
                   M.~Carena, M.~Olechowski, S.~Pokorski and C.~Wagner,
                   {\em Nucl.\ Phys.}\ {\bf B 426} (1994) 269
                   [arXiv:hep-ph/9402253].

\bibitem{mhiggsEP5} G.~Degrassi, A.~Dedes and P.~Slavich,
                    {\em Nucl. Phys.} {\bf B 672} (2003) 144
                    [arXiv:hep-ph/0305127].

\bibitem{fullEP2l} S.~Martin, 
                    {\em Phys. Rev.} {\bf D 65} (2002) 116003
                    [arXiv:hep-ph/0111209];
                    {\em Phys. Rev.} {\bf D 66} (2002) 096001
                    [arXiv:hep-ph/0206136];
                    Phys. Rev. {\bf D 67} (2003) 095012
                    [arXiv:hep-ph/0211366];
                    {\em Phys. Rev.} {\bf D 68} 075002 (2003)
                    [arXiv:hep-ph/0307101]; 
                    {\em Phys. Rev.} {\bf D 70} (2004) 016005
                    [arXiv:hep-ph/0312092];
                    {\em Phys. Rev.} {\bf D 71} (2005) 016012
                    [arXiv:hep-ph/0405022];
                    {\em Phys. Rev.} {\bf D 71} (2005) 116004
                    [arXiv:hep-ph/0502168];\\
                    S.~Martin and D.~Robertson,
                    {\em Comput.\ Phys.\ Commun.} {\bf 174} (2006) 133
                    [arXiv:hep-ph/0501132].

\bibitem{mhiggs3l} S.~Martin, 
                   {\em Phys.\ Rev.} {\bf D 75} (2007) 055005
                   [arXiv:hep-ph/0701051].

\bibitem{mhiggsFD3l} R.~Harlander, P.~Kant, L.~Mihaila and M.~Steinhauser,
                     arXiv:0803.0672 [hep-ph].

\bibitem{mhiggsFDalbals} S.~Heinemeyer, W.~Hollik, H.~Rzehak  and G.~Weiglein,
                    {\em Eur. Phys. J.} {\bf C 39} (2005) 465
                    [arXiv:hep-ph/0411114];
                    arXiv:hep-ph/0506254.

\bibitem{mhiggsWN} B.~Allanach, A.~Djouadi, J.~Kneur, W.~Porod and P.~Slavich,
                   {\em JHEP} {\bf 0409} (2004) 044
                   [arXiv:hep-ph/0406166].

\bibitem{tbexcl} S.~Heinemeyer, W.~Hollik and G.~Weiglein, 
                 {\em JHEP} {\bf 0006} (2000) 009
                 [arXiv:hep-ph/9909540].

\bibitem{LEPHiggsMSSM} LEP Higgs working group,
                       {\em Eur.\ Phys.\ J.} {\bf C 47} (2006) 547
                       [arXiv:hep-ex/0602042].

\bibitem{LEPHiggsSM} LEP Higgs working group,
                     {\em Phys. Lett.} {\bf B 565} (2003) 61
                     [arXiv:hep-ex/0306033].

\bibitem{asbs1} S.~Ambrosanio, A.~Dedes, S.~Heinemeyer, S.~Su and
                G.~Weiglein,
                {\em Nucl. Phys.} {\bf B 624} (2001) 3
                [arXiv:hep-ph/0106255].

\bibitem{ehow1} J.~Ellis, S.~Heinemeyer, K.~Olive and G.~Weiglein,
                {\em Phys. Lett.} {\bf  B 515 } (2001) 348
                [arXiv:hep-ph/0105061].

\bibitem{teslatdr} J.~Aguilar-Saavedra et al.,
                   TESLA TDR Part~3: 
                   ``Physics at an $e^+e^-$ Linear Collider'',
                   arXiv:hep-ph/0106315,
                   see: {\tt tesla.desy.de/tdr/} .

\bibitem{orangebook} T.~Abe et al.
                     [American Linear Collider Working Group Collaboration],
                     {\it Resource book for Snowmass 2001}, 
                     arXiv:hep-ex/0106055. 

\bibitem{acfarep} K.~Abe et al. 
                  [ACFA Linear Collider Working Group Collaboration],
                  arXiv:hep-ph/0109166.

\bibitem{Snowmass05Higgs} S.~Heinemeyer et al.,
                          arXiv:hep-ph/0511332.

\bibitem{bsgtheonew} M.~Misiak et al.,
                     {\em Phys.\ Rev.\ Lett.} {\bf 98} (2007) 022002 
                     [arXiv:hep-ph/0609232].

\bibitem{hulupo} T.~Hurth, E.~Lunghi and W.~Porod,
                 {\em Nucl.\ Phys.} {\bf B 704} (2005) 56
                 [arXiv:hep-ph/0312260].

\bibitem{bsgneubert} M.~Neubert,
                     {\em Eur. Phys. J.} {\bf C 40} (2005) 165
                     [arXiv:hep-ph/0408179].

\bibitem{bsgexp} R.~Barate et al.\ [ALEPH Collaboration],
                 {\em Phys. Lett.} {\bf B 429} (1998) 169;\\
                 S.~Chen et al.\ [CLEO Collaboration],
                 {\em Phys. Rev. Lett.} {\bf 87} (2001) 251807
                 [arXiv:hep-ex/0108032];\\
                 P.~Koppenburg et al.\ [Belle Collaboration],
                 {\em Phys. Rev. Lett.} {\bf 93} (2004) 061803
                 [arXiv:hep-ex/0403004];\\
                 K.~Abe et al.\ [Belle Collaboration],
                 {\em Phys. Lett.} {\bf B 511} (2001) 151
                 [arXiv:hep-ex/0103042];\\
                 B.~Aubert et al.\ [BABAR Collaboration],
                 arXiv:hep-ex/0207074;
                 arXiv:hep-ex/0207076.

\bibitem{bsgGH} P.~Cho, M.~Misiak and D.~Wyler,
                {\em Phys. Rev.} {\bf D 54}, 3329 (1996)
                [arXiv:hep-ph/9601360];\\
                A.~Kagan and M.~Neubert,
                {\em Eur. Phys. J.} {\bf C 7} (1999) 5
                [arXiv:hep-ph/9805303];\\
                A.~Ali, E.~Lunghi, C.~Greub and G.~Hiller,
                {\em Phys. Rev.}  {\bf D 66} (2002) 034002
                [arXiv:hep-ph/0112300];\\
                G.~Hiller and F.~Kr\"uger,
                {\em Phys. Rev.} {\bf D 69} (2004) 074020
                [arXiv:hep-ph/0310219];\\
                M.~Carena, D.~Garcia, U.~Nierste and C.~Wagner,
                {\em Phys. Lett.} {\bf B 499} (2001) 141
                [arXiv:hep-ph/0010003];\\
                D.~Demir and K.~Olive,
                {\em Phys.\ Rev.} {\bf D 65} (2002) 034007
                [arXiv:hep-ph/0107329];\\
                T.~Hurth, 
                arXiv:hep-ph/0212304. 

\bibitem{ali} K.~Adel and Y.~Yao,
                {\em Phys. Rev.} {\bf D 49} (1994) 4945
                [arXiv:hep-ph/9308349];\\
                C.~Greub, T.~Hurth and D.~Wyler,
                {\em Phys. Lett.} {\bf B 380} (1996) 385
                [arXiv:hep-ph/9602281];
                {\em Phys. Rev.} {\bf D 54} (1996) 3350
                [arXiv:hep-ph/9603404];\\
                A.~Ali,
                talk given at ICHEP04, Beijing, August 2004, appeared in
                the proceedings, 
                see: {\tt ichep04.ihep.ac.cn/db/paper.php} .

\bibitem{ali2}  K.~Chetyrkin, M.~Misiak and M.~M\"unz,
                {\em Phys. Lett.} {\bf B 400}, (1997) 206
                [Erratum-ibid.\ {\bf B 425} (1998) 414]
                [arXiv:hep-ph/9612313].

\bibitem{bsgMicro} G.~Belanger, F.~Boudjema, A.~Pukhov and A.~Semenov,
                   {\em Comput. Phys. Commun.} {\bf 149} (2002) 103
                   [arXiv:hep-ph/0112278];
                   arXiv:hep-ph/0405253.

\bibitem{bsgKO1} C.~Degrassi, P.~Gambino and G.~ Giudice,
                 {\em JHEP} {\bf 0012} (2000) 009
                 [arXiv:hep-ph/0009337].

\bibitem{bsgKO2} P.~Gambino and M.~Misiak,
                {\em Nucl. Phys.} {\bf B 611} (2001) 338
                [arXiv:hep-ph/0104034].

\bibitem{bsmmtheosm} G.~Buchalla and A.~Buras,
                     {\em Nucl. Phys.} {\bf B 400} (1993) 225;\\
                     M.~Misiak and J.~Urban,
                     {\em Phys. Lett.} {\bf B 451} (1999) 161
                     [arXiv:hep-ph/9901278];\\
                     G.~Buchalla and A.~Buras,
                     {\em Nucl. Phys.} {\bf B 548} (1999) 309
                     [arXiv:hep-ph/9901288];\\
                     A.~Buras,
                     {\em Phys. Lett.} {\bf B 566} (2003) 115
                     [arXiv:hep-ph/0303060].

\bibitem{bsmmexp}  [CDF Collaboration],
                   arXiv:0712.1708 [hep-ex];\\
                   K.~Tollefson
                   talk given at {\em Lepton Photon 07}, 
                   August 2007, Daegu, Korea, see:\\
                   {\tt chep.knu.ac.kr/lp07/htm/S4/S04\_14.pdf}~.

\bibitem{bsmmexpfuture} CDF Collaboration,
    see: {\tt www-cdf.fnal.gov/physics/projections/}~.

\bibitem{lhcb} P.~Ball et al.,
               arXiv:hep-ph/0003238.

\bibitem{bsmumu} K.~Babu and C.~Kolda,
                 {\em Phys. Rev. Lett.}  {\bf 84} (2000) 228
                 [arXiv:hep-ph/9909476];\\
                 S.~Choudhury and N.~Gaur,
                 {\em Phys. Lett.}  {\bf B 451} (1999) 86
                 [arXiv:hep-ph/9810307];\\
                 C.~Bobeth, T.~Ewerth, F.~Kr\"uger and J.~Urban, 
                 {\em Phys. Rev.}  {\bf D 64} (2001) 074014
                 [arXiv:hep-ph/0104284];\\
                 A.~Dedes, H.~Dreiner and U.~Nierste,
                 {\em Phys. Rev. Lett.}  {\bf 87} (2001) 251804
                 [arXiv:hep-ph/0108037];\\
                 G.~Isidori and A.~Retico,
                 {\em JHEP} {\bf 0111} (2001) 001
                 [arXiv:hep-ph/0110121];\\
                 A.~Dedes and A.~Pilaftsis,
                 {\em Phys. Rev.}  {\bf D 67} (2003) 015012
                 [arXiv:hep-ph/0209306];\\
                 A.~Buras, P.~Chankowski, J.~Rosiek and L.~Slawianowska,
                 {\em Nucl. Phys.}  {\bf B 659} (2003) 3
                 [arXiv:hep-ph/0210145];\\
                 A.~Dedes,
                 {\em Mod. Phys. Lett.}  {\bf A 18} (2003) 2627
                 [arXiv:hep-ph/0309233].

\bibitem{ourBmumu} J.~Ellis, K.~Olive and V.~Spanos,
                   {\em Phys.\ Lett.} {\bf B 624} (2005) 47
                   [arXiv:hep-ph/0504196].

\bibitem{atlastdr} ATLAS Collaboration,
        {\em Detector and Physics Performance Technical Design Report},
        CERN/LHCC/99-15 (1999),\\ see:
        {\tt atlasinfo.cern.ch/Atlas/GROUPS/PHYSICS/TDR/access.html} .

\bibitem{cmstdr} CMS Collaboration,
        {\em Physics Technical Design Report, Volume 2. CERN/LHCC
          2006-021}, 
        see: {\tt cmsdoc.cern.ch/cms/cpt/tdr/} .

\bibitem{jakobs} V.~B\"uscher and K.~Jakobs,
                 {\em Int.\ J.\ Mod.\ Phys.} {\bf A 20} (2005) 2523
                 [arXiv:hep-ph/0504099].

\bibitem{schumi} M.~Schumacher,
                 {\em Czech. J. Phys.} {\bf 54} (2004) A103;
                 arXiv:hep-ph/0410112.

\bibitem{cmshiggsOrg} S.~Abdullin et al.,
                      {\em Eur. Phys. J.} {\bf C 39S2} (2005) 41.

\bibitem{cmsHiggs} S.~Gennai, S.~Heinemeyer, A.~Kalinowski, R.~Kinnunen, 
                   S.~Lehti, A.~Nikitenko and G.~Weiglein,
                   {\em Eur.\ Phys.\ J.} {\bf C 52} (2007) 383
                   [arXiv:0704.0619 [hep-ph]];\\
                   M.~Hashemi, S.~Heinemeyer, R.~Kinnunen, 
                   A.~Nikitenko and G.~Weiglein,
                   arXiv:0804.1228 [hep-ph].

\bibitem{AMSBpheno} J.~Feng, T.~Moroi, L.~Randall, M.~Strassler and S.~Su,
                    {\em Phys.\ Rev.\ Lett.} {\bf 83} (1999) 1731
                    [arXiv:hep-ph/9904250];\\
                    S.~Asai, T.~Moroi and T.~Yanagida,
                    arXiv:0802.3725 [hep-ph].

\bibitem{slhc} F.~Gianotti et al.,
               {\em Eur.\ Phys.\ J.} {\bf C 39} (2005) 293
               [arXiv:hep-ph/0204087].

\bibitem{WMAPstrips} J.~Ellis, K.~Olive, Y.~Santoso and V.~Spanos,
                     {\em Phys. Lett.} {\bf B 565} (2003) 176
                     [arXiv:hep-ph/0303043].
                     
\bibitem{wmapothers} U.~Chattopadhyay, A.~Corsetti and P.~Nath,
                     {\em Phys. Rev.} {\bf D 68} (2003) 035005
                     [arXiv:hep-ph/0303201];\\
                     H.~Baer and C.~Balazs,
                     {\em JCAP} {\bf 0305} (2003) 006
                     [arXiv:hep-ph/0303114];\\
                     A.~Lahanas and D.~Nanopoulos,
                     {\em Phys. Lett.} {\bf B 568} (2003) 55
                     [arXiv:hep-ph/0303130];\\
                     R.~Arnowitt, B.~Dutta and B.~Hu
                     arXiv:hep-ph/0310103.

\bibitem{deschi} K.~Desch, E.~Gross, S.~Heinemeyer, G.~Weiglein and 
                 L.~Zivkovic,
                 {\em JHEP} {\bf 0409} (2004) 062
                 [arXiv:hep-ph/0406322].


\end{thebibliography}
\end{document}
